\documentclass[aps,preprint,superscriptaddress,nofootinbib,preprintnumbers,prd,eqsecnum]{revtex4-1}

\usepackage{bm,amsmath,amsfonts,amssymb,slashed,graphicx,xspace,placeins,rotating}
\usepackage{xcolor,colortbl}
\usepackage{multirow,array,makecell}
\usepackage[inline]{enumitem}
\usepackage{subfigure}
\allowdisplaybreaks
\DeclareMathAlphabet{\mathpzc}{OT1}{pzc}{m}{it}

\definecolor{nicered}{rgb}{0.7,0.1,0.1}
\definecolor{nicegreen}{rgb}{0.1,0.5,0.1}
\definecolor{niceblue}{rgb}{0.1,0.1,0.5}
\usepackage[bookmarks=false]{hyperref}
\hypersetup{colorlinks,citecolor= nicegreen,linkcolor= niceblue,urlcolor= niceblue}
\usepackage{breakurl}

\newcommand{\nn}{\nonumber}
\newcommand{\GeV}{\text{GeV}}

\newcommand{\g}{\gamma}

\newcommand{\ab}{\alpha\beta}
\newcommand{\ceq}{\stackrel{\to}{=}}

\newcommand{\mubc}{\mu_{bc}}

\newcommand{\genbar}[1]{\,\overline{\!#1}{}}
\newcommand{\Overrightarrow}[1]{{%
	#1
}}
\newcommand{\Overleftarrow}[1]{{%
	#1
}}

\newcommand{\Dx}{D^{(*)}}
\newcommand{\Bx}{B^{(*)}}
\newcommand{\bbar}{\bar{b}}
\newcommand{\cbar}{\bar{c}}
\newcommand{\Bbar}{\genbar{B}}

\newcommand{\Bxbar}{\Bbar^{(*)}}

\newcommand{\cbvp}{\cbar^{v'}_+}
\newcommand{\cvp}{c^{v'}_+}
\newcommand{\bv}{b^v_+}
\newcommand{\bbv}{\bbar^v_+}

\newcommand{\Dslash}{\slashed{D}}
\newcommand{\vslash}{\slashed{v}}
\newcommand{\ccdot}{\!\cdot\!}
\newcommand{\vcD}{v \ccdot D}
\newcommand{\Qbar}{\genbar{Q}}
\newcommand{\Jbar}{\genbar{J}}
\newcommand{\mJ}{\mathcal{J}}
\newcommand{\mJbar}{\genbar{\mJ}}
\newcommand{\Hc}{H_c}
\newcommand{\Hb}{H_b}

\newcommand{\Hbar}{\genbar{H}}
\newcommand{\Pbar}{\genbar{P}}
\newcommand{\Vbar}{\genbar{V}}

\newcommand{\Cv}[1]{C_{V#1}}
\newcommand{\Ca}[1]{C_{A#1}}

\newcommand{\hL}[2]{\hat{L}_{#1}^{(#2)}}
\newcommand{\hM}[1]{\hat{M}_{#1}}
\newcommand{\hDg}[1]{\hat{\Delta}_{#1}}

\newcommand{\lqcd}{\Lambda_{\text{QCD}}}
\newcommand{\rD}{r_{D}}
\newcommand{\rDs}{r_{D^*}}
\newcommand{\LamB}{\bar{\Lambda}}
\newcommand{\lam}[1]{\lambda_{#1}}
\newcommand{\aS}{\alpha_s}
\newcommand{\haS}{{\hat{\alpha}_s}}
\newcommand{\ec}{\varepsilon_c}
\newcommand{\eb}{\varepsilon_b}
\newcommand{\eQ}{\varepsilon_Q}

\newcommand{\quot}{\natural}
\newcommand{\hvph}{\hat\varphi_1}
\newcommand{\hvphq}{\hat\varphi^\quot_1}
\newcommand{\hvphp}{\hat\varphi'_1}

\newcommand{\mbS}{m_b^{1S}}
\newcommand{\dmbc}{\delta m_{bc}}

\newcommand{\amp}[3]{\frac{\langle #1(p') |\, #2\, | #3(p) \rangle}{\sqrt{m_{\let\overline\relax#3} m_{#1}}}}

\newcommand{\rhossq}{\rho_*^2}
\newcommand{\cs}{c_*}
\newcommand{\trhossq}{\tilde{\rho}_*^2}
\newcommand{\tcs}{\tilde{c}_*}

\newcommand{\nax}{\text{---}}

\newcommand{\Tr}{\text{Tr}}
\newcommand{\sqrtb}[1]{\sqrt{\smash[b]{#1}}\vphantom{b}}

\newcommand{\BaseFit}[1][]{$L^{D;D^*}_{w\ge1;=1}${#1}\xspace}
\newcommand{\ShapeFit}{$L^{D;D^*}_{w\ge1;=1}$Shape\xspace}
\newcommand{\AllLQCDFit}{$L^{D;D^*}_{w\ge1;\ge1}$\xspace}
\newcommand{\NoLQCDFit}{NoL\xspace}
\newcommand{\hAOneFit}{$L^{D;D^*}_{w\ge1;\ge1[h_{A1}]}$\xspace}

\makeatletter
\g@addto@macro\bfseries{\boldmath}
\makeatother

\makeatletter
\let\tmp@footnote\footnote
\renewcommand{\footnote}[1]{\tmp@footnote{\linespread{0.9}\selectfont{}#1}}
\makeatother

\makeatletter
\let\temp@caption\caption
\renewcommand{\caption}[2][]{\temp@caption[#1]{\linespread{1.2}\selectfont{}#2}}
\makeatother

\allowdisplaybreaks
\begin{document}

\preprint{CALT-TH-2022-022}

\title{Constrained second-order power corrections in HQET: \\ $R(D^{(*)})$, $|V_{cb}|$, and new physics}

\author{Florian U.\ Bernlochner}
\affiliation{Physikalisches Institut der Rheinischen Friedrich-Wilhelms-Universit\"at Bonn, 53115 Bonn, Germany}

\author{Zoltan Ligeti}
\affiliation{Ernest Orlando Lawrence Berkeley National Laboratory, 
University of California, Berkeley, CA 94720, USA}
\affiliation{Berkeley Center for Theoretical Physics, 
Department of Physics,
University of California, Berkeley, CA 94720, USA}

\author{Michele Papucci}
\affiliation{Walter Burke Institute for Theoretical Physics, California Institute of Technology, Pasadena, CA 91125, USA}

\author{Markus T.\ Prim}
\affiliation{Physikalisches Institut der Rheinischen Friedrich-Wilhelms-Universit\"at Bonn, 53115 Bonn, Germany}

\author{Dean J.\ Robinson}
\affiliation{Ernest Orlando Lawrence Berkeley National Laboratory, 
University of California, Berkeley, CA 94720, USA}
\affiliation{Berkeley Center for Theoretical Physics, 
Department of Physics,
University of California, Berkeley, CA 94720, USA}

\author{Chenglu Xiong}
\affiliation{Physikalisches Institut der Rheinischen Friedrich-Wilhelms-Universit\"at Bonn, 53115 Bonn, Germany}

\begin{abstract}
We postulate a supplemental power counting within the heavy quark effective theory, 
that results in a small, highly-constrained set of second-order power corrections,
compared to the standard approach. 
We determine all $\bar{B} \to D^{(*)}$ form factors, 
both within and beyond the standard model to $\mathcal{O}(\alpha_s/m_{c,b}, 1/m_{c,b}^2)$, under truncation by this power counting.
We show that the second-order power corrections to the zero-recoil normalization of the $\bar{B} \to D^{(*)} l \nu$ 
matrix elements ($l = e$, $\mu$, $\tau$) are fully determined by hadron mass parameters,
and are in good agreement with lattice QCD (LQCD) predictions.
We develop a parametrization of these form factors under the postulated truncation, 
that achieves excellent fits to the available LQCD predictions and experimental data,
and we provide precise updated predictions for the $\bar{B} \to D^{(*)} \tau \bar\nu$ decay rates, 
lepton flavor universality violation ratios $R(D^{(*)})$, and the CKM matrix element $|V_{cb}|$.
We point out some apparent errors in prior literature concerning the $\mathcal{O}(1/m_cm_b)$ corrections,
and note a tension between commonly-used simplified dispersive bounds and current data.
\end{abstract}

\maketitle

\twocolumngrid
{
\fontsize{10}{8}\selectfont 
\columnsep20pt
	\tableofcontents
}
\onecolumngrid

\section{Introduction}

The Heavy Quark Effective Theory (HQET)~\cite{Isgur:1989vq, Isgur:1989ed, Eichten:1989zv,Georgi:1990um} 
underpins key foundations in our theoretical understanding of exclusive semileptonic $b \to c l \bar\nu$ decays ($l = e$, $\mu$, $\tau$).
HQET allows for a hadronic model-independent and high-precision
determination of the CKM matrix element $|V_{cb}|$ from fits to exclusive semileptonic decay measurements to light leptons, $\Bbar \to \Dx \ell \bar\nu$ ($\ell = e$, $\mu$).
Furthermore, one may obtain model-independent precision predictions for ratios testing lepton flavor universality violation (LFUV),
\begin{equation}
	R(\Dx) = \frac{\Gamma[\Bbar \to \Dx \tau \bar\nu]}{\Gamma[\Bbar \to \Dx \ell \bar\nu]}\,,
\end{equation}
both within and beyond the standard model (SM). The current HFLAV arithmetic-averaged SM predictions are $R(D) = 0.299(3)$ and $R(D^*) = 0.258(5)$~\cite{Amhis:2019ckw, Bigi:2016mdz, Bernlochner:2017jka, Bigi:2017jbd, Jaiswal:2017rve}.

These ratios have provided tantalizing hints for lepton flavor universality violation over the past decade: 
When combined, they currently exhibit $3\sigma$~\cite{Amhis:2019ckw} 
(or more~\cite{Bernlochner:2021vlv}) tension with SM predictions.
Anticipating future measurement precision at the percent level for $R(\Dx)$ 
(see Ref.~\cite{Bernlochner:2021vlv} for a review), similarly precise SM predictions are warranted.
Moreover, $|V_{cb}|$ recovered from exclusive $\Bbar \to D^*\ell\bar\nu$ measurements currently exhibits 
a $3\sigma$ tension compared to the measured value from inclusive $\Bbar \to X_c l \bar\nu$ decays~\cite{Amhis:2019ckw},
with the magnitude of the deviation near the $\mathcal{O}(10\%)$ level (also see~\cite{Bernlochner:2022ucr}).
Because the extraction of $|V_{cb}|$ (currently) relies on extrapolation to the zero-recoil point, 
at which the hadron velocities are equal and phase space vanishes, 
the exclusive measurement of $|V_{cb}|$ is particularly sensitive to the parametrization of the $\Bbar \to \Dx$ form factors.

In the SM, $\Bbar \to D$ ($\Bbar \to D^*$) transitions are described by two (four) form factors, for a total of six.
(There are 20 form factors for $\Bx  \to \Dx$ decays in the SM and $34$ if one further includes all possible NP interactions.)
The $\mathcal{O}(1/m_{c,b})$ and $\mathcal{O}(\aS)$ HQET corrections to the $\Bbar \to \Dx$ matrix elements in the SM
have been known for three decades~\cite{Luke:1990eg, Neubert:1991xw, Neubert:1993mb,Manohar:2000dt},
and explicit $\mathcal{O}(1/m_{c,b},\aS)$ results for all NP matrix elements were recently derived~\cite{Bernlochner:2017jka}.
At first order in the heavy quark (HQ) expansion, six first-order wavefunctions are described by three subleading Isgur-Wise functions, 
while the one-loop $\mathcal{O}(\aS)$ perturbative corrections are calculable
(see, e.g., Ref.~\cite{Bernlochner:2017jka} for their closed-form expressions).
The $\mathcal{O}(1/m^2_{c,b})$ corrections have also been known for three decades~\cite{Falk:1992wt}.
Considering $\mathcal{O}(1/m^2_{c})$ alone (all corrections at $\mathcal{O}(1/m^2_{c,b})$), 
$6$ ($30$) possible second-order wavefunctions
are described by an overcomplete basis of $20$ ($32$) subsubleading Isgur-Wise functions
(hereafter, $m_{c,b}^2$ denotes $m_c^2$, $m_b^2$, and $m_cm_b$).
The counting is summarized in Table~\ref{tab:HQETFFs}.
Finally, the $\mathcal{O}(\aS \times 1/m_{c,b})$ corrections are also long known 
(see, e.g., Ref.~\cite{Neubert:1993mb} for a review).

\begin{table}[b!]
\renewcommand*{\arraystretch}{0.9}
\newcolumntype{C}{ >{\centering\arraybackslash } m{2.5cm} <{}}
\begin{tabular}{ccCCC}
	\hline\hline
	\multirow{2}{*}{HQET order} & \multirow{2}{*}{\makecell[c]{Fixed-order \\[-5pt] wavefunctions}} & \multicolumn{3}{c}{Isgur-Wise functions} \\
	&& All& RC Expansion & VC Limit\\
	\hline
	$1/m_{c,b}^0$ 	 & 1 & 1 & 1 & 1\\
	$1/m_{c,b}^1$	  & 6 & 3 & 3 & 1\\
	$1/m_c^2$  & 6 & 20  & 1 & 3\\
	$1/m_{c,b}^2$ & 30 & 32 & 3 & 4\\
 	\hline\hline
\end{tabular}
\caption{Number of $\Bbar \to \Dx$ form factors and Isgur-Wise functions entering at each fixed order in HQET. 
The residual chiral (RC) expansion is developed in this work,
which we compare to the vanishing chromomagnetic (VC) limit. 
(For details of the counting rules see Sec.~\ref{sec:summff}.)}
\label{tab:HQETFFs}
\end{table}

The expected size of the $\mathcal{O}(1/m_c)$, $\mathcal{O}(1/m_b)$, and $\mathcal{O}(\alpha_s)$ corrections are about $20\%$, $5\%$, and $10\%$, respectively
(as the approximate small parameter is $\sim \lqcd/m_{c,b}$ or $\sim \aS/\pi$).
The second-order corrections at $\mathcal{O}(1/m^2_{c})$ can also be expected to contribute at the $\sim 5\%$ level.
Moreover, in the zero-recoil limit
the $\Bbar \to D^*$ form factor, 
$\mathcal{F}(1)$ (defined below), has vanishing first-order corrections,
but its resulting value at $\mathcal{O}(\aS, 1/m_{c,b})$ differs at the $5\%$ level from lattice QCD (LQCD) predictions~\cite{Aoki:2021kgd}.
These observations lead to the following possibilities:
(a) given that second-order or higher corrections must fix the $\mathcal{F}(1)$ tension, 
it is possible the HQ expansion of the 
$\Bbar \to D^{(*)}$ matrix elements could be `badly behaved', such that $1/m_c^2$ terms may be unexpectedly large;
or (b), while second-order power corrections must be important at zero recoil because of the vanishing first-order corrections at the phase space point,
they are otherwise subdominant and the data beyond zero recoil will be predominantly described by first-order corrections.
The latter is the approach used in Ref.~\cite{Bernlochner:2017jka},
that performed the first combined and self-consistent analysis of $\Bbar \to \Dx \ell \nu$ decays at $\mathcal{O}(\aS, 1/m_{c,b})$.
In this approach, only the shape of the differential distributions was used to constrain the subleading Isgur-Wise functions.

Recent analyses that attempt to quantify the effect of second-order power corrections~\cite{Jaiswal:2017rve, Bordone:2019vic,Bordone:2019guc}
treat the six $\mathcal{O}(1/m_c^2)$ wavefunctions (or a subset of them) as nuisance parameters in phenomenological fits.
These analyses further make use of theoretical inputs from model-dependent calculations, 
such as QCD sum rules (QCDSR) or light cone sum rules (LCSR).
Such studies typically estimated that the HQ expansion appears well-behaved at $\mathcal{O}(1/m^2_{c})$.
In addition, the constrained structure of the HQET for $\Lambda_b \to \Lambda_c l \nu$ decay 
has permitted its far simpler $\mathcal{O}(1/m_{c}^2)$ contributions~\cite{Falk:1992ws} to be extracted
from combined fits to data and LQCD results~\cite{Bernlochner:2018kxh, Bernlochner:2018bfn}.
These were found to be compatible with a well-behaved HQ expansion, too.

In this paper, we propose a supplemental power counting within the HQ expansion 
that allows one to truncate, in a well-defined manner, the number of subsubleading Isgur-Wise functions to $3$ instead of $32$ in the $\Bbar \to \Dx$ system,
and to just a single subsubleading Isgur-Wise function at $\mathcal{O}(1/m_{c}^2)$ alone.
This approach provides a predictive set of second-order power corrections in $B \to \Dx l \nu$ decays
that can be tested with available data,
without using additional model-dependent QCDSR or LCSR inputs.
Specifically, we show that one may formally
power count in insertions of the transverse residual momentum operator of the HQET mass-subtracted states, $\slashed{D}_\perp$,
and we develop a conjecture that terms entering at third order or higher in this power counting should be suppressed.
We refer to the resulting expansion as the `residual chiral' (RC) expansion.

The goal of this work is to derive the set of $\mathcal{O}(\aS \times 1/m_{c,b}, 1/m_{c,b}^2)$ corrections 
to the $\Bbar \to \Dx$ form factors, in both the SM and beyond, under truncation at second order in the RC expansion.
As a point of comparison to the RC expansion, we also consider the vanishing chromomagnetic (VC) interaction
limit~\cite{Neubert:1992hb, Neubert:1992wq, Neubert:1992pn, Falk:1992wt},
which also dramatically simplifies the number of subleading and subsubleading Isgur-Wise functions.
We then confront the RC expansion and VC limit results with available experimental measurements and LQCD predictions,
and obtain precise results for $|V_{cb}|$ and SM predictions for $R(\Dx)$.

These objectives require the assembly of a wide range of theoretical and phenomenological components.
First, in order to ensure self-consistent conventions 
we carefully (re)develop the formal elements of the general HQ expansion
that are required when working at second order,
and then develop the RC conjecture,
showing how it constrains and simplifies the structure of the power corrections
(Sec.~\ref{sec:RCexp} and Apps~\ref{app:hadmass}, \ref{app:SDrel}, and \ref{app:genradcorr}).
Second, we proceed to apply this to the $\Bbar \to \Dx$ system, 
deriving the corrections up to and including $\mathcal{O}(\aS \times 1/m_{c,b}, 1/m_{c,b}^2)$
under truncation by the RC expansion,
incorporating zero-recoil and normalization constraints and removing redundant higher-order terms
(Sec.~\ref{sec:BDDsFFs} and Apps.~\ref{app:firstsecondorder} and~\ref{app:asmcb}).
We note apparent sign errors or inconsistencies for several $\mathcal{O}(1/m_cm_b)$ wavefunctions derived in Ref.~\cite{Falk:1992wt}.
Third, we construct a parametrization of these corrections, 
implementing the $1S$ short-distance mass scheme for heavy quark masses 
and an analytic structure for the leading Isgur-Wise function that respects the HQ expansion at second order (Sec.~\ref{sec:parampres}).
These results are encoded in the \texttt{Hammer} library~\cite{bernlochner_florian_urs_2022_5828435, Bernlochner:2020tfi}.
We show that the tremendous simplification of the second-order power corrections under the RC expansion constrains most zero-recoil corrections
to be a combination of the hadron mass parameters, $\lam{1}$ and $\lam{2}$.
We investigate the zero-recoil predictions for various form-factors and their ratios, and 
find that the value of $\mathcal{F}(1)$ is in good agreement with LQCD results (Sec.~\ref{sec:zrp}).
Finally, in Sec.~\ref{sec:fitres}, 
the parametrizations of both the RC expansion and the VC limit are fitted against all available experimental measurements and LQCD data,
examining various fit scenarios that consider different combinations of experimental and LQCD inputs and different assumptions.
The latter includes fits that truncate at lower order in HQET, 
and fits that constrain only the shape of the distributions, as done in Ref.~\cite{Bernlochner:2017jka}.
To properly identify optimal parameter subsets that describe the data 
and avoid potential overfitting, we employ a nested hypothesis test (NHT) prescription.

Several important observations ensue from our analysis:
(i) Under the NHT prescription, we identify optimal parameter sets for the RC expansion 
including $\mathcal{O}(1/m_c^2)$ terms,
that achieve excellent agreement with the data with relatively few parameters, and without using any QCDSR (or LCSR) inputs.
We obtain
\begin{equation}
		R(D) = 0.288(4)	\,, \quad R(D^*) = 0.249(3)\,, \quad 
		|V_{cb}| = 38.7(6)\times 10^{-3}\,.
\end{equation}
These results can be compared to other recent results~\cite{Bordone:2019vic, BaBar:2019vpl,Gambino:2019sif, Iguro:2020cpg, Bazavov:2021bax,Aoki:2021kgd}.
(ii) While the inclusion of zero-recoil second-order power corrections in the RC expansion is crucial to good fits, 
the inclusion of second-order power corrections beyond zero recoil is not.
This supports the approach used in Ref.~\cite{Bernlochner:2017jka};
(iii) The slope-curvature relation developed by Ref.~\cite{Caprini:1997mu} is in tension with the data,
and leads to large upward biases in $R(D)$;
(iv) The VC limit, in contrast to the RC expansion, produces poor fits because of its structure at zero recoil,
but using only shape information yields good fits.

\section{The residual chiral expansion}
\label{sec:RCexp}

\subsection{General HQET preliminaries}
\label{sec:HQE}

The standard construction of HQET follows from a reorganization 
of the QCD Lagrangian for a heavy quark $Q$ with mass $m_Q$, 
in terms of the mass-subtracted fields 
\begin{equation}
	\label{eqn:masssub}
	Q^v_{\pm}(x) = e^{im_Q v \cdot x}\, \Pi_{\pm} Q(x)\,.
\end{equation}
The parameter $v$ is a heavy quark velocity---defined up to reparametrization freedom~\cite{Luke:1992cs} via $p_Q = m_Q v + k$, 
in which $k \sim \lqcd$ is a residual momentum---and 
the projectors $\Pi_{\pm} = (1 \pm \vslash)/2$. 
This yields
\begin{equation}
	\label{eqn:fullL}
	\mathcal{L}_{\text{QCD}} 
	= \Qbar^v_+ i \vcD Q^v_+ + \Qbar^v_+ i \Dslash_\perp Q^v_-  
	+ \Qbar^v_- i \Dslash_\perp Q^v_+ - \Qbar^v_- (i \vcD + 2m_Q) Q^v_-\,.
\end{equation}
Here $D^\mu$ is a gauge covariant derivative of QCD, 
and the transverse derivative $D^\mu_\perp = D^\mu - (\vcD)v^\mu$.
Because of the mass subtraction in the phase of $Q \sim e^{-ip_Q\cdot x}$, the derivative $D \sim k \sim \lqcd$,
so that in the heavy quark regime, $m_Q \gg \lqcd$, one may integrate out the double heavy field $Q^v_-$ 
yielding an effective theory for the light field $Q_+^v$ with order-by-order corrections in $1/m_Q$.
This HQET Lagrangian reads
\begin{equation}
	\label{eqn:eftL}
	\mathcal{L}_{\text{HQET}} = \Qbar^v_+ i \vcD Q^v_+ +  \Qbar^v_+ i\Dslash_\perp \frac{1}{i \vcD + 2m_Q}  i\Dslash_\perp Q^v_+\,.
\end{equation}
Writing $\mathcal{L}_{\text{HQET}}  = \sum_{n = 0}\mathcal{L}_n/(2m_Q)^n$ to second order,
\begin{subequations}
\begin{align}
	\mathcal{L}_0 & = \Qbar^v_+ i \vcD Q^v_+ \,,\\
	\mathcal{L}_1 & = -\Qbar^v_+ \Dslash_{\perp} \Dslash_{\perp} Q^v_+  =  -\Qbar^v_+ \bigg[D^2 + a_Q(\mu)\frac{g}{2} \sigma_{\ab} G^{\ab} \bigg]Q^v_+ \,,\\
	\mathcal{L}_2 & =  \Qbar^v_+  [\Dslash_{\perp} i\vcD \Dslash_{\perp} ] Q^v_+  =  g\Qbar^v_+ \bigg[ v_\beta  D_\alpha G^{\ab}  - i v_\alpha \sigma_{\beta\gamma} D^\gamma G^{\ab} \bigg]Q^v_+ \,.
\end{align}
\end{subequations}	
Here the field strength $ig\, G^{\ab} = [D^\alpha, D^\beta]$, $\sigma^{\ab} \equiv \frac{i}{2}[\g^\alpha, \g^\beta]$, 
and we have made use of the equation of motion, $i \vcD\, Q_+^v  = 0$ in the free effective theory.\footnote{%
In the full effective theory, the equation of motion receives corrections, 
such that beyond leading order $i \vcD Q_+^v  \simeq 1/(2m_Q)\Dslash_\perp \Dslash_\perp Q_+^v + \ldots$.
As usual in any perturbation theory, consistent power counting in $1/m_Q$ mandates use of the free equation of motion at each order in $1/m_Q$
(see e.g. \cite{Politzer:1980me}).}
The coefficient of the $\mathcal{L}_1$ chromomagnetic operator $a_Q(\mu)$ is renormalized by the strong interactions,
where $\mu$ is an arbitrary matching scale of QCD onto HQET.
Its deviation from unity is important when considering corrections at $\mathcal{O}(\aS/m_{Q})$ and higher. 
Therefore it can be consistently neglected everywhere except when discussing the $\mathcal{O}(\aS/m_Q)$ radiative corrections, 
for which we use 
\begin{equation}
	\label{eqn:aQ}
	a_Q(\mu) \equiv 1 + \frac{\aS}{\pi} \, C_g^Q(\mu)\,,
\end{equation}
with $C_g^Q(\mu) = -(3/2)\, [\ln(m_Q/\mu)-13/9]$~\cite{Eichten:1990vp} 
(see also Refs.~\cite{Falk:1990pz, Neubert:1993mb}),
and except for explicit evaluation of the $\lam2$ parameter 
(see Eqs.~\eqref{eqn:deflam12} and \eqref{eqn:masslam12}).
The renormalization of the coefficients of the $\mathcal{L}_2$ terms can be neglected at $\mathcal{O}(\aS/m_Q)$.

At any order in $1/m_Q$, one may compute Lagrangian corrections 
to a particular HQET correlator via an operator product involving the $\mathcal{L}_i$.
In addition, the quark source term $\Jbar Q$ for a QCD correlator 
can be expressed with respect to mass-subtracted quantities via
$\Jbar Q = \Jbar^v (Q_+ + Q_-) \equiv \Jbar^v \mJ_{\text{HQET}} Q^v_+$, with $J^v = e^{im_Q v\cdot x}J$. 
The time-ordered correlators of full QCD will then match onto time-ordered HQET correlators 
determined by functional derivatives with respect to $J^v$, order-by-order in $1/m_Q$.
Applying the equations of motion from Eq.~\eqref{eqn:fullL}, 
the source term becomes
\begin{equation}
	\label{eqn:JQexp}
	\Jbar Q  = \Jbar^v \bigg[1 + \frac{1}{i \vcD + 2m_Q}i\Dslash_\perp \bigg] Q^v_+  =  \Jbar^v \bigg[1 + \Pi_-\bigg(\frac{i\Dslash}{2m_Q} - \frac{\Dslash\Dslash}{4m_Q^2} + \ldots\bigg)\bigg] Q^v_+\,, 
\end{equation}
keeping terms to second order, 
and noting the second order term arises via $\Pi_- \vcD \Dslash_\perp Q_+^v = -\Pi_- \Dslash\Dslash Q_+^v$ 
because of the $Q_+^v$ equation of motion.
Expanding the current factor 
\begin{equation}
	\mJ_{\text{HQET}} = 1 + \Pi_-\sum_{n = 1}\mJ_n/(2m_Q)^n\,,
\end{equation} 
then $\mJ_1 = i \Dslash$, and $\mJ_2 = -\Dslash \Dslash$.
We define the conjugate 
\begin{equation}
	\mJbar_n \equiv \gamma^0 \overleftarrow{\mJ}_n^\dagger \gamma^0\,.
\end{equation}
Here $\overleftarrow{\mJ}$ ($\overrightarrow{\mJ}$) indicates action of the derivatives to the left and right, respectively:
it is always the case that $\mJbar$ ($\mJ$) acts to the left (right).
 
In this work, we are interested in computing the matrix elements for exclusive heavy-quark hadron transitions, via matching onto HQET.
For a $b \to c$ transition (i.e., $Q = c$ or $b$) involving hadrons $\Hb \to \Hc$, 
the QCD matrix elements of interest are $\langle \Hc | \cbar \,\Gamma\, b | \Hb \rangle$,
where $\Gamma$ is any Dirac matrix.
The matching onto HQET corresponds to equating the QCD matrix element to the path integral computed in HQET,
\begin{align}
	\frac{\langle \Hc | \cbar \,\Gamma\, b | \Hb \rangle}{\sqrt{m_{\Hc} m_{\Hb}}}
	& = \big\langle \Hc^{v'} \big|\frac{1}{\mathcal{Z}}\int \mathcal{D} \cbvp \mathcal{D} c^{v'}_+ \mathcal{D} \bbar^{v}_+ \mathcal{D} \bv \nn \\*
	& \quad \times 
	\exp\bigg\{i \!\int\! d^4x \big[\mathcal{L}'_{\text{HQET}} + \mathcal{L}_{\text{HQET}} \big](x) \bigg\} 
	\cbvp \mJbar_{\text{HQET}}^{\prime} \Gamma \mJ_{\text{HQET}} \bv\big| \Hb^v \big\rangle\,, \label{eqn:QCDmatch}
\end{align}
in which $\mathcal{Z}$ is the partition function of the free theory generated by $\mathcal{L}_0$.
As is the usual practice, we use a notational convention
that labels charm parameters with primes while beauty parameters are unprimed.
At any order in $1/m_{c,b}$, one may read off from Eq.~\eqref{eqn:QCDmatch} 
the HQET correlators that contribute to the hadronic matrix element.
Note the corrections to the source term induce corrections to the HQET current $\cbvp \Gamma \bv$: 
the current corrections.

A particularly important application of Eq.~\eqref{eqn:QCDmatch} is the matching 
of the QCD correlator involving the HQET Hamiltonian $\langle H \big| \Qbar^v_+ i \vcD Q^v_+ \big| H \rangle$ onto HQET,
from which one may derive the hadron mass expansion,
\begin{equation}
	\label{eqn:HmassHQE}
	m_{H} = m_Q + \LamB + \frac{\Delta m^{H}_2}{2m_Q} + \ldots\,, \qquad \Delta m^{H}_2 = - \lam1 - d_{H}a_Q(\mu)\lam2(\mu)\,.
\end{equation}
In Appendix~\ref{app:hadmass} we present this derivation from first principles,
including precise definitions of the HQ mass parameters $\LamB$, $\lam1$ and $\lam2$
and pertinent conventions used in this work,
that are important to a self-consistent derivation of the second-order power corrections.
In Eq.~\eqref{eqn:HmassHQE} we have explicitly restored the renormalization factor for the chromomagnetic operator.
For a pseudoscalar ($P$) and vector ($V$) meson,
furnishing a heavy quark doublet with brown muck spin-parity $s^{\pi_\ell} = \frac{1}{2}^-$, 
the factor $d_P = 3$ and $d_V = -1$, respectively.

The HQET eigenstates of $\mathcal{L}_0$, $|H^v\rangle$, are normalized 
such that 
\begin{equation}
	\label{eqn:HQETnorm}
	\langle H^{v'}(k') | H^v(k) \rangle = 2v^0 \delta_{v v'}(2\pi)^3 \delta^3(k-k')\,.
\end{equation}
Note this normalization choice differs from that in Ref.~\cite{Falk:1992wt}, which normalized the HQET states with respect to an HQ mass scale, $m_Q + \LamB$.
Similarly the matching~\eqref{eqn:QCDmatch} is defined with respect to normalized QCD states.
To second-order, Eq.~\eqref{eqn:QCDmatch} then becomes
\begin{align}
	\frac{\langle \Hc | \cbar \,\Gamma\, b | \Hb \rangle}{\sqrt{m_{\Hc} m_{\Hb}}}  
	& \simeq \big\langle \Hc^{v'} \big| \cbvp \, \Gamma \, \bv \big| \Hb^v \big\rangle  \label{eqn:QCDmatchexp}\\
	& + \frac{1}{2m_c} \big\langle \Hc^{v'} \big|  \big(\cbvp\Overleftarrow{\mJbar}^{\prime}_1+ \mathcal{L}'_1 \circ \cbvp \big) \Gamma \, \bv \big| \Hb^v \big\rangle \nn \\
	& + \frac{1}{2m_b} \big\langle \Hc^{v'} \big| \cbvp  \, \Gamma  \big(\Overrightarrow{\mJ}_1\bv + \bv \circ \mathcal{L}_1 \big)  \big| \Hb^v \big\rangle \nn \\
	& + \frac{1}{4m_c^2} \big\langle \Hc^{v'} \big| \big(\cbvp\Overleftarrow{\mJbar}^{\prime}_2\Pi_-' + \mathcal{L}'_2 \circ \cbvp 
		+ \mathcal{L}'_1 \circ \cbvp \Overleftarrow{\mJbar}^{\prime}_1\Pi_-'
		+ \frac{1}{2}\,\mathcal{L}'_1 \circ \mathcal{L}'_1 \circ \cbvp\big) \Gamma \, \bv \big| \Hb^v \big\rangle \nn \\
	& + \frac{1}{4m_b^2} \big\langle \Hc^{v'} \big| \cbvp \, \Gamma \big(\Pi_-\Overrightarrow{\mJ}_2\bv  + \bv \circ \mathcal{L}_2
		+ \Pi_-\Overrightarrow{\mJ}_1 \bv \circ \mathcal{L}_1 
		+ \frac{1}{2}\,\bv \circ \mathcal{L}_1 \circ \mathcal{L}_1 \big) \big| \Hb^v \big\rangle \nn \\
	&+ \frac{1}{4m_c m_b} \big\langle \Hc^{v'} \big| \big(\cbvp\Overleftarrow{\mJbar}^{\prime}_1 + \mathcal{L}'_1 \circ \cbvp \big)
		\Gamma \big(\Overrightarrow{\mJ}_1\bv + \bv \circ \mathcal{L}_1 \big) \big| \Hb^v \big\rangle\,. \nn
\end{align}
The $\Pi_-$ projectors on the $\mathcal{J}_1$ terms have been eliminated in some terms by the $Q^{v^{(\prime)}}_+$ equation of motion.
Here, the $\circ$ operator denotes an operator product.
For instance, 
\begin{align}
	 \mathcal{L}'_1 \circ \cbvp(z)  
	 & = i\int d^4 x \,  \mathcal{L}'_1(x) \cbvp(z) \nn \\*
	 & = -i\int d^4 x \, \cbvp(x)\big [D^2 + \frac{g}{2} \sigma_{\ab} G^{\ab} ] \mathcal{P}'_2(x- z)\,,
\end{align}
in which $\mathcal{P}_2(x-z)$ is the (dressed) heavy quark two-point function.

Of particular utility when working at second order is the observation that the two-point function is a Green's function
\begin{equation}
	\label{eqn:cont}
	i \vcD^z\,  \mathcal{P}_2(x-z) = -i \Pi_+\delta^4(x-z)\,,
\end{equation}
i.e., it generates a contact term.
Such terms play a particularly important role within Schwinger-Dyson relations, 
as discussed in Appendix \ref{app:SDrel}.
At first order, these relations constrain matrix elements arising from the $\mathcal{O}(1/m_Q)$ current corrections 
with respect to $\LamB$ times the leading order matrix element, i.e., the well-known relation in Eq.~\eqref{eqn:SDrel}.
At second order, one obtains Eq.~\eqref{eqn:subSDrelall},
which relates matrix elements arising from $\mathcal{O}(1/m_Q^2)$ current corrections and the $\mathcal{O}(1/m_Q m_{Q'})$ mixed corrections, 
with $\Delta m^{H}_2$ times the leading order matrix element and $\LamB$ times the matrix element from $\mathcal{O}(1/m_Q)$ Lagrangian corrections.
In Appendix \ref{app:SDrel} we derive these relations from first principles (cf. Appendix C of Ref.~\cite{Falk:1992wt}).

It is important to keep in mind that the notation $Q = c$ or $b$ 
here and throughout is merely a convenient reminder 
of which HQET operator acts on the ingoing and outgoing states.
From the point of view of HQET, there is no distinction between $b^v_+$ and $c^v_+$: 
The heavy quark flavor symmetry is only broken by the masses $m_b \not= m_c$.
We will therefore switch as convenient between writing $\Qbar^{v'}_+ \Gamma Q^v_+$ and $\cbvp \Gamma \bv$.

\subsection{Interaction operator basis}

 Writing the QCD current $J_\Gamma = \cbar \,\Gamma\, b$,
 then a full operator basis entering the QCD matrix elements $\langle \Hc | \cbar \,\Gamma\, b | \Hb \rangle$ is
\begin{equation}
	\label{eqn:Gdef}
	J_S = \bar c\, b\,, \quad  J_P = \bar c\, \g^5\, b\,, \quad
	J_V = \bar c\,\g^\mu\, b\,, \quad  J_A = \bar c\, \g^\mu\g^5\, b\,, \quad
	J_T = \bar c\, \sigma^{\mu\nu}\, b\,,
\end{equation}
where, again, $\sigma^{\mu\nu} \equiv (i/2)[\g^\mu,\g^\nu]$.  
The pseudotensor contribution is determined by the identity $\sigma^{\mu\nu} \g^5 \equiv \pm(i/2)\epsilon^{\mu \nu \rho \sigma}
\sigma_{\rho \sigma}$, in which the sign is subject to a convention choice.
For $\Bbar \to \Dx$, the sign convention most often chosen is such that $\sigma^{\mu\nu} \g^5 \equiv -(i/2)\epsilon^{\mu \nu \rho \sigma}
\sigma_{\rho \sigma}$, which implies $\text{Tr}[\g^\mu\g^\nu\g^\rho\g^\sigma\g^5] = +4i \epsilon^{\mu\nu\rho\sigma}$.
This is the opposite of the sign convention often used for $\Bbar \to D^{**}$ or $\Lambda_b \to \Lambda_c^{(*)}$.

Perturbative corrections to the currents~\eqref{eqn:Gdef} may be computed by
matching QCD onto HQET local operators~\cite{Falk:1990yz, Falk:1990cz, Neubert:1992qq} 
at a suitable matching scale $\mu$.
We present the general derivation of these corrections in Appendix~\ref{app:genradcorr}.

\subsection{Modified power counting}

We are interested in exploring whether it is possible to develop a supplemental power counting, 
on top of the heavy quark expansion, 
that may systematically reorganize the second order power corrections into a small set of dominant terms,
plus a larger set of subdominant contributions that can be truncated.

The heavy quark expansion arises from a reorganization of the QCD Lagrangian 
into the $\mathcal{L}_0$ term that obeys heavy quark spin-flavor symmetry, plus 
symmetry-breaking corrections suppressed by powers of $1/m_Q$ (and by $\alpha_s$).
The order of any given correction in the $1/m_Q$ expansion is effectively 
determined by the number of insertions of $Q^v_- \Qbar^v_-$ into a QCD correlator of interest,
which are then integrated out to form the corresponding HQET matrix element.
This expansion does not assign any relative importance to the local current corrections versus non-local Lagrangian insertions, that enter at each fixed order.
However, with respect to the structure of $1/m_{c,b}$ corrections in $\Bbar \to \Dx$ decays, 
it has been hypothesized (albeit based on model-dependent calculations, such as QCDSR),
that corrections from the chromomagnetic operator in the $\mathcal{L}_1$ Lagrangian may be
numerically small compared to the current corrections from $\mJ_1$~\cite{Neubert:1992pn, Neubert:1992hb, Neubert:1992wq}.
These expectations are also supported somewhat by fits to $\Bbar \to \Dx$ data at $\mathcal{O}(1/m_{c,b})$~\cite{Bernlochner:2017jka},
and compatible with fits to $\Bbar \to D^{**}$ data, 
which find that first-order chromomagnetic contributions are consistent with zero~\cite{Bernlochner:2016bci,Bernlochner:2017jxt}.

With this in mind, a distinguishing feature between a current and a Lagrangian correction is the number of $\Dslash_\perp$ insertions involved:
there is one for the former, and two for the latter,
as follows immediately from Eqs.~\eqref{eqn:JQexp} and~\eqref{eqn:eftL}.
Thus, one may contemplate an additional expansion that resembles counting in $\Dslash_\perp/\lqcd$,
in which each Lagrangian insertion amounts to two powers of $\Dslash_\perp/\lqcd$, 
while a current insertion involves just a single power of $\Dslash_\perp/\lqcd$.
As we will see in this section and the next, this counting can be related to an expansion in the number of operator products inserted along the heavy quark line,
with the additional counting rule that a current insertion counts for half that of a Lagrangian one. 

Before discussing further such an expansion, which can be fully defined within HQET (the low energy effective field theory), 
it is useful to consider first the origin of the difference in the number of $\Dslash_\perp$'s entering the current and Lagrangian insertions. 
This is better understood by looking at the matching between QCD and HQET.
In particular, apart from counting the number of insertions of $Q^v_- \Qbar^v_-$,
one may additionally count the number of insertions of the cross term $\Qbar^v_+ i \Dslash_\perp Q^v_-$ into a QCD correlator 
(after which $Q^v_-$ is integrated out to form an HQET correlator).
This counting is not the same as for $1/m_Q$, because of the equation of motion for $Q^v_-$.
The cross terms break the accidental $U(1)^2$ chiral symmetry, respected 
by the $Q^v_{\pm}$ kinetic terms, to a diagonal $U(1)$.
Therefore, although there is no small parameter in the 
QCD Lagrangian~\eqref{eqn:fullL} that parametrizes this chiral symmetry breaking,
one may nonetheless systematically organize the contributions to any matrix element
by power counting in the number of insertions of the chiral symmetry breaking cross-term 
that enter into each correlator.

Referring to Eq.~\eqref{eqn:fullL}, one may implement this power counting 
by introducing a chiral symmetry breaking parameter $\theta$, such that
\begin{equation}
	i \Dslash_{\perp} \to \theta \, i \Dslash_{\perp}\,,
\end{equation}
and determining the degree of $\theta$ in any HQET correlator
after $Q^v_-$ is integrated out.
With respect to the heavy quark expansion of the Lagrangian, 
it follows from Eq.~\eqref{eqn:eftL}
that the leading term $\mathcal{L}_0 \sim \theta^0$, 
while all the $\mathcal{L}_{n\geq1} \sim \theta^2$.
In the heavy quark expansion of the source term~\eqref{eqn:JQexp}, however, 
all the current correction terms $\mathcal{J}_{n\geq1} \sim \theta$.
Moreover, all product current correction terms from $\mJbar^{\prime}_m \mJ_n$---i.e., 
terms at order $1/(m_{Q'}^mm_Q^n)$---are then $\sim \theta^2$.
The $\theta$ power counting is summarized in Table~\ref{tab:pcs}.

\begin{table}[bt]
\newcolumntype{C}{ >{\centering\arraybackslash $} m{3cm} <{$}}
\begin{tabular}{CCC}
	\hline\hline
	\makecell{\text{Correction or} \\[-10pt] \text{Parameter}}  & \makecell{\text{Associated} \\[-10pt] \text{HQ order}} & \makecell{\Dslash_\perp \\[-10pt] \text{power counting}}\\
	\hline
	\mathcal{L}_{0} 	& 1/m_Q^0		& \theta^0 \\
	\hline
	\mathcal{L}_{n} 	& 1/m_Q^n 		&  \theta^2 \\
	\mJ_{n} 			& 1/m_Q^n 		&  \theta \\
	\hline
	\LamB 			& 1/m_Q^1		& \theta \\ 
	\lambda_{1,2} 		& 1/m_Q^2		& \theta^2 \\ 
	\rho_{1} 		& 1/m_Q^3		& \theta^2 \\ 
	\hline\hline
\end{tabular}
\caption{Orders at which current and Lagrangian corrections ($n\geq1$), as well as hadron mass parameters, enter in the heavy quark and residual chiral expansions.
The RC power counting for $\LamB$ follows from the Schwinger-Dyson relation~\eqref{eqn:SDrel} and for $\lam{1,2}$ from Eq.~\eqref{eqn:HmassHQEapp}.}
\label{tab:pcs}
\end{table}

\subsection{Operator product conjecture}
\label{sec:OPconj}

Because pure current corrections act on one of the external states, 
the single $\Dslash_\perp$ term that is inserted by these corrections
amounts to inserting a factor $\slashed{k}_\perp(\ldots) \sim \lqcd (\ldots)$ in the matrix element,
where $k_\perp = k - (v \cdot k) v$ and $(\ldots)$ denotes powers of $i v \cdot k/(2m_Q) \sim \lqcd/m_Q$.\footnote{%
With some abuse of notation we track here only the transverse momentum contributions originating from the Fourier transform of $\partial_\perp$. 
The same power counting would also apply to the soft gluon interactions contained in the covariant derivative $D_\perp$.} 
We similarly expect a current-current product correction to involve a factor $(\ldots) \slashed{k}_\perp \slashed{k}_\perp(\ldots) \sim (\ldots)\lqcd^2(\ldots)$.
By contrast, a pure Lagrangian correction involves an operator product with two $\Dslash_\perp$ insertions,
producing a factor of the form $Q_+^v(z) \circ \mathcal{L}_n \sim \int d^4 x \mathcal{P}_2(x-z) \Dslash_\perp (\ldots) \Dslash_\perp Q_+^v(x)$.
This, in turn, entails an integral of the form $\int d^4k\, \slashed{k}_\perp (\ldots) \slashed{k}_\perp/(v \cdot k) Q_+^v(k) e^{ik \cdot z}$,
which can be thought of as a second moment of the (dressed) two-point function, with respect to the transverse residual momentum,
plus higher-order HQET corrections.

Recalling, as mentioned in the previous section, that corrections in $\Bbar \to \Dx$ decays from the first-order chromomagnetic operator are hypothesized to be
numerically small compared to the current corrections~\cite{Neubert:1992pn, Neubert:1992hb, Neubert:1992wq},
and that chromomagnetic contributions are consistent with zero in fits to $\Bbar \to D^{**}$ data~\cite{Bernlochner:2016bci,Bernlochner:2017jxt},
one might hypothesize that, generally,
\begin{equation}
	\label{eqn:L1int}
	\int d^4k\, \slashed{k}_\perp (\ldots) \slashed{k}_\perp/(v \cdot k) Q_+^v(k) e^{ik \cdot z} \ll \lqcd^{5/2}(\ldots)\,.
\end{equation}
A mixed current and Lagrangian correction involves three $\Dslash_\perp$'s, 
yielding a factor $\sim \int d^4k\slashed{k}_\perp \ldots  \,\slashed{k}_\perp (\ldots) \slashed{k}_\perp/(v \cdot k) Q_+^v(k) e^{i k \cdot z}$,
while a Lagrangian-Lagrangian operator product correction yields a factor 
$\sim \int d^4k \,\big[\slashed{k}_\perp (\ldots) \slashed{k}_\perp/(v \cdot k)\big]^2 Q_+^v(k) e^{i k \cdot z}$.
Given Eq.~\eqref{eqn:L1int}, this leads us to the conjecture regarding the magnitudes of integrals of the form
\begin{equation}
	\label{eqn:RCconj}
	\int d^4k\, \slashed{k}_\perp^l\ldots \big[\slashed{k}_\perp (\ldots) \slashed{k}_\perp/(v \cdot k)\big]^m Q_+^v(k) e^{ik \cdot z} \sim \epsilon^{2m+l} \lqcd^{m+l+3/2}(\ldots)\,, 
\end{equation}
with $l = 0$ or $1$ and $m \ge 1$, and treating $\epsilon$ as a small parameter.
I.e., the greater the number of operator products in a correlator, the smaller its value.

The conjectured $\epsilon$ expansion in Eq.~\eqref{eqn:RCconj} requires at least one operator product,
and is formally different from that of the $\theta$ expansion,
because at $\mathcal{O}(\theta^2)$ the product current corrections enter, that do not involve an operator product. 
Further, radiative corrections in HQET may induce mixing under the renormalization group evolution (RGE), 
such that the Wilson coefficient of an operator containing $n$ time-ordered operator products may 
induce a ($\aS/\pi$-suppressed) correction to one containing $m \le n$~\cite{Gilman:1982ap}.\footnote{We thank Mike Luke for pointing this out.}
In the context of the $\epsilon$ power counting, this amounts to higher-order operators generating contributions to lower order ones. 
This, however, is not a problem (and not dissimilar to what happens with conventional perturbative expansions) as long as $\epsilon$ is small,
which is the basic assumption motivating this expansion:
it is based on empirical evidence at $\mathcal{O}(1/m_Q)$ and ultimately involves a question about non-perturbative QCD dynamics, 
that can only be determined by comparing this constrained expansion to experimental or lattice data.

The first occurrence of this phenomenon---higher-order operators generating contributions to lower order 
ones---is at second order in the heavy quark expansion, 
at which for example the operator product $\mathcal{L}_1 \circ \mathcal{L}_1$ 
induces an $\mathcal{O}(\aS/m_Q^2)$ correction to the $\mathcal{L}_2$ Wilson coefficient~\cite{Blok:1996iz,Bauer:1997gs}. 
However, the $\mathcal{L}_2$ Wilson coefficient or that of any other term up to and including $\mathcal{O}(\theta^2)$ 
cannot radiatively generate contributions to $\mathcal{O}(\epsilon^3)$ or higher-order operators.
Moreover, at $\mathcal{O}(\theta^3)$ and beyond, the $\theta$ and tree-level $\epsilon$ power countings coincide,
so that the $\theta$ expansion becomes a convenient tool for tracking, within HQET, 
the conjectured $\epsilon$ suppressions.
That is, from the conjectured $\epsilon$ expansion in HQET, 
one may deduce that $\mathcal{O}(\theta^2)$ and lower-order terms dominate those at $\mathcal{O}(\theta^3)$ and higher,
while any RGE-induced counterterms from the latter will be captured by $\mathcal{O}(\theta^2)$ terms that are already present.
Thus we may truncate the expansion at $\mathcal{O}(\theta^2)$.
We refer to this as the residual chiral (RC) expansion.

\subsection{Modified heavy quark expansion}
\label{sec:truncexp}
If one keeps terms up to and including $\mathcal{O}(\theta^2)$, 
then all the usual $\mathcal{O}(1/m_Q)$ terms are retained in Eq.~\eqref{eqn:QCDmatch}. 
However, at $\mathcal{O}(1/m_Q^2)$, 
while the second-order $\mJ_2$ corrections and $\mathcal{L}_2$ Lagrangian corrections are retained, 
the mixed corrections involving $\mJ_1 \mathcal{L}_1 \sim \theta^3$ 
and $\mathcal{L}_1$ double insertions $\mathcal{L}_1 \mathcal{L}_1 \sim \theta^4$ are neglected.
At $\mathcal{O}(1/m_{c,b}^2, \theta^2)$, Eq.~\eqref{eqn:QCDmatchexp} then simplifies to
\begin{align}
	\frac{\langle \Hc | \cbar \,\Gamma\, b | \Hb \rangle}{\sqrt{m_{\Hc} m_{\Hb}}}  
	& \simeq \big\langle \Hc^{v'} \big| \cbvp \, \Gamma \, \bv \big| \Hb^v \big\rangle \label{eqn:QCDmatchRC}\\*
	& + \frac{1}{2m_c} \big\langle \Hc^{v'} \big|  \big(\cbvp\Overleftarrow{\mJbar}^{\prime}_1 + \mathcal{L}'_1 \circ \cbvp \big) \Gamma \, \bv \big| \Hb^v \big\rangle
	 + \frac{1}{2m_b} \big\langle \Hc^{v'} \big| \cbvp  \, \Gamma  \big(\Overrightarrow{\mJ}_1\bv + \bv \circ \mathcal{L}_1 \big)  \big| \Hb^v \big\rangle \nn \\
	& + \frac{1}{4m_c^2} \big\langle \Hc^{v'} \big| \big(\cbvp\Overleftarrow{\mJbar}^{\prime}_2\Pi_-' + \mathcal{L}'_2 \circ \cbvp \big) \Gamma \, \bv \big| \Hb^v \big\rangle \nn \\
	 & + \frac{1}{4m_b^2} \big\langle \Hc^{v'} \big| \cbvp \, \Gamma \big(\Pi_-\overrightarrow{\mJ}_2\bv  + \bv \circ \mathcal{L}_2 \big) \big| \Hb^v \big\rangle \nn \\*
	&+ \frac{1}{4m_c m_b} \big\langle \Hc^{v'} \big| \cbvp\Overleftarrow{\mJbar}^{\prime}_1 
		\Gamma \Overrightarrow{\mJ}_1 \bv \big| \Hb^v \big\rangle\,. \nn
\end{align}
Because the two neglected types of $\mathcal{O}(1/m_Q^2)$ corrections 
are the sources of a large number of subsubleading Isgur-Wise functions in $\Bbar \to \Dx$,
tremendous simplification of the second-order power corrections in $\Bbar \to \Dx$ ensues.

Similarly, for the hadron mass expansion parameters, the Schwinger-Dyson relation~\eqref{eqn:SDrel} implies $\LamB \sim \theta$, 
while by definition $\Delta m^H_2 \sim \theta^2$ and thus $\lambda_{1,2} \sim \theta^2$. 
Thus the Schwinger-Dyson relations~\eqref{eqn:subSDrelall} at $\mathcal{O}(\theta^2)$ simplify to
\begin{align}
	\label{eqn:subSDrelRC}
	\big\langle \Hc^{v'} \big|  \cbvp(z) \Overleftarrow{\mJbar}^{\prime}_2 \, \Pi'_+ \Gamma \, \bv(z)  \big| \Hb^v \big\rangle 
	\ceq -\Delta m^{\Hc}_2 \big\langle \Hc^{v'} \big| J_{\Gamma+}(z) \big| \Hb^v \big\rangle\,\nn\\*
	\big\langle \Hc^{v'} \big|  \cbvp(z) \Gamma \, \Pi_+ \Overrightarrow{\mJ}_2 \, \bv(z)  \big| \Hb^v \big\rangle 
	\ceq -\Delta m^{\Hb}_2 \big\langle \Hc^{v'} \big| J_{\Gamma+}(z) \big| \Hb^v \big\rangle\,,
\end{align}
writing the HQET current operator $J_{\Gamma+}(z) = \cbvp(z) \Gamma \bv(z)$.
These allow us to relate the IW functions associated with second order current corrections 
to $\lambda_{1,2}$ times the leading IW function.

One additional point of importance is that in this expansion the $\mathcal{O}(\theta^3)$ terms at second order---the terms
corresponding to mixed current and Lagrangian corrections $\sim \mJ_1 \mathcal{L}_1$---vanish at zero recoil~\cite{Falk:1992wt}.
Thus, if the residual chiral expansion is a good approximation,
we may expect it to be particularly useful at zero recoil, 
because only $\mathcal{O}(\theta^4)$ corrections enter.

\section{$\Bbar \to \Dx$ form factors}
\label{sec:BDDsFFs}
\subsection{HQET matrix elements}

The $D$ and $D^*$ (or $B$ and $B^*$) mesons belong to a HQ spin-symmetry doublet, 
formed by the tensor product of a spin-$1/2$ heavy quark 
with brown muck in the $s^{\pi_\ell} = \frac{1}{2}^-$ spin-parity state.
This doublet, containing the pseudoscalar~($P$) and vector~($V$) mesons with a single heavy quark,
can be represented as~\cite{Falk:1990yz, Bjorken:1990rr, Falk:1991nq} 
\begin{equation}
	H^v \mapsto H(v) = \Pi_+ \big[V^v \slashed{\epsilon} - P^v\g^5 \big]\,, \qquad \Hbar^v \mapsto \Hbar(v) \equiv  \g^0 H^{v\dagger} \g^0 =  \big[\Vbar^v \slashed{\epsilon}^* + \Pbar^v \g^5 \big] \Pi_+ \,,
\end{equation}
in which $\epsilon^{\mu}$ denotes the polarization vector of the spin-1 state with velocity $v$.
Here and hereafter $\genbar{X} = \g^0 X^\dagger \g^0$ for any Dirac object $X$.

With reference to the reduced terms in Eq.~\eqref{eqn:QCDmatchRC} 
at $\mathcal{O}(1/m_{c,b}^2, \theta^2)$
the matching of HQET to the QCD matrix elements becomes
\begin{align}
	\frac{\langle \Dx | \cbar \,\Gamma\, b | \Bxbar \rangle}{\sqrt{m_{\Hc} m_{\Hb}}}  
	& = -\xi(w) \bigg\{ \Tr\big[ \Hbar_c(v') \Gamma H_b(v) \big] \nn \\*
	& + \sum_{n=1}^2 \ec^n\, \Tr\big[ \Hbar_c^{(n)}(v',v) \Gamma H_b(v) \big] + \sum_{n=1}^2 \eb^n\, \Tr\big[ \Hbar_c(v') \Gamma H_b^{(n)}(v,v') \big]  \nn\\*
	& + \ec\eb \Tr\big[   \Gamma H_{bc}^{(1,1)}(v,v')\big]\bigg\}\,, \label{eqn:BDxmatch}
\end{align}
defining
\begin{subequations}
\label{eqn:reps}
\begin{align}
	H^{(n)}(v,v') & = \Pi_+\Big\{ P^{v} \hL1n (-\g^5) + V^{v} \big(\hL2n \slashed{\epsilon} + \hL3n \epsilon \ccdot v'\big) \Big\}  \nn \\*
		& \qquad \qquad \qquad  + \Pi_-\Big\{ P^{v} \hL4n (-\g^5) + V^{v} \big(\hL5n\slashed{\epsilon} + \hL6n \epsilon \ccdot v'\big) \Big\}\,,\\
	\Hbar^{(n)}(v',v) & = \Big\{ \Pbar^{v'} \hL1n \g^5 + \Vbar^{v'} \big(\hL2n \slashed{\epsilon}^{\prime*} + \hL3n \epsilon^{\prime*} \ccdot v\big) \Big\}\Pi'_+  \nn \\*
		& \qquad \qquad \qquad  + \Big\{ \Pbar^{v'} \hL4n \g^5 + \Vbar^{v'} \big(\hL5n \slashed{\epsilon}^{\prime*} + \hL6n \epsilon^{\prime*} \ccdot v\big) \Big\}\Pi'_-\,,\\
	H^{(1,1)}_{bc}(v,v') & = 	\Pi_-\Big\{P^v \Pbar^{v'}\hM{8}(-\g^5)\g^5   + P^v\Vbar^{v'}(-\g^5) \big[ \hM{9}\slashed{\epsilon}^{\prime*} + \hM{10}\epsilon^{\prime*} \ccdot v \big] \nn \\*
				 & \qquad \qquad + V^{v}\Pbar^{v'}\big[\hM{9} \slashed{\epsilon} + \hM{10}\epsilon \ccdot v'\big] \g^5  
				   + V^{v}\Vbar^{v'}\big[ \hM{11} \slashed{\epsilon}\slashed{\epsilon}^{\prime*}  + \hM{12} \epsilon \ccdot \epsilon^{\prime*}  \nn\\*
	 			& \qquad \qquad  \qquad + \hM{13}\slashed{\epsilon}\epsilon^{\prime*} \ccdot v + \hM{13}\slashed{\epsilon}^{\prime*}\epsilon \ccdot v' + \hM{14} \epsilon^{\prime*} \ccdot v\, \epsilon \ccdot v'\big] \Big\}\Pi'_-\,.
\end{align}
\end{subequations}
Here we have included in $H^{(1,1)}_{bc}$ only those terms relevant for matching at $\mathcal{O}(1/m_{c,b}^2, \theta^2)$. 
The full expressions are given in Appendix~\ref{app:mbmc-ffs}.
The recoil parameter is defined as
\begin{equation}
	w = v \ccdot v' = \frac{m_{\Bx}^2 + m_{\Dx}^2 - q^2}{2 m_{\Bx} m_{\Dx}}\,, \qquad q^2 = (p -p')^2\,,
\end{equation}
and the HQ expansion parameters,
\begin{equation}
	\varepsilon_{c,b} = \frac{\LamB}{2m_{c,b}}\,.
\end{equation}
(As mentioned above in Sec.~\ref{sec:HQE}, the $c$, $b$ subscripts on the field representations in Eq.~\eqref{eqn:BDxmatch} 
are mere reminders of the flavor of the hadron: 
the HQET is agnostic to this distinction except via $\eb \not= \ec$ and within perturbative corrections.)
Since we have normalized the HQET states according to Eq.~\eqref{eqn:HQETnorm},
then Eq.~\eqref{eqn:BDxmatch} need not include the $Z_M$ factors present in Eq.~(4.34) of Ref.~\cite{Falk:1992wt}.

Following convention, in Eq.~\eqref{eqn:BDxmatch} we have factored out the leading Isgur-Wise function, $\xi(w)$, from all terms.
Matching onto HQET, at leading order, 
the normalization of the QCD matrix element for the conserved vector current in the equal mass, zero recoil limit
\begin{equation}
	\label{eqn:massnormcon}
	\langle H | \Qbar \g^\mu Q | H \rangle = 2 m_H v^\mu\,,
\end{equation}
implies that $\xi(1) = 1$. 
We discuss further zero-recoil constraints below in Sec.~\ref{sec:zerorec}.
As done in, e.g., Refs.~\cite{Neubert:1993mb, Grinstein:2001yg, Bernlochner:2017jka} (but not in
Refs.~\cite{Luke:1990eg, Manohar:2000dt}) we have further normalized terms in the expansion with respect to $\LamB$, 
such that all Isgur-Wise functions are dimensionless.

We use the notation that hatted functions of $w$ are normalized to the leading Isgur-Wise function,
\begin{equation}
	\label{eqn:hatdef}
	\hat{W}(w) \equiv W(w)/\xi(w)\,,
\end{equation}
for any Isgur-Wise function or form factor.
In particular, the $\hL{i}{n}$ and $\hM{i}$ denote linear combinations of higher-order Isgur-Wise functions\footnote{%
In the notation of Ref.~\cite{Falk:1992wt}, $\LamB \hL{i}1 = L_i /\xi$, $\LamB^2\hL{i}2 = \ell_i /\xi$, and $\LamB^2\hM{i} = m_i /\xi$.
We have added the superscript index to the $\hat{L}$'s 
in order to make clearer at which order they enter into the power expansion. 
We use the standard numbering for the subscripts of the $\hL{i}{n}$,
while our numbering for the $\hM{i}$ is the same as those used for the $m_i$ in Ref.~\cite{Falk:1992wt}.} normalized by $\xi$.

\subsection{Form factor matching}

We use the standard HQET definitions for the $\Bbar \to \Dx$ form factors. 
The $\Bbar \to D$ matrix elements are
\begin{subequations}
\begin{flalign}
	\amp{D}{\cbar\,b}{\Bbar} & = h_S\, (w+1)\,, \label{eqn:DS} &\\*
	\amp{D}{\cbar\g^5 b}{\Bbar} & = \amp{D}{\cbar \g^\mu\g^5 b}{\Bbar} = 0\,, \label{eqn:DPA} &\\
	\amp{D}{\cbar \g^\mu b}{\Bbar}  & = \big[ h_+(v+v')^\mu + h_-(v-v')^\mu\big]\,, &\\*
	\amp{D}{\cbar \sigma^{\mu\nu} b}{\Bbar}  & = i\big[ h_T\, (v'^\mu v^\nu - v'^\nu v^\mu )\big]\,,&
\end{flalign}
\end{subequations}
and for $\Bbar \to D^*$, 
\begin{subequations}
\begin{flalign}
	\amp{D^*}{\cbar\,b}{\Bbar} & = 0\,, \label{eqn:DsS}&\\*
	\amp{D^*}{\cbar \g^5 b}{\Bbar} & = -h_P\, (\epsilon^* \cdot v)\,, &\\
	\amp{D^*}{\cbar \g^\mu b}{\Bbar} & = i h_V\, \varepsilon^{\mu\nu\alpha\beta}\, 
			\epsilon^*_{\nu}v'_\alpha v_\beta \,,\label{eqn:DsV} &\\
	\amp{D^*}{\cbar \g^\mu \g^5 b}{\Bbar} & = \big[h_{A_1} (w+1)\epsilon^{*\mu}
  		- h_{A_2}(\epsilon^* \cdot v)v^\mu 
 		- h_{A_3}(\epsilon^* \cdot v)v'^\mu \big]\,,\label{eqn:DsA} &\\
	\amp{D^*}{\cbar \sigma^{\mu\nu} b}{\Bbar} & = -\varepsilon^{\mu\nu\alpha\beta} \big[ h_{T_1} \epsilon^*_{\alpha}
  		(v+v')_\beta + h_{T_2} \epsilon^*_{\alpha} (v-v')_\beta + h_{T_3}(\epsilon^*\cdot v) v_\alpha v'_\beta\big]\,. \!\!\!\! \label{eqn:DsT} &
\end{flalign}
\end{subequations}
Here the $h_{\Gamma_i}$ are functions of $w$.
While typically we are not interested in $\Bbar^* \to \Dx$ decays, because the $B^*$ decays to $B\gamma$, 
the vector current matrix element for $\Bbar^* \to D^*$ is important for 
mass normalization constraints at second order: a generalization of Luke's theorem.
In particular, we need also consider
\begin{equation}
	\amp{D^*}{\cbar \g^\mu b}{\Bbar^*}  = -\epsilon \ccdot \epsilon^{\prime*}\, h_1 (v + v')^\mu + \ldots\,,
\end{equation}
which is the only form factor that contributes in the zero recoil, equal mass limit.

For the sake of writing the form factors in terms of $\hL{i}{n}$ and $\hM{i}$ in a compact manner,
it is convenient to define
\begin{equation}
	\label{eqn:hLcb}
	\hL{i}{Q} = \hL{i}1 + \eQ \hL{i}2\,,\qquad Q = c,b\,.
\end{equation}
In this notation, the hatted $\Bbar \to D$ form factors (see Eq.~\eqref{eqn:hatdef}) 
at $\mathcal{O}(\aS, 1/m_{c,b}^2, \theta^2)$ are
\begin{align}
	\label{eqn:BDFFs}
	\hat h_S & = 1 + \haS\, C_S + \sum_{Q = c,b} \eQ \bigg(\hL{1}{Q} - \hL{4}{Q}\, \frac{w-1}{w+1} \bigg) + \ec\eb \hM{8}\,, \nn\\*
	\hat h_+ & = 1 + \haS\Big[C_{V_1} + \frac{w+1}2\, (C_{V_2}+C_{V_3})\Big] + \sum_{Q = c,b} \eQ \hL{1}{Q} - \ec\eb \hM{8}\,, \nn\\
	\hat h_- & = \haS\, \frac{w+1}2\, (C_{V_2}-C_{V_3}) + \ec \hL{4}{c} - \eb\hL{4}{b}  \,, \nn\\
	\hat h_T & = 1 + \haS \big(C_{T_1}-C_{T_2}+C_{T_3}\big) + \sum_{Q=c,b} \eQ \big[ \hL{1}{Q} - \hL{4}{Q}\big] + \ec\eb \hM{8} \,.
\end{align}
The hatted $\Bbar \to D^*$ form factors at this order are
\begin{align}
	\label{eqn:BDsFFs}
	\hat h_P 		& = 1 + \haS\, C_P + \ec \big[\hL{2}{c} + \hL{3}{c} (w-1)  + \hL{5}{c} - \hL{6}{c} (w+1)\big]  + \eb \big[ \hL{1}{b} - \hL{4}{b} \big] \nn\\*
				& \qquad -\ec\eb\big[\hM{9} - (w-1)\hM{10}\big]\,, \nn\\
	\hat h_V 		& = 1 + \haS\, C_{V_1} + \ec \big[\hL{2}{c} - \hL{5}{c}\big]  + \eb \big[ \hL{1}{b} - \hL{4}{b} \big] +\ec\eb \hM{9} \,,\nn\\
	\hat h_{A_1} 	& = 1 + \haS\, C_{A_1} 
  					+ \ec \bigg( \hL{2}{c} - \hL{5}{c}\, \frac{w-1}{w+1} \bigg)
  					+ \eb \bigg( \hL{1}{b} - \hL{4}{b}\, \frac{w-1}{w+1} \bigg) +\ec\eb \hM{9}\,,\nn\\
	\hat h_{A_2} 	& = \haS\, C_{A_2} + \ec \big[\hL{3}{c} + \hL{6}{c}\big] -\ec\eb\hM{10} \,,\nn\\
	\hat h_{A_3} 	& = 1 + \haS \big(C_{A_1} + C_{A_3}\big) + \ec \big[\hL{2}{c} - \hL{3}{c} + \hL{6}{c} - \hL{5}{c} \big] + \eb \big[\hL{1}{b} - \hL{4}{b}\big]  \nn\\*
				& \qquad + \ec\eb\big[\hM{9} + \hM{10} \big]\,,\nn\\
	\hat h_{T_1} 	& = 1 + \haS \Big[ C_{T_1}  + \frac{w-1}2\, \big(C_{T_2}-C_{T_3}\big) \Big] + \ec \hL{2}{c} + \eb \hL{1}{b} -\ec\eb \hM{9}\,,\nn\\
	\hat h_{T_2} 	& = \haS\, \frac{w+1}2\, \big(C_{T_2}+C_{T_3}\big) + \ec \hL{5}{c} - \eb \hL{4}{b} \,,\nn\\
	\hat h_{T_3} 	& = \haS\, C_{T_2} + \ec \big[\hL{6}{c} - \hL{3}{c}\big] - \ec\eb\hM{10}\,.
\end{align}
We have included here the leading perturbative corrections in $\haS = \aS/\pi$, that are given in Eq.~\eqref{eqn:ascurrent}.
The higher-order $\haS/m_{c,b}$ corrections are discussed in Appendix~\ref{app:asmcb}.
Finally, the $B^* \to D^*$ vector form factor
\begin{equation}
	\label{eqn:BsDsFFs}
	\hat h_1  = 1 + \haS\Big[C_{V_1} + \frac{w+1}2\, (C_{V_2}+C_{V_3})\Big] + \sum_{Q = c,b} \eQ \hL{2}{Q} - \ec\eb \big[\hM{11}+ \hM{12}\big]\,.
\end{equation}
In Eqs.~\eqref{eqn:BDFFs}, \eqref{eqn:BDsFFs} and~\eqref{eqn:BsDsFFs} we have included only those $\hM{i}$ terms 
relevant for matching at $\mathcal{O}(\theta^2)$ in the RC expansion;
the additional $1/m_cm_b$ terms that enter when all second-order 
power corrections are considered are given in Eq.~\eqref{eqn:compmcmb}.

In Appendix~\ref{app:firstsecondorder} we present a derivation of the first and second-order power corrections 
entering the $\Bbar \to \Dx$ form factors, 
making use of the formalism and conventions presented in Appendices~\ref{app:hadmass} and~\ref{app:SDrel}.
For the first-order power corrections in Eq.~\eqref{eqn:Lhat1def},
we use the standard set of Isgur-Wise functions,
namely $\hat\chi_{1,2,3}$ and $\hat\eta$.
The full expressions for the second-order power corrections are shown in Eqs.~\eqref{eqn:Lhat2def} and \eqref{eqn:Mhatdef}.
These involve: 
\begin{enumerate}[wide, labelwidth=0pt, labelindent=0pt, label = \textbf{(\roman*)}, noitemsep, topsep =0pt, itemjoin =\quad, series = fits]
\item one Isgur-Wise function, $\hvph$, arising from current corrections;
\item the mass parameters $\lam{1,2}$, 
that enter the current corrections via the Schwinger-Dyson relations~\eqref{eqn:subSDrelRC};
\item three Isgur-Wise functions $\hat\beta_{1,2,3}$ that enter via second-order Lagrangian corrections.
\end{enumerate}

The Isgur-Wise function $\hvph$ is constrained at zero recoil by the Schwinger-Dyson relations,
such that it is convenient to write the form factors in terms of $[\hvph(w) - \hvph(1)]/(w-1)$.
This combination must be regular because of the analyticity of the matrix elements 
(see discussion leading to Eqs.~\eqref{eqn:vphzerorec} and~\eqref{eqn:phpdef}).
To express this explicitly,
we define the \emph{quotient} of an Isgur-Wise function with respect to $w=1$,
\begin{equation}
	\label{eqn:quotdef}
	W^\quot(w) \equiv [W(w) - W(1)]/(w-1)\,,
\end{equation}
which must be regular.
By definition $W^\quot(1) = W'(1)$, the gradient at zero recoil.

\subsection{Chiral corrections}

A full assessment of potential percent-level corrections to the form factors requires 
consideration of chiral corrections,
which originate from strong dynamics of the brown muck at momentum scales below that of chiral symmetry breaking, 
sensitive to the light mesons spectrum and the heavy mesons mass differences.
This dynamics may therefore be represented using Heavy Hadron Chiral Perturbation Theory (HH$\chi$PT)~\cite{Wise:1992hn, Burdman:1992gh},
under which the dominant chiral corrections to the HQET matrix elements are generated by loops containing a light pseudoscalar,
$P = \pi, K, \eta$. 
Schematically the structure of the chiral corrections can be expanded in powers of the heavy mass scale $M_H$ (e.g., a heavy hadron mass) as 
\begin{equation}
    \sum_n \frac{A_{n}(w) \log(m_P^2/\mu^2) + B_n(w,\mu)}{M_H^n}
\end{equation}
where $\log(m_P^2/\mu^2)$ denotes chiral logarithms of the light meson masses, 
and $B_n$ contain finite and counterterm contributions. 
The scale $\mu \sim {\cal O}(1\,{\rm GeV})$ is where HQET is matched onto HH$\chi$PT. 
The expressions for $\Bbar \to \Dx$ decays have been known for a long time~\cite{Goity:1992tp,Randall:1993qg,Boyd:1995pq}. 
The terms $A_n$, $B_n$ are in general different for $B^0$, $B^+$ and $B_s$ decays.

At zero recoil, the leading and subleading corrections vanish, and the leading nonzero contribution is proportional 
to the hyperfine mass splitting $\Delta m_H^2 \sim \lambda_2^2$~\cite{Randall:1993qg}.
Because $\lambda_2^2 \sim \mathcal{O}(\theta^4)$ in the RC expansion, the chiral corrections can be neglected at zero recoil.
Similarly, in the VC limit they vanish because $\lam2 \to 0$.

Away from zero recoil, both leading and subleading corrections in powers of $1/M_H$ are present. 
Parametrically the size of $A_n$, $B_n$ is controlled by the chiral loop factor $(g_P m_P/4\pi f_P)^2 \lesssim 1\%$, 
where $g_P$ is the coupling of the light meson $P$ to the heavy hadrons and $m_P$ and $f_P$ are its mass and decay constant. 
At the order of precision we are interested in, only the leading corrections are important, 
because the subleading ones contribute $\sim 1\% \times \lqcd/m_Q$ or $\sim 1\% \times \alpha_s/\pi$, which are both $\ll 1\%$~\cite{Boyd:1995pq}.

Importantly, the leading chiral corrections are universal for any HQ current. 
As a result, they can be reabsorbed via a redefinition of the leading order Isgur-Wise function, 
up to induced corrections of order $\sim 1\% \times \lqcd/m_Q$ or $\sim 1\% \times \alpha_s/\pi$ that can be neglected.
Because the leading order chiral corrections are flavor diagonal, but not flavor universal,
this reabsorption further induces isospin and flavor $SU(3)$ breaking,
which distinguishes the leading order Isgur-Wise functions $\xi_{+,0}(w)$ in $B^{+,0}$  and $\xi_s(w)$ in $B_s$ decays, respectively.

The size of isospin corrections enter proportional to the isospin mass splittings of light or heavy mesons $\Delta m_{\pi,K,H}/m_{\pi,K,H}$
on top of the chiral loop factor suppression, i.e., at the $\sim 10^{-4}$ level.
Therefore we need only consider a single isospin-invariant leading order Isgur-Wise function $\xi_{+,0}(w) \simeq \xi(w)$ for the $B^{+,0}$ decays.
Flavor $SU(3)$ breaking from the chiral corrections may be sizeable enough, however, to distinguish $\xi_s(w)$ for $B_s$ decays from $\xi(w)$ in $B$ decays.
Explicit expressions for the leading $SU(3)$-breaking contributions can be found in Ref.~\cite{Goity:1992tp}.
Additional $SU(3)$ breaking contributions, parametrically as large as $m_s/m_Q \sim \LamB/m_Q$ will be present at $\mathcal{O}(1/M_H)$ in the HQET matrix elements themselves,
and must also be considered when relating $B_s \to D_s^{(*)}$ Isgur-Wise functions to those for $\Bbar \to \Dx$.
Since $B_s \to D_s^{(*)}$ decays are not considered in this paper, we do not discuss them further.

\subsection{Zero-recoil normalization constraints and redefinitions}
\label{sec:zerorec}

The mass normalization condition~\eqref{eqn:massnormcon} implies that in the equal mass limit the vector current matrix elements satisfy, to all orders,
\begin{equation}
	\label{eqn:hp1zr}
	h_{+,1}(1)\Big|_{m_c = m_b} = 1\,.
\end{equation}
In this limit, both the perturbative corrections  
and the power corrections to $h_{+,1}(1)$ vanish order-by-order.
At first order, Eqs.~\eqref{eqn:BDFFs} and~\eqref{eqn:BsDsFFs} then 
imply $L_1(1) = L_2(1) = 0$,
which is a part of Luke's theorem~\cite{Luke:1990eg}
\begin{equation}
	\label{eqn:Lthm}
	\hat\chi_1(1) = \hat\chi_3(1) = 0\,.
\end{equation}
At second order, the mass normalization constraints for pseudoscalar and vector mesons respectively require
\begin{align}
	2\hL12(1) - \hM8(1) = -\frac{\lam1 + 3\lam2}{\LamB^2} + 4\hat\beta_1(1) + 24 \hat\beta_3(1) & = 0\,, \nn\\
	2\hL22(1) - \hM{11}(1) - \hM{12}(1) = -\frac{\lam1 - \lam2}{\LamB^2} + 4\hat\beta_1(1) - 8 \hat\beta_3(1) &= 0\,,
\end{align}
which results in the zero recoil constraints
\begin{equation}
	\label{eqn:betazerorec}
	\hat\beta_1(1) = \frac{\lam1}{4\LamB^2}\,,\qquad \hat\beta_3(1) = \frac{\lam2}{8\LamB^2}\,.
\end{equation}
Just as in Eq.~\eqref{eqn:phpdef}, we therefore write
\begin{equation}
	\label{eqn:betaredef}
	\hat\beta_{1}(w) = \frac{\lam1}{4\LamB^2} + (w-1)\hat\beta^\quot_{1}(w)\,,\qquad 
	\hat\beta_{3}(w) = \frac{\lam2}{8\LamB^2} + (w-1)\hat\beta^\quot_{3}(w)\,,
\end{equation}
in which the quotient functions $\beta^\quot_{1,3}$ are regular near zero recoil.

The three Isgur-Wise functions $\beta_1$, $\chi_1$ and $\xi$ arise from the same leading-order trace, 
as can be seen by comparing Eqs.~\eqref{eqn:BDxmatch}, \eqref{eqn:chromoNLO} and~\eqref{eqn:chromoNNLO}.
This trace conserves heavy quark spin symmetry, 
therefore these three Isgur-Wise functions always enter 
in the same linear combination.
Based on this observation, when working at $\mathcal{O}(\aS,1/m_Q)$ in the heavy quark expansion, 
it is common (see e.g. Refs~\cite{Manohar:2000dt,Bernlochner:2017jka}) 
to reabsorb $\chi_1$ into the leading-order Isgur-Wise function via the replacement
\begin{equation}
	\label{eqn:replNLO}
	\xi + 2(\ec + \eb)\chi_1 \to \xi\,.
\end{equation}
The constraint $\chi_1(1) = 0$ ensures that the normalization condition $\xi(1) =1$ is preserved.
The replacement~\eqref{eqn:replNLO} induces $\mathcal{O}( 1/m_{c,b}^2)$ and $\mathcal{O}(\aS/m_{cb})$ corrections,
that must be incorporated consistently when working at second order in the power expansion.
In the power counting of the residual chiral expansion, however,
because $\chi_1$ enters at $\mathcal{O}(\theta^2)$,
then the $\mathcal{O}(1/m_{c,b}^2)$  corrections induced by 
Eq.~\eqref{eqn:replNLO} enter only at $\mathcal{O}(\theta^3)$ or higher,
and can therefore be neglected at $\mathcal{O}(\theta^2)$.
Furthermore, reparametrization invariance ensures that the replacement~\eqref{eqn:replNLO} 
induces only $\mathcal{O}(\aS/m_{c,b}^2)$ or higher-order corrections in the $\mathcal{O}(\aS/m_{c,b})$ terms (see Appendix~\ref{app:asmcb}).

Since the second terms in Eq.~\eqref{eqn:betaredef} vanish at zero recoil by construction,
then at $\mathcal{O}(\aS/m_Q, 1/m_Q^2,\theta^2)$ we can therefore extend Eq.~\eqref{eqn:replNLO} to
\begin{equation}
	\label{eqn:replNNLO}
	\xi + 2(\ec + \eb)\chi_1 + 2(\ec^2 + \eb^2)(w-1)\beta^\quot_1 \to \xi\,,
\end{equation}
while preserving $\xi(1) = 1$.
Note, however, one cannot generally redefine $\chi_{2,3}$ to absorb $\beta_{2,3}$,
because although $\beta_{2}$ and $\chi_2$ enter via an identical trace, as do $\beta_3$ and $\chi_3$, 
each trace violates heavy quark spin symmetry.
As a result $\chi_{2,3} + \ec \beta_{2,3}$ enters into the $\Bbar \to D^*$ form factors 
with a different prefactor than $\chi_{2,3} + \eb \beta_{2,3}$: Specifically, $d_V=-1$ versus $d_P=3$, respectively.
With this in mind, if one neglects $\mathcal{O}(1/m_cm_b, 1/m_b^2)$ or $\mathcal{O}(\aS/m_{c,b})$ corrections,
then one can absorb $\beta_{2,3}$ via the redefinitions
\begin{equation}
	\label{eqn:b2b3abs}
	\chi_{2} + \ec \beta_{2} \to \chi_{2}\,,\qquad \chi_{3} + \ec (w-1) \beta^\quot_{3} \to \chi_{3}\,,
\end{equation}
the latter of which preserves $\chi_3(1) = 0$.

\subsection{Summary of constrained form factors}
\label{sec:summff}
Including all the zero recoil constraints in Eqs.~\eqref{eqn:phpdef} and~\eqref{eqn:betaredef},
and the redefinition~\eqref{eqn:replNNLO},
one finds finally from Eqs.~\eqref{eqn:Lhat1def}, \eqref{eqn:Lhat2def} and~\eqref{eqn:Mhatdef}

\begin{subequations}
\label{eqn:Lhat12def}
\begin{align}
	\hL11 &=  - 4(w-1) \hat\chi_2 + 12 \hat\chi_3\,, \nn\\*
 	\hL21 & = - 4 \hat\chi_3\,, \nn\\*
	\hL31  & = 4 \hat\chi_2\,, \nn\\*
	\hL41 &= 2 \hat\eta - 1\,, \nn\\*
	\hL51 & = -1\,, \nn\\*
	\hL61 &= - 2 (1 + \hat\eta)/(w+1)\,, \\ \nn\\
	\hL12 &= \frac{\lam1 + 3 \lam2}{2\LamB^2} + (w-1)\big[ - 4\hat\beta_2 + 12 \hat\beta^\quot_3\big]\,, \nn\\*
 	\hL22 & = \frac{\lam1 - \lam2}{2\LamB^2} + (w-1)\big[- 4 \hat\beta^\quot_3\big]\,,\nn\\*
	\hL32 & = 4 \hat\beta_2\,, \nn\\*
	\hL42 & = \frac{\lam1(w+1)}{3\LamB^2} - \frac{\lam2(w-5)}{2\LamB^2} + 2(w^2-1)\hvphq\,,\nn\\*
	\hL52 &  = \frac{\lam1(w+1)}{3\LamB^2} - \frac{\lam2(w-1)}{2\LamB^2} + 2(w^2-1)\hvphq\,,\nn\\* 
	\hL62 & =\frac{2\lam1}{3\LamB^2} - \frac{\lam2}{\LamB^2} + 4(w-1)\hvphq\,, \\ \nn \\
	\hM{8} & = \frac{\lam1(4-w)}{3\LamB^2} +\frac{\lam2 (w^2 +11)}{2\LamB^2(w+1)} - 2(w-1)^2\hvphq - \frac{2 (2 \hat\eta-1) (w-1)}{w+1}\,,\nn\\*
	\hM{9} & = \frac{\lam1}{3\LamB^2} + \frac{\lam2(5-w)}{2\LamB^2(w+1)} + 2 (w-1) \hvphq - \frac{(2\hat\eta-1)(w-1)}{w+1}\,,\nn\\*
	\hM{10} & =\frac{\lam1}{3\LamB^2} -\frac{\lam2 (w+4)}{2\LamB^2(w+1)} + 2 (w+2) \hvphq - \frac{2 \hat\eta-1}{w+1}\,\nn\\*
	\hM{11} & = \frac{\lam1(w-2)}{3\LamB^2} + \frac{\lam2(3-w)}{2\LamB^2} +2 (w^2-1) \hvphq\,,\nn\\*
	\hM{12} & = \frac{2 \lam1(3-w)}{3\LamB^2} - \frac{\lam2  (w^2 - 3w -2)}{\LamB^2(w+1)} - 4 w(w-1) \hvphq  +\frac{2 (w-1)}{w+1}\,,\nn\\*
	\hM{13} & = \frac{\lam1}{3\LamB^2} -\frac{\lam2 w}{2\LamB^2(w+1)} + 2 (w+2) \hvphq + \frac{2 \hat\eta+1}{w+1}\,,\nn\\*
	\hM{14} & = \frac{2 \lam1 (w-2)}{3\LamB^2 (w+1)} -  \frac{\lam2 (w^2-2 w-4)}{\LamB^2(w+1)^2}+ \frac{4 (w^2+2) \hvphq}{w+1} + \frac{4(2\hat\eta + 1) - 2w}{(w+1)^2}\,.
\end{align}
\end{subequations}
Additional wave functions (see Eq.~\eqref{eqn:deldef}) entering the $\mathcal{O}(\aS/m_{c,b})$ corrections are presented in Appendix~\ref{app:asmcb}.
Up to and including second order in the power expansion and in the residual chiral expansion, 
the $\Bxbar \to \Dx$ form factors are determined fully by seven Isgur-Wise functions: 
$\xi$ at leading order; $\hat\chi_{2,3}$, and $\hat\eta$ at first-order; and $\hvphq$, $\hat\beta_2$ and $\hat\beta^\quot_3$ at second order.
The relevant Isgur-Wise functions in the residual chiral expansion are shown in Table~\ref{tab:expansions},
including the reduced set in the case that one truncates at $\mathcal{O}(1/m_c^2)$ 
and reabsorbs redundant Isgur-Wise functions via Eq.~\eqref{eqn:b2b3abs}.
In the latter case the $\Bxbar \to \Dx$ form factors are determined fully by just a single Isgur-Wise function at second order.

Both of these countings---at $\mathcal{O}(1/m_c^2)$ and $\mathcal{O}(1/m_{c,b}^2)$---are also reflected in Table~\ref{tab:HQETFFs}.
In this table, moving from leading order to first order to second order, 
we only count the new, independent Isgur-Wise functions entering at each order
(though $1/m_{c,b}^2$ counts functions also counted at $1/m_c^2$).
The counting of Isgur-Wise functions is performed after redefinition of lower-order functions 
to absorb all possible redundant higher-order ones, such as in Eq.~\eqref{eqn:b2b3abs}.
Some redundant higher-order functions can be reabsorbed in this manner only up to their value at zero-recoil:
these vestigial constant terms are counted as independent functions unless they are fully determined with respect to known quantities, 
such as $\lam{1,2}$, as happens for $\hat\beta_{1,3}(1)$ in Eq.~\eqref{eqn:betazerorec} (cf. Eq.~\eqref{eqn:xivcreabs} below).

\begin{table}[t]
\begin{tabular}{cc|c|c|cc}
\hline\hline
\multicolumn{2}{c|}{Expansions}  &  $1/m_{c,b}^0$  &  $1/m_{c,b}$  &  $1/m_c^2$ only  &  $1/m_{c,b}^2$  \\
\hline\hline
\multirow{2}{*}{Ref.~\cite{Bernlochner:2017jka}}  &  Form factors  &  $\xi(w)$  &  $\eta(w)$, $\chi_2(w)$, $\chi_3(w)$  &  --- & ---  \\
  &  Parameters  &  $\rho^2$  &  \makecell{$\LamB$, $\hat\eta(1)$, $\hat\eta'(1)$, \\[-4pt] $\hat\chi_2(1)$, $\hat\chi'_2(1)$, $\hat\chi_3(1)$}  &  --- & --- 
\\  \hline
\multirow{2}{*}{RC}  &  Form factors  &  $\xi(w)$  &  $\eta(w)$, $\chi_2(w)$, $\chi_3(w)$  & $\varphi_1^\quot(w)$ & $\varphi_1^\quot(w)$, $\beta_2(w)$, $\beta^\quot_3(w)$  \\
  &  Parameters  &  $\rhossq$, $\cs$  & 
  		\makecell{ $\LamB$, $\hat\eta(1)$, $\hat\eta'(1)$, \\[-4pt] $\hat\chi_2(1)$, $\hat\chi'_2(1)$, $\hat\chi_3'(1)$}  & $\lam1$, $\lam2$, $\hvphp(1)$  & 
		\makecell{$\lam1$, $\lam2$,\\[-4pt] $\hvphp(1)$, $\hat\beta_2(1)$, $\hat\beta'_3(1)$}
\\  \hline
\multirow{2}{*}{VC}  &  Form factors  &  $\xi(w)$  &  $\eta(w)$, $\chi_1(w)$  &  $\varphi^\quot_0(w)$, $c_0(1)$ &  $\varphi^\quot_0(w)$, $c_0(1)$, $e_3(w)$\\
  &  Parameters   &   $\rhossq$, $\cs$  &  $\LamB$, $\hat\eta(1)$, $\hat\eta'(1)$, $\hat\chi_1'(1)$   &  $\hat\varphi_0'(1)$, $c_0(1)$  &  $\hat\varphi_0'(1)$, $c_0(1)$, $e_3(1)$  \\
\hline\hline
\end{tabular}
\caption{Isgur-Wise functions and their parametrizations used in our fits, order by order in the HQ expansion, 
as used in Ref.~\cite{Bernlochner:2017jka}, and in the RC and VC expansions in this work.
If one includes additional input on the quark masses, $m_{c,b}$, 
certain combinations of $\LamB$ and $\lam{1,2}$ can be eliminated.
In the VC case, $c_0(w)$ enters proportional to the leading Isgur-Wise function and can be absorbed, except for $c_0(1)$ 
(see Eq.~\eqref{eqn:xivcreabs}).
Further, $\chi_1$ enters only when second-order terms are included, that prevent its reabsorption at $\mathcal{O}(1/m_{c,b})$.
(With no constraints, 11 unknown functions parametrize the SM form factors at order $1/m_{c,b}^2$ for $B\to D^{(*)}\ell\bar\nu$, and 6 when restricted to $1/m_c^2$.)}
\label{tab:expansions}
\end{table}

\subsection{Vanishing chromomagnetic limit}
\label{sec:FFsVClimit}

Besides the RC expansion truncated at $\mathcal{O}(\theta^2)$, 
we also consider another Ansatz, in which the field strength $G_{\ab}$ is set to zero.\footnote{%
Prior literature~\cite{Mannel:1994kv} considered the limit in which HQ spin symmetry-violating 
matrix elements involving the field strength $G^{\ab}$ are neglected,
which implies that some spin symmetry conserving terms must also vanish for self-consistency (see Appendix~\ref{app:vclimit}).
At second order in HQET, the difference between this limit and $G_{\ab} \to 0$ 
amounts to contributions from $B_0$ (see Eq.~\eqref{eqn:chromoNNLO}),
which may be reabsorbed into $C_0$ (see Eq.~\eqref{eqn:L1L1chromo}), and is thus unphysical at this order.}
This vanishing chromomagnetic (VC) limit, already considered in Ref.~\cite{Falk:1992wt}, 
also significantly reduces the number of Isgur-Wise functions at $\mathcal{O}(1/m_{c,b}^2)$.
It is motivated by the smallness of $\hat\chi_{2,3}$ calculated using QCD sum rules~\cite{Neubert:1992wq, Neubert:1992pn}, 
and is also consistent with $\mathcal{O}(1/m_{c,b})$ fits (see, e.g., Ref.~\cite{Bernlochner:2017jka}). 
For the sake of completeness and consistency of notation, 
we revisit the derivation of this limit in Appendix~\ref{app:vclimit},
noting a few differences with respect to Ref.~\cite{Falk:1992wt}.

\enlargethispage{-2\baselineskip}

The expressions for the nonvanishing $\hL{i}{1,2}$ and $\hM{i}$ in the VC limit are\footnote{\label{ft:mcorrs}%
Though our definitions of the $\hM{i}$ form factors in Eq.~\eqref{eqn:fullMi} are the same as in Ref.~\cite{Falk:1992wt},
we find with respect to Eqs.~(A5)--(A7) of Ref.~\cite{Falk:1992wt}  a different sign for $\hM{10}$ (as above), 
and further $\hM{17}$ is swapped with $\hM{19}$, and $\hM{22}$ is swapped with $\hM{23}$.
The latter two swaps are also present in Appendix~B of Ref.~\cite{Falk:1992wt}. 
}
\begin{subequations}
\label{eqn:hLhMVClim}
\begin{alignat}{3}
 \hL11  & =  \hL21  = 2\hat\chi_1 \,, \\*
 \hL41 & = 2\hat\eta - 1 \,, 
 && \hL51  = -1 \,,   
 & \hL61  & = -2\frac{1+\hat\eta}{w+1}\,, \nn\\  \nn\\
 \hL12  & = \hL22 = 2 \hat c_0\,,  \\*
 \hL42  & = -2(\hat\chi_1+w\hat\varphi^\quot_0 -\hat e_3)\,, 
&& \hL52   = -2(\hat\chi_1+w\hat\varphi^\quot_0)\,,
 & \hL62  & = -\frac{4w \hat\varphi^\quot_0+ 2\hat e_3 + 4\hat \chi_1}{w+1}\,, \nn \\ \nn\\
 \hM{1}  & = \hM{2} = \hM{4} = 2\hat d_0 \,,  \\ \nn\\
 \hM{8}  & = 2(w-1)\bigg[\hat\varphi^\quot_0-\frac{2\hat\eta-1}{w+1}\bigg]\,,  
&& \hM{9}   = -\frac{w-1}{w+1}(2\hat\eta-1)\,, & \\
 \hM{10}  & = -2\hat\varphi^\quot_0-\frac{2\hat\eta-1}{w+1}\,, 
&& \hM{11}  = -2(w-1)\hat\varphi^\quot_0\,, \quad 
& \hM{12}  & = 2(w-1)\bigg[2\hat\varphi^\quot_0+\frac{1}{w+1}\bigg]\,, \nn \\
 \hM{13}  & = -2\hat\varphi^\quot_0+\frac{2\hat\eta+1}{w+1}\,, 
&& \hM{14}  = -\frac{4 w \hat\varphi^\quot_0}{w+1}+\frac{4(2\hat\eta+1)-2w}{(w+1)^2}\,, \hspace{-50pt}&\nn \\ \nn\\
 \hM{15} &  = \hM{18} = 2(\hat e_3+\hat\varphi^\quot_0-\hat\chi_1)\,, 
&& \hM{16}  = \hM{20} = 2(\hat\varphi^\quot_0-\hat\chi_1)\,,  \\*
 \hM{17}  & = \hM{22} = -2\frac{\hat e_3 -2\hat\varphi^\quot_0 + 2\hat\chi_1}{w+1}\,. \nn
\end{alignat}
\end{subequations}
Concerning the perturbative corrections,
in the vanishing chromomagnetic limit the $\mathcal{O}(\aS)$ expressions remain the same, 
while at $\mathcal{O}(\aS/m_{c,b})$ the terms proportional to $C_g^{c,b}$ vanish (see Appendix~\ref{app:asmcb}),
as they correspond to insertions of the chromomagnetic operator.
Finally, the effects of the chiral corrections in the VC limit are the same as for the RC expansion, as the zero recoil effects proportional to $\lambda_2^2$ vanish altogether instead of being higher order.

The zero-recoil constraints in the equal mass limit, $h_+(1)=1$ and $h_1(1)=1$, now impose a relation between $c_0(1)$ and $d_0(1)$, namely
\begin{equation}
2 c_0 (1) + d_0(1) = 0\,.
\end{equation}
The Isgur-Wise functions $\chi_1$, $c_0$, $d_0$ all involve the same trace structure as the leading-order matrix element parametrized by $\xi(w)$. 
On the one hand, because of the relation Eq.~\eqref{eqn:EEp}, $\chi_1$ enters indirectly in terms originating from different tensor structures. 
Therefore, even though $\chi_1$ vanishes at zero recoil, 
it can no longer be fully reabsorbed via a redefinition of the leading Isgur-Wise function $\xi$. 
On the other hand, one can reabsorb $c_0$ and $d_0$ into $\xi$, 
up to their contribution at zero recoil, $2(\ec-\eb)^2 \hat{c}_0(1)$,
in order to preserve the normalization condition $\xi(1) =1$.
Explicitly this can be achieved via the shift
\begin{equation}
	\label{eqn:xivcreabs}
   \xi(w) + 2 (w-1) \big[(\eb^2 +\ec^2) \, c_0^\quot(w)  +  \eb\ec  \, d_0^\quot(w) \big] \to \xi(w)\,,
\end{equation}
which induces only higher-order corrections at $\mathcal{O}(1/m_{c,b}^3)$.
In addition, if one is only interested in second-order corrections at $\mathcal{O}(1/m_c^2)$,
 then in a similar manner to Eq.~\eqref{eqn:b2b3abs},
$\hat e_3$ may be reabsorbed into $\hat \eta$ via 
\begin{equation}
	\label{eqn:e3etaabs}
	\hat\eta + \ec \hat e_3 \to \hat\eta\,,
\end{equation}
up to induced $\mathcal{O}(1/m_cm_b)$ and $\mathcal{O}(\aS/m_c)$ corrections.

After all these constraints and redefinitions, 
the form factors now depend on five Isgur-Wise functions and one zero-recoil constant only:
$\xi$ at leading order;
$\hat\eta$ and $\hat\chi_1$ at first order;
and at second order, $\hat\varphi^\quot_0$ and $\hat{e}_3$ together with the zero-recoil constant $ \hat{c}_0(1)$.
The relevant Isgur-Wise functions for the VC limit are shown in Table~\ref{tab:expansions},
including the reduced set in the case that one truncates at $\mathcal{O}(1/m_c^2)$ 
and reabsorbs the redundant Isgur-Wise function via Eq.~\eqref{eqn:e3etaabs}.

\section{Parametrizations and prescriptions}
\label{sec:parampres}
\subsection{$1S$ scheme and numerical inputs}\label{sec:mb_dmbc}

Cancellation of the leading renormalon ambiguities~\cite{Bigi:1994em, Beneke:1994sw}
from the mass parameter $\LamB$ 
against those in the factorially growing coefficients in the $\aS$ perturbative power series
can be achieved by use of a short distance mass scheme.
We use the $1S$ scheme~\cite{Hoang:1998ng, Hoang:1998hm, Hoang:1999ye}, which defines $\mbS$ 
as half of the perturbatively computed $\Upsilon(1S)$ mass.
It is related to the pole mass via $\mbS = m_b\, (1 - 2\aS^2/9 + \ldots)$. 
This may be inverted to express the pole mass 
\begin{equation}
	\label{eqn:mb1S}
	m_b(\mbS) \simeq \mbS(1 +2\aS^2/9 + \ldots)\,.
\end{equation}
The splitting of the bottom and charm quark pole mass $\dmbc \equiv m_b - m_c$ 
is subject to a renormalon ambiguity only at third order when one computes just the leading $n_f$-dependence at high orders~\cite{Beneke:1994bc, Neubert:1994wq, Luke:1994xd},
so we fix $m_c=m_b - \dmbc$.
Thus, when working at second order in the HQ expansion, 
we may parametrize observables in terms of $\mbS$ and $\dmbc$.

In practice, however, because $\mbS$ and $\dmbc$ are extracted from fits to inclusive spectra at $\mathcal{O}(1/m_Q^3)$~\cite{Bauer:2002sh,Bauer:2004ve,Ligeti:2014kia},
third-order terms must be retained numerically in the expansion of the hadron mass, 
even though we formally work to second order in the expressions for the form factors.
In particular, the spin-averaged mass of the HQ pseudoscalar-vector doublet, $\overline{m} \equiv (m_P + 3m_V)/4$, 
can be written in the $1S$ scheme, defining $m_c(\mbS) \equiv m_b(\mbS) - \dmbc$,
\begin{align}
	\overline{m}_B  & \simeq m_b(\mbS) + \LamB - \frac{\lam1}{2m_b(\mbS)} + \frac{\rho_1}{4 [m_b(\mbS)]^2}\,,\nn\\ 
	\overline{m}_D  & \simeq m_c(\mbS) + \LamB - \frac{\lam1}{2m_c(\mbS)} + \frac{\rho_1}{4[m_c(\mbS)]^2}\,,
\end{align}
noting from Eq.~\eqref{eqn:masslam12} that the $\lam2$ dependence cancels.
Here we have included only the $\mathcal{O}(\theta^2)$ contribution to the hadron mass from $\Delta m_3^H$, proportional to the parameter $\rho_1$,
as defined in Ref.~\cite{Gremm:1996df}.
This leads to
\begin{align}
	\label{eqn:1Slams}
	\LamB & = \frac{m_b(\mbS) \overline{m}_B - m_c(\mbS)\overline{m}_D}{\dmbc} - \big[m_b(\mbS)  + m_c(\mbS)\big] + \frac{\rho_1}{4m_b(\mbS) m_c(\mbS)}\,, \nn \\
	\lam1 & = \frac{2 m_b(\mbS)m_c(\mbS)}{\dmbc}\big[ \overline{m}_B - \overline{m}_D - \dmbc\big] + \frac{\rho_1\big[m_b(\mbS)  + m_c(\mbS)\big]}{2m_b(\mbS) m_c(\mbS)} \,.
\end{align}
That is, truncating the hadron masses at third order in the HQ power expansion and second order in the residual chiral expansion, 
$\LamB$ and $\lam1$ are parametrized in terms of $\mbS$, $\dmbc$, and $\rho_1$.

The fits to inclusive $B\to X_c l \bar\nu$ spectra and other determinations of $\mbS$, find that~\cite{Ligeti:2014kia} 
\begin{equation}
	\label{eqn:1Sinputs}
	\mbS = (4.71 \pm 0.05)\,\GeV\,, \qquad \dmbc \equiv m_b - m_c = (3.40 \pm 0.02)\,\GeV\,,
\end{equation}	
which we use as inputs to our fits.
For $\rho_1$ we use
\begin{equation}
	\label{eqn:rho1input}
	\rho_1 = (-0.1 \pm 0.2)\,\GeV^3\,,
\end{equation}
corresponding numerically to the range $\lam1 = (-0.3 \pm 0.1)\,\GeV^2$, 
commensurate with the ranges quoted in Ref.~\cite{Ligeti:2014kia}.\footnote{%
These somewhat arbitrary uncertainties, to be used in fit \emph{inputs}, should not be confused 
with the \emph{recovered} uncertainties from fit results,
that determine the uncertainties in predictions of, e.g., $R(\Dx)$.
\label{foot:uncert}}
We choose the HQET to QCD matching scale
\begin{equation}
	\mubc = \sqrt{m_bm_c} \simeq 2.5\,\GeV\,.
\end{equation}
One may apply these inputs to Eq.~\eqref{eqn:mb1S}, combined with the renormalization group evolution $\aS = \aS(\aS(m_Z),m_Z; \mu)$
computed at four-loop order~\cite{Czakon:2004bu,VANRITBERGEN1997379}.
One finds 
\begin{equation}
	\aS(\mubc) \simeq 0.27\,, 
\end{equation}
up to small uncertainties that are negligible when working at $\mathcal{O}(1/m_{c,b}^2, \aS/m_{c,b})$.
We shall therefore treat $\aS(\mubc)$ as a fixed external parameter.\footnote{%
In Ref.~\cite{Bernlochner:2017jka} we used $\aS = 0.26$ 
due to the slightly higher scale choice. %
Both formally and numerically, the difference between these choices only enters at higher order compared to $\mathcal{O}(1/m_c^2, \aS/m_c)$.}

At first order in the HQ expansion, cancellation of the leading renormalon ambiguity amounts 
to replacing the pole mass $m_b(\mbS)$ by $\mbS$ everywhere, 
except in $\mathcal{O}(1/m_{c,b})$ terms originating from the Schwinger-Dyson relation~\eqref{eqn:SDrel},
i.e., that enter via the `$\ceq$' relations defined in Eq.~\eqref{eqn:diffeq}.
In order to compare with the results of Ref.~\cite{Bernlochner:2017jka}, we similarly enforce this prescription here. 
Cancellation of higher-order renormalon ambiguities is relevant only with the inclusion of $\aS^2$ and higher terms, that we do not consider.\footnote{%
Order $\aS^2$ corrections will be included at zero recoil only for $\mathcal{F}(1)$ in Sec.~\ref{sec:zrrce} because $1/m_{c,b}$ corrections vanish there. 
While at that order one should consider also second-order corrections in the $1S$ expansion, these would multiply $\varepsilon_{c,b}$ and therefore vanish in $\mathcal{F}(1)$.}
There is, in addition, a leading renormalon cancellation between the Lagrangian $\mathcal{O}(\aS/m_{c,b})$ corrections and certain terms at $\mathcal{O}(1/m_{c,b}^2)$. 
It would affect terms originating from $\LamB \times \mathcal{O}(1/m_Q)$ matrix elements in the second-order Schwinger-Dyson relations, 
but such terms vanish at $\mathcal{O}(\theta^2)$.

Cancelling the third-order power corrections from $\Delta m_3^H$, one may express $\lam2$ via~\cite{Bauer:2002sh}
\begin{align}
	\lam2(\mubc) 
	& \simeq \frac{m_b^2(m_{B^*} - m_B) - m_c^2 (m_{D^*} - m_D)}{2 m_b a_b(\mubc) - 2 m_c a_c(\mubc)}\,,\nn\\*
	& \simeq 0.11 \pm 0.02\, \GeV^2\,, \label{eqn:lam2input}
\end{align}
using the $1S$ prescription and Eq.~\eqref{eqn:aQ},
in which we have assigned an inflated $\sim20\%$ uncertainty to $\lam2$
to absorb possible higher-order renormalon effects. 
The fits to inclusive $B\to X_c\ell\bar\nu$ decays~\cite{Bauer:2002sh,Bauer:2004ve,Ligeti:2014kia},
from which we obtain the $1S$ inputs, 
use the leading log approximation for $a_Q(\mu)$. 
This leads to an enhancement of the extracted $\lam2$ by approximately a factor $(1 + 13\aS/6\pi)\simeq 1.2$, 
which is formally higher order, $\mathcal{O}(\aS/m_{c,b}^2)$.
This difference is also covered by the assigned uncertainties in $\lam2$.

\subsection{Leading order Isgur-Wise function}

Parametrizations that make use of unitarity plus dispersion relations for the QCD matrix elements typically apply the conformal transformation
\begin{equation}
	\label{eqn:zqq}
	z(q^2, q_0^2) = \frac{\sqrtb{q_+^2 - q^2}- \sqrtb{q_+^2 - q_0^2}}{\sqrtb{q_+^2 - q^2} + \sqrtb{q_+^2 - q_0^2}}\,,
\end{equation}
which maps the pair-production threshold $q^2 > q_{+,J}^2 \equiv (m_{H^J_b} + m_{H^J_c})^2$ 
to the boundary of the unit circle $|z| = 1$, centered at $q^2 = q_0^2$,
while the interval $q_-^2  < q^2 < q_+^2$ is mapped to the real axis $(0\ge) z(q_-^2, q_0^2) > z  > -1$.
Here $H^J_{c}H^J_{b}$ denotes the lightest pair of hadrons that couple to current $J$, 
generating the QCD matrix element $\langle H_c | J | H_b\rangle$.
Thus there is in principle a different $z$ for each current $J$.
At second order in the HQ expansion, the $\lambda_2$-induced splitting of the pseudoscalar and vector meson masses in the spin symmetry doublet
means that the branch point $q_{+,V}^2 = (m_B + m_D)^2$ for the $B \to D$ and $B \to D^*$ vector current, 
while for the $B \to D^*$ axialvector current it is $q_{+,A}^2 = (m_{B^*} + m_D)^2$.
That is, second-order corrections to the hadron masses enter into the conformal transformation~\eqref{eqn:zqq}.
The choice $q_0^2 = q_{+,J}^2 (1 - [1- q_-^2/q_{+,J}^2]^{1/2}) \equiv q_{\text{opt},J}^2$, minimizes $|z(q^2 = 0)|$ and hence the range of $z$,
while for $q_0^2 = q_-^2$ simply $z(w=1) = 0$.  
We define $z_* \equiv z(q^2, q_\text{opt}^2)$.

In the HQET representation of the matrix elements, the Isgur-Wise functions 
are functions of the variable $w = v \cdot v'$. 
One may reexpress Eq.~\eqref{eqn:zqq} as
\begin{equation}
	z(w,w_0) = \frac{\sqrtb{w - w_+}- \sqrtb{w_0 - w_+}}{\sqrtb{w - w_+}- \sqrtb{w_0 - w_+}}\,.
\end{equation}
In this form, the branch point $w_{+,V} = -1$ for $B \to D$, which is the minimal possible branch point,
but for $B \to D^*$ both $w_{+,V}$ and $w_{+,A} > -1$.
This leads to different $z$'s not only for different currents but also for different hadrons in the same HQ multiplet.
However, because the Isgur-Wise functions are universal by construction, 
a conformal transformation $z(w,w_0)$ consistent with the HQET 
matching in Eq.~\eqref{eqn:BDxmatch} should respect HQ symmetry,
while higher-order corrections in the HQ expansion itself 
reabsorb the differences in the branch points of different matrix elements.
Adopting this construction, we therefore choose to parametrize the leading order Isgur-Wise function 
as a polynomial in the optimized $z_*$ for $B \to D$, with the minimal branch point at $w_+ = -1$.
Thus
\begin{equation}
	\label{eqn:xiexp}
	\frac{\xi(w)}{\xi(w_0)} = 1 - 8 a^2 \rhossq z_* + 16(2\cs a^4 - \rhossq a^2)z_*^2  + \ldots
\end{equation}
in which the optimized conformal variable
\begin{equation}
	z_*(w) = \frac{\sqrt{w+1}-\sqrt{2}a}{\sqrt{w+1}+\sqrt{2}a}\,, \quad \text{with} \quad a^2  \equiv \frac{w_0 + 1}{2} = \frac{1+r_D}{2\sqrt{r_D}}\,.
\end{equation}
By construction $\xi'(w_0)/\xi(w_0) = -\rhossq$, $\xi''(w_0)/\xi(w_0) = \cs$, and $z_*(w_0) = 0$. 
Numerically $|z_*| < 0.032$. 
Because $\xi(1) = 1$, 
it follows from Eq.~\eqref{eqn:xiexp} that $\xi(w_0) = \big[1 - 8 a^2 \rhossq z_*(1) + 16(2\cs a^4 - \rhossq a^2) z_*(1)^2 \big]^{-1}$.

Another convention in the literature is to expand the leading-order Isgur-Wise function about zero recoil, 
via the choice $z \equiv z(q^2, q_-^2)$,
such that $\xi(w) = 1 - 8 \rho^2 z + 16(2 c - \rho^2)z^2  + \ldots$ and $z(w) = (\sqrt{w+1}-\sqrt{2})/(\sqrt{w+1}+\sqrt{2})$.
By construction $\xi'(1) = -\rho^2$ and $\xi''(1) = c$. 
Differentiating Eq.~\eqref{eqn:xiexp} at $w=1$, one may relate $\rho^2$ and $c$ to $\rhossq$ and $\cs$.

In the fits below, we keep $\rhossq$ and $\cs$ as free parameters.
This differs from the approach of Ref.~\cite{Caprini:1997mu} (see also Ref.~\cite{Bernlochner:2017jka}), 
in which one expands the $B \to D$ form factor $\mathcal{G}(w)$ with respect to $z_*$ (using the same $w_+ = -1$ branch point),
\begin{equation}
	\label{eqn:Gexp}
	\frac{\mathcal{G}(w)}{\mathcal{G}(w_0)} = 1 - 8 a^2 \trhossq z_* +16(2\tcs a^4 - \trhossq a^2) z_*^2  + \ldots\,,
\end{equation}
and then applies dispersive bounds to constrain the curvature $\tcs$ and the slope $\trhossq$ (and higher coefficients) to lie in an elongated ellipsoidal region.
Further fixing $\tcs$ to the central value---the major axis---of the allowed region yields~\cite{Caprini:1997mu}
\begin{equation}
	\label{eqn:tcscons}
	\tilde{c}_* \simeq [(V_{21} + 16 a^2) \trhossq - V_{20}]/32a^4\,,
\end{equation}
with $V_{21} \simeq 57.$ and $V_{20} \simeq 7.5$.
As we discuss further in Sec.~\ref{sec:biases} below, 
this approach leads to fit biases when applied to current data.
Because $\mathcal{G}(w) = \xi(w)\big[\hat h_+(w) - \rho_D \hat h_-(w)\big]$, 
one may relate the coefficients of Eqs.~\eqref{eqn:xiexp} and~\eqref{eqn:Gexp} directly,
\begin{align}
     \rhossq & = \trhossq + \frac{\hat h_+'(w_0) - \rho_D \hat h_-'(w_0)}{\hat h_+(w_0) - \rho_D \hat h_-(w_0)}\,, \\
      \cs & = \tcs +  2\rhossq(\rhossq - \trhossq) -  \frac{\hat h_+''(w_0) - \rho_D \hat h_-''(w_0)}{\hat h_+(w_0) - \rho_D \hat h_-(w_0)}\,, \nn
\end{align}
in which $\hat h_{\pm}^{(\prime, \prime\prime)}(w_0)$ can be expanded to arbitrary order in HQET as desired.
These relations allow the fitted $\rhossq$ and $\cs$ parameters to be compared to the dispersive bounds for $\trhossq$ and $\tcs$ 
(see Sec.~\ref{sec:biases}).

\subsection{Sub(sub)leading Isgur-Wise functions}

We approximate the subleading Isgur-Wise functions (as in Ref.~\cite{Bernlochner:2017jka} and elsewhere), as
\begin{equation}
	\hat{\chi}_2(w)  \simeq \hat{\chi}_2(1) + \hat{\chi}'_2(1)(w-1)\,,\qquad \hat{\chi}_3(w)  \simeq \hat{\chi}'_3(1)(w-1)\,,\qquad
	\hat\eta(w)  \simeq \hat\eta(1) + \hat\eta'(1)(w-1)\,,
\end{equation}
expanding to linear order in $w-1$.
The remaining subsubleading Isgur-Wise functions 
after the redefinitions in Eqs.~\eqref{eqn:phpdef} and~\eqref{eqn:betaredef} are $\hvphq$, $\hat\beta_{2}$, and $\hat\beta^\quot_{3}$. 
To limit the number of fit parameters, given the precision of the available data,
we treat these functions as constants
\begin{equation}
	\hvphq(w) \simeq \hvphp(1)\,, \qquad  \hat\beta_{2}(w) \simeq \hat\beta_{2}(1)\, \qquad \hat\beta^\quot_{3}(w) \simeq \hat\beta'_{3}(1)\,.
\end{equation}
The relevant parameters for the residual chiral expansion are shown in Table~\ref{tab:expansions}.
Applying the $1S$ scheme, the full set of Isgur-Wise parameters in our parametrization of the form factors are
\begin{equation}
	 \rhossq\,, \quad \cs\,; \qquad \hat{\chi}_2(1)\,, \quad \hat{\chi}'_{2,3}(1)\,, \quad \hat\eta(1)\,, \quad \hat\eta'(1)\,; \qquad \hvphp(1)\,, \quad \hat\beta_{2}(1)\,, \quad \hat\beta'_{3}(1)\,,
\end{equation}
in addition to the constrained parameters $\mbS$,  $\dmbc$,  $\rho_1$, and $\lam2$  
per Eqs.~\eqref{eqn:1Sinputs}, \eqref{eqn:rho1input}, and~\eqref{eqn:lam2input}. 

Though $\beta_2$ and $\beta_3$ may be reabsorbed into $\chi_2$ and $\chi_3$ via Eq.~\eqref{eqn:b2b3abs}
if one neglects $\mathcal{O}(1/m_cm_b)$ or $\mathcal{O}(\aS/m_{c,b})$ corrections,
at zero recoil many lower-order corrections vanish or are constrained, 
such that higher-order corrections could still have large effects (see Sec.~\ref{sec:zrp}, below).
However, the gradient $\beta'_3(1)$ does not contribute at $w=1$, 
and one may explicitly see from Eqs.~\eqref{eqn:BDFFs}, \eqref{eqn:BDsFFs}, \eqref{eqn:BDFFs_asmcb}, and~\eqref{eqn:BDsFFs_asmcb}  
that at zero recoil $\chi_2$ and $\beta_2$ enter only via $\ec \hL{3}{1,2}$ terms, respectively.
Thus, $\beta_2(1)$ is redundant with $\chi_2(1)$ at zero recoil: 
the induced $\mathcal{O}(1/m_cm_b)$ or $\mathcal{O}(\aS/m_{c,b})$ corrections from applying Eq.~\eqref{eqn:b2b3abs}
must be suppressed by $w-1$.
Since the precision of current or near-future data is too low for sensitivity to
$\mathcal{O}(1/m_cm_b)$ or $\mathcal{O}(\aS/m_{c,b})$ corrections beyond zero recoil, 
where lower-order corrections do not vanish,
as a practical matter both $\hat\beta_2(1)$ and $\hat\beta'_3(1)$ may therefore be neglected in fits.

\section{Zero-recoil predictions}
\label{sec:zrp}
\subsection{Form factors and ratios}

The SM differential rates for $\Bbar \to \Dx l \nu$ with respect to $w$ have the well-known expressions
in the massless lepton limit
\begin{subequations}
\begin{align}
\frac{\mathrm{d} \Gamma(\Bbar \to D \ell \bar\nu)}{\mathrm{d} w} & = \frac{G_F^2|V_{cb}|^2 \, \eta_{\rm EW}^{2} \, m_B^5}{48 \pi^3}\, (w^2-1)^{3/2}\, \rD^3\, (1 + \rD)^2\, \mathcal{G}(w)^2\,,\\
\frac{\mathrm{d} \Gamma(\Bbar \to D^* \ell \bar\nu)}{\mathrm{d} w} & = \frac{G_F^2|V_{cb}|^2 \, \eta_{\rm EW}^{2} \,  m_B^5}{48 \pi^3}\, (w^2-1)^{1/2}\, (w + 1)^2\, \rDs^3 (1- \rDs)^2 \nn \\
  & \quad \times \bigg[1 + \frac{4w}{w+1}\frac{1- 2 w\rDs + \rDs^2}{(1 - \rDs)^2} \bigg] \mathcal{F}(w)^2\,,
\end{align}
\end{subequations}
where $r_{\Dx} = m_{\Dx}/m_B$ and $\eta_{\rm EW} \simeq 1.0066$~\cite{Sirlin:1981ie}  is the electroweak correction
(see e.g. Ref.~\cite{Bernlochner:2021vlv} for full expressions including the lepton mass).
The form factors
\begin{subequations}
\label{eqn:GFdef}
\begin{align}
\mathcal{G}(w) & = h_+ - \frac{1-\rD}{1 + \rD} h_- \,,\\
\mathcal{F}(w)^2 & = h_{A_1}^2 \bigg\{ 2(1 - 2 w \rDs + \rDs^2)  \bigg(1 + R_1^2\, \frac{w-1}{w+1}\bigg) 
  + \big[(1 - \rDs) + (w-1)\big( 1 - R_2\big) \big]^2 \bigg\} \nn \\
	& \qquad\quad \times \bigg[(1 - \rDs)^2 + \frac{4w}{w+1}\big(1- 2 w\rDs + \rDs^2\big) \bigg]^{-1}\,,
\end{align}
\end{subequations}
in which the form factor ratios are
\begin{equation}
	\label{eqn:R1R2Def}
	R_1(w) = \frac{h_V}{h_{A_1}}\,, \qquad 
	R_2(w) = \frac{h_{A_3} + \rDs\, h_{A_2}}{h_{A_1}}\,.
\end{equation}
In addition,
\begin{align}
	\mathcal{S}_1(w) & = h_+ - \frac{1+\rD}{1 - \rD}\frac{w-1}{w+1} h_-\,,\nn\\
 	R_0(w) & = \frac{h_{A_1}(w+1) - h_{A_3}(w -\rDs) - h_{A_2}(1 - w\rDs)}{(1+\rDs)\, h_{A_1}}
\end{align}
parametrize the contributions that enter only proportional to the lepton mass in the $\Bbar \to D l \nu$ and $\Bbar \to D^* l \nu$ rates, respectively.

In HQET at $\mathcal{O}(1/m_{c,b}^2, \aS/m_{c,b})$, the zero-recoil $\mathcal{G}(1)$ and $\mathcal{F}(1)$ form factors have explicit expressions, 
defining $\rho_D \equiv (1-\rD)/(1+\rD)$,
\begin{align}
	\mathcal{G}(1)_{\text{HQET}} & \simeq 1 + \haS [\Cv1 + \Cv2(1-\rho_D) + \Cv3(1+\rho_D)] - (\ec-\eb)\rho_D\hL{4}{1} \nn\\*
	& \qquad + \ec^2 [\hL{1}{2} - \rho_D\hL{4}{2}] + \eb^2 [\hL{1}{2} + \rho_D\hL{4}{2}]  + \ec\eb[\hM{1} - \hM{8}]\nn \\*
	& \qquad - \haS\rho_D \big[(\ec -\eb) \Cv1\hL{4}{1} + 2 (\ec \Cv3 - \eb \Cv2)\hL{5}{1}\big]\,,\nn\\
	\mathcal{F}(1)_{\text{HQET}} 
	& \simeq 1 + \haS \Ca1+ \ec^2 \hL{2}{2} + \eb^2 \hL{1}{2} + \ec\eb[\hM{2}+ \hM{9}] \,, \label{eqn:F1G1HQETall}
\end{align}
in which all functions are evaluated at $w=1$. 
Note $\mathcal{S}_1(1) = h_+(1) = \mathcal{G}(1)|_{\rho_D \to 0}$.
Similarly, the zero-recoil form factor ratios at $\mathcal{O}(1/m_{c,b}^2, \aS/m_{c,b})$
\begin{align}
	R_1(1)_{\text{HQET}} & \simeq 1 + \haS[\Cv1 - \Ca1]  - \ec \hL{5}{c} - \eb\hL{4}{b}  \\* 
					& \qquad -\ec\eb[\hM{16} - \hM{18}]\,,\nn\\*
					& \qquad  + \haS\big\{ \hL{4}{1}\big[\eb(\Ca1 - \Cv1 -\Cv2) -\ec \Cv3\big] \nn\\*
					& \qquad \qquad + \hL{5}{1}\big[\eb\Cv2+\ec(\Ca1 - \Cv1 + \Cv3)\big]\big\}\,, \nn\\
	R_2(1)_{\text{HQET}} & \simeq 1 + \haS\big(\Ca3 + \rDs \Ca2\big) - \eb \hL{4}{b} \\*
					& \qquad	+ \ec\big[ (1 + \rDs)\hL{6}{c} - (1 - \rDs) \hL{3}{c} - \hL{5}{c}\big] \nn\\*
					& \qquad + \ec\eb\big\{\hM{16} + \hM{18} + (1 + \rDs)\big[\hM{17} - \hM{19}\big] - (1-\rDs)\big[\hM{3} - \hM{10}\big] \big\} \nn\\*
					& \qquad + \haS\ec\big\{-(1-\rDs)C_g^c\hL{3}{1} + \Ca2 \rDs( \hL{5}{1} - 2\hL{6}{1}) \nn\\*
					& \qquad \qquad + \Ca3\big[(1-\rDs)\hL{4}{1} + (1-3\rDs)\hL{5}{1} + 4\hL{6}{1}\big]/2\big\} \nn\\
					& \qquad + \haS\eb\big\{ \Ca2\big[ (1-3\rDs)\hL{4}{1} +(1-\rDs)\hL{5}{1} \big]/2 - \Ca3\hL{4}{1}\big\}\,, \nn\\
	R_0(1)_{\text{HQET}} & \simeq 1 + \frac{1 - \rDs}{1 + \rDs}\bigg\{ -\haS[\Ca2 + \Ca3] +\ec[\hL{5}{c} - 2\hL{6}{c} ] + \eb \hL{4}{b} \\
					& \qquad + \ec\eb\big[ \hM{16} + \hM{18} -2 (\hM{17} - \hM{19})\big]  \nn \\
					& \qquad + \haS \big(\Ca2 + \Ca3 \big)\big[\eb \hL{4}{1} - \ec\big( \hL{5}{1} - 2\hL{6}{1}\big) \big]\bigg\}\,,\nn
\end{align}
in which all functions are evaluated at $w=1$ and we use the notation $\hL{i}{Q} = \hL{i}{1} + \eQ \hL{i}{2}$ as in Eq.~\eqref{eqn:hLcb}.

The most precise LQCD predictions provide that~\cite{Aoki:2021kgd}
\begin{equation}
	\label{eqn:F1G1}
	\mathcal{G}(1)_{\text{LQCD}} = 1.054(9)\,, \qquad 
	\mathcal{F}(1)_{\text{LQCD}} = 0.906(13)\,.
\end{equation}
(The latter does not include the recent result $\mathcal{F}(1)=0.909(17)$~\cite{Bazavov:2021bax}.)
Fits at $\mathcal{O}(\aS, 1/m_{c,b})$, allowing the form factor ratios at and beyond zero recoil to self-consistently float,
obtain~\cite{Bernlochner:2017jka}
\begin{equation}
	\label{eqn:R1R2fits}
	R_1(1) \simeq 1.32(3)\,,\qquad R_2(1) \simeq 0.88(3)\,,
\end{equation}
with a correlation of $-0.7$, which can be compared to our second-order fits in Sec.~\ref{sec:fitres}.

\subsection{Residual chiral expansion}
\label{sec:zrrce}
At zero-recoil, as noted in Sec.~\ref{sec:truncexp}, the RC expansion receives only $\mathcal{O}(\theta^4)$ corrections,
so that the convergence of the expansion might be tested
by examining its zero recoil predictions at $\mathcal{O}(\theta^2)$ versus LQCD data.
At the same time, the $\Bbar \to \Dx$ form factors necessarily have a higher HQ symmetry at $w=1$
(because the initial and final states are in the same $s^{\pi_\ell} = \frac{1}{2}^-$ HQ doublet),
so that the structure of the form factors is more tightly constrained than for $w >1$.
As a result, one can expect higher sensitivity to $\mathcal{O}(1/m_c m_b)$ and $\mathcal{O}(\aS/m_{c,b})$ corrections, which must therefore be included.

At $\mathcal{O}(\theta^2)$, the full $\mathcal{O}(1/m_{c,b}^2, \aS/m_{c,b})$ expressions for the $\hL{i}{1,2}(1)$ and $\hM{i}(1)$ form factors 
can be read off from Eqs.~\eqref{eqn:Lhat12def} and Appendix~\ref{app:asmcb}.
This leads to
\begin{subequations}
\label{eqn:FGhqetrc}
\begin{align}
	\mathcal{G}(1)_{\text{RC}}
		& \simeq 1 + \haS \big[\Cv1 + \Cv2(1-\rho_D) + \Cv3(1+\rho_D)\big] \\
		& \qquad - (\ec-\eb)\rho_D\big[2\hat\eta(1) -1\big] \nn\\
		& \qquad + \Big[\ec^2(3 - 4\rho_D) + \eb^2(3 + 4\rho_D) -6\ec\eb\Big]\bigg[\frac{\lam1 + 3\lam2}{6\LamB^2}\bigg] \nn\\
		& \qquad -\haS\ec\rho_D\big[ (2\hat\eta(1) -1)\Cv1 - 2\Cv3\big] \nn\\
		& \qquad + \haS\eb\rho_D\big[(2\hat\eta(1) -1) \Cv1 - 2\Cv2\big] \,,\nn\\
	\mathcal{F}(1)_{\text{RC}}
		& \simeq 1 + \haS \Ca1 + \ec^2\bigg[\frac{\lam1  -\lam2}{2\LamB^2}\bigg] + \big(2\ec\eb+3\eb^2\big)\bigg[\frac{\lam1 + 3\lam2}{6\LamB^2}\bigg]\,,\\
	R_1(1)_{\text{RC}}
		& \simeq 1 + \haS \big[\Cv1 - \Ca1\big] + \ec - \eb(2\hat\eta(1) -1)  \\
		& \qquad -\ec^2 \frac{2 \lam1}{3\LamB^2} -\eb^2\frac{2(\lam1 + 3\lam2)}{3\LamB^2}\nn\\
		& \qquad + \haS\ec\big[ \Cv1 - \Ca1 - 2\hat\eta(1)\Cv3\big] \nn\\
		& \qquad + \haS\eb\big[ (2\hat\eta(1) - 1)\big( \Ca1 - \Cv1\big) - 2\hat\eta(1) \Cv2 \big]\,, \nn\\
	R_2(1)_{\text{RC}}
		& \simeq	1 + \haS\big[\Ca3 + \rDs \Ca2\big]  \\
		& \qquad -\ec\big[(1+\rDs) \hat\eta(1) + (1 - \rDs)4\hat\chi_2(1) + \rDs\big] - \eb(2\hat\eta(1) -1) \nn\\
		& \qquad + \ec^2\bigg[\frac{ (2\lam1 - 3\lam2)\rDs - 3\lam2}{3\LamB^2} - 4\hat\beta_2(1)(1-\rDs)\bigg] -2\eb^2\frac{\lam1 + 3\lam2}{3\LamB^2}\nn\\
		& \qquad - \ec\eb(1-\rDs)\bigg[ \frac{15 \lam2 - 4\lam1}{12\LamB^2} + \frac{2\hat\eta(1) -1}{2} - 6 \hvphp(1))\bigg] \nn\\
		& \qquad + \haS\ec\big\{(2\hat\eta(1) +1)\rDs\Ca2 \nn\\
		& \qquad \qquad + [3\hat\eta(1) + 2 - \rDs(\hat\eta(1) + 1)]\Ca3 - 4(1-\rDs)\hat\chi_2(1)C_g^c \big\} \nn\\*
		& \qquad + \haS\eb\big\{ [\rDs(2-3\hat\eta(1)) +\hat\eta(1) - 1]\Ca2 -(2\hat\eta(1) -1)\Ca3  \big\}\,, \nn\\
	R_0(1)_{\text{RC}}
		& \simeq 1 + \frac{1 - \rDs}{1 + \rDs}\bigg\{ -\haS[\Ca2 + \Ca3]	+ \ec(2\hat\eta(1) +1) + \eb(2\hat\eta(1) -1) \\
		& \qquad - 2\ec^2\frac{\lam1 - 3\lam2}{3\LamB^2} + 2\eb^2\frac{\lam1 + 3\lam2}{3\LamB^2} \nn \\
		& \qquad + \haS \big(\Ca2 + \Ca3 \big)\big[\eb(2\hat\eta(1) -1) - \ec(2\hat\eta(1) +1) \big]\bigg\}\,,\nn
\end{align}
\end{subequations}
in which all $C_X$ functions are evaluated at $w=1$.
In $R_2(1)$, two additional subleading Isgur-Wise functions, $\hvphp$ and $\hat\beta_2$, enter that 
may absorb any sensitivity to the higher-order perturbative corrections.
We see from $\mathcal{G}(1)_{\text{RC}}$, $\mathcal{F}(1)_{\text{RC}}$ and $R_0(1)_{\text{RC}}$ that the second-order power corrections
to the $\Bbar \to \Dx l \nu$ rates at $w=1$ are fully determined by $\lam{1,2}$.

The expressions in Eqs.~\eqref{eqn:FGhqetrc} are useful for building intuition concerning the dominant second-order power corrections.
By Luke's theorem the first-order corrections to $\mathcal{F}(1)$ vanish, 
and one finds
\begin{equation}
	\label{eqn:F1NLO}
	\mathcal{F}(1)^{\text{NLO}}_{\text{HQET}} \simeq 0.966(2)\,,
\end{equation}
where the quoted uncertainty arises from the parameter $z = m_c/m_b$, on which $\Ca1$ depends.
Since $\lam1$ is expected to be $\sim -0.3\,\GeV^2$, and $\lam2(\mu) \simeq 0.11\,\GeV^2$, 
then the $\lam1 - \lam2$ second-order correction to $\mathcal{F}(1)$ induces a few percent negative shift, 
towards LQCD expectations, 
while the $\lam1 + 3\lam2$ second-order contribution in $\mathcal{G}(1)$ and elsewhere approximately cancels.

Regarding $\hat\eta(1)$, prior fits at $\mathcal{O}(\aS, 1/m_{c,b})$~\cite{Bernlochner:2017jka}
obtained approximately
\begin{equation}
	\label{eqn:eta1m}
	\hat\eta(1) = 0.3 \pm 0.05\,,
\end{equation}
mainly driven by the LQCD data for $\mathcal{G}(1)$. 
In our fits in Sec.~\ref{sec:fitres} below we do not constrain $\hat\eta(1)$ \emph{a priori}.
However, assuming for the moment that the recovered value of $\eta(1)$ 
will not be overly perturbed by the inclusion of higher-order corrections---this is the 
expected behavior in a well-behaved expansion; we will see that this assumption is justified in Sec.~\ref{sec:fitres}---we 
may use this value of $\hat\eta(1)$ to inform zero-recoil predictions,
when combined with the $1S$ inputs~\eqref{eqn:1Sinputs}, \eqref{eqn:rho1input} and~\eqref{eqn:lam2input}, 
plus the relations~\eqref{eqn:1Slams}.

Some further insight can be gained from examining the numerical forms,
\begin{subequations}
\begin{align}
	\mathcal{G}(1)_{\text{RC}}
		& \simeq 1.105 + [0.006- 0.024\hat\eta(1)\big](\lam1/\GeV^2) -0.170 \hat\eta(1)\,,\\
	\mathcal{F}(1)_{\text{RC}}
		& \simeq 0.963 + 0.092 (\lam1/\GeV^2)\,,\\
	R_1(1)_{\text{RC}}
		& \simeq 1.400 - \big[0.063 + 0.018\hat\eta(1)\big](\lam1/\GeV^2) -0.129\hat\eta(1)\,,\\
	R_2(1)_{\text{RC}}
		& \simeq 0.955 +\big[0.036 - 0.100 \hat\chi_2(1) -0.062\hat\eta(1) -0.010\hvphp(1)\big](\lam1/\GeV^2) \nn\\
		& \qquad -0.443\hat\eta(1) -0.717\hat\chi_2(1) -0.129 \hat\beta_2(1)\,, \\
	R_0(1)_{\text{RC}}
		& \simeq 1.119  +(0.038 \hat\eta(1)-0.030)(\lam1/\GeV^2) + 0.272 \hat\eta(1)\,,	
\end{align}
\end{subequations}
in which we have kept the $\lam1$ and subleading Isgur-Wise function dependence explicit:
the former arises from the $\lam1$ dependence in Eqs.~\eqref{eqn:FGhqetrc}, 
as well as from $\eQ$, via the relations~\eqref{eqn:1Slams},
from which one may express $\rho_1$ in terms of $\lam1$.
In these expressions, we have taken the central values $\mbS = 4.71\,\GeV$ and $\dmbc = 3.40\,\GeV$.
The $\hat\eta(1)$-induced corrections to $\mathcal{G}(1)$
shift it down towards the LQCD predicted range while the $\lam1$-dependent terms approximately cancel.
In $R_1(1)$ the $\hat\eta(1)$-induced corrections tend to cancel against the $\lam1$ terms.
In $R_2(1)$ the $\hat\eta(1)$-induced corrections dominate, shifting it downwards.
Thus the second-order corrections at $\mathcal{O}(\theta^2)$ have a structure 
naively compatible with the data in Eqs.~\eqref{eqn:F1G1} and~\eqref{eqn:R1R2fits}.

The form factor $\mathcal{F}(1)$ is fully constrained at second order by $\lam{1,2}$, 
and its first-order corrections vanish.
Because of this, in addition to the $\mathcal{O}(1/m_{c,b}^2, \aS/m_{c,b})$ corrections, 
we also include
the full $\mathcal{O}(\aS^2)$ correction to the axial-vector current~\cite{Czarnecki:1996gu, Czarnecki:1997cf, Franzkowski:1997vg},
that amounts to
\begin{equation}
	\label{eqn:F1as2shift}
	\delta \mathcal{F}(1) \simeq -0.944 C_F \haS^2 \simeq -0.009\,,
\end{equation}
where $C_F = (N_c^2 -1)/(2N_c)$. 
We do not include such corrections in SM form factors that do not have vanishing first-order corrections.
In practice, in our fits in Sec.~\ref{sec:fitres}, we implement Eq.~\eqref{eqn:F1as2shift} via an overall shift in $\hat{h}_{A_1}$, such that
\begin{equation}
	\label{eqn:hA1as2shift}
	\hat{h}_{A_1} \to \hat{h}_{A_1} -0.944 C_F \haS^2\,.
\end{equation}
This additional $\mathcal{O}(\aS)$ term should be included in $\hat{h}_{A1}$ when using the fit results in Sec.~\ref{sec:fitres}.

\begin{figure}[t]
	\includegraphics[width = .6\textwidth]{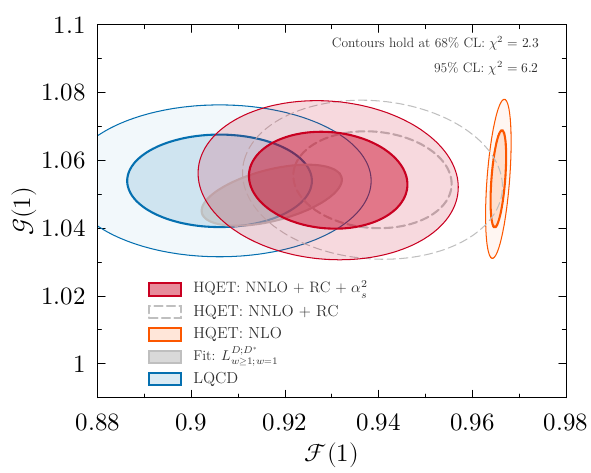}
	\caption{Second-order HQET predictions for $\mathcal{F}(1)$ and $\mathcal{G}(1)$ 
	at $\mathcal{O}(\theta^2)$ in the RC expansion (red ellipse),
	using the inputs~\eqref{eqn:1Sinputs}, \eqref{eqn:rho1input} and~\eqref{eqn:lam2input}, and Eq.~\eqref{eqn:eta1m},
	and including the $\mathcal{O}(\aS^2)$ corrections~\cite{Czarnecki:1996gu, Czarnecki:1997cf} in Eq.~\eqref{eqn:F1as2shift}.
	These are compared to LQCD predictions (blue ellipse) and predictions at first order in HQET (orange ellipse).
	The $68\%$ CL region from a fit of the RC expansion at $\mathcal{O}(1/m_{c,b}^2, \aS/m_{c,b}, \theta^2)$ to LQCD 
	and experimental data (see Sec.~\ref{sec:fitres}) is shown by the gray solid ellipse.
	The dashed gray ellipse shows RC expansion predictions without the $\mathcal{O}(\aS^2)$ terms.
}
	\label{fig:F1G1}
\end{figure}

In Fig.~\ref{fig:F1G1} we show the CLs (red ellipse) in the $\mathcal{F}(1)$--$\mathcal{G}(1)$ plane
determined by the \emph{fit inputs}~\eqref{eqn:1Sinputs}, \eqref{eqn:rho1input} and~\eqref{eqn:lam2input}, 
as well as Eqs~\eqref{eqn:eta1m} and~\eqref{eqn:F1as2shift} and the relations~\eqref{eqn:1Slams},
imposed on their $\mathcal{O}(1/m_{c,b}^2, \aS/m_{c,b}, \theta^2)$ expressions in Eqs.~\eqref{eqn:F1G1HQETall}.
This range is in agreement with the LQCD predictions (blue ellipse) at the $0.85\sigma$ level $(p =0.40)$.
If the $\mathcal{O}(\aS^2)$ correction in Eq.~\eqref{eqn:F1as2shift} is not included, 
the agreement is at the $1.4\sigma$ level $(p=0.16)$, indicated by the dashed gray ellipse.
By comparison, the first-order HQET CL (orange ellipse), 
whose small $\mathcal{F}(1)$ uncertainty is determined by the $1S$ inputs as in Eq.~\eqref{eqn:F1NLO},
is approximately $4.2\sigma$ from the LQCD values.
Fits at first-order must therefore consider nuisance parameters for higher-order terms, 
or consider shape-only fits as in Ref.~\cite{Bernlochner:2017jka};
see Sec.~\ref{sec:fitscens}.

\subsection{Vanishing chromomagnetic limit}
\label{sec:zrVClimit}
Because $\lam{1,2} \to 0$ in the VC limit, we fix $\lam{1,2} = 0$ in the $1S$ scheme, 
with $\mbS = 4.71(5)\,\GeV$ and $\dmbc = 3.40(2)\,\GeV$.
Although the mass degeneracy in the HQ doublet is formally not lifted at $\mathcal{O}(1/m_{c,b}^2)$,
we use the physical masses of the hadrons in the kinematics,
just as in the approach used for fits at $\mathcal{O}(1/m_{c,b},\aS)$ in Ref.~\cite{Bernlochner:2017jka}, at which order $\lam{1,2}$ also formally vanish.
However, note Ref.~\cite{Bernlochner:2017jka} used the nonzero value $\lam{1} = -0.3\,\GeV^2$, 
which amounts to a correction of about $5\%$ in the value of $\LamB$ used here. 

In the $G_{\ab} \to 0 $ limit, the expressions for the $\mathcal{F}(1)$ and $\mathcal{G}(1)$ form factors 
and the ratios $R_1(1)$ and $R_2(1)$ become, 
\begin{subequations}
\label{eqn:FGhqetnc}
\begin{align}
	\mathcal{G}(1)_{\text{VC}}
		& \simeq 1 + \haS [\Cv1 + \Cv2(1-\rho_D) + \Cv3(1+\rho_D)] \\*
		& \qquad - (\ec-\eb)\rho_D\big[2\hat\eta(1) -1\big] \nn\\*
		& \qquad + 2(\ec - \eb)^2 \hat{c}_0(1) + 2(\ec^2 -\eb^2)\rho_D[\hat\varphi_0'(1)-\hat{e}_3(1)] \nn \\
		& \qquad -\haS\ec\rho_D\big[ (2\hat\eta(1) -1)\Cv1 - 2\Cv3\big] \nn\\*
		& \qquad + \haS\eb\rho_D\big[(2\hat\eta(1) -1) \Cv1 - 2\Cv2\big] \,,\nn\\
	\mathcal{F}(1)_{\text{VC}}
		& \simeq 1 + \haS \Ca1 +  2 (\ec-\eb)^2 \hat c_0(1)  \,, \label{eqn:FGhqetncF1}\\
	R_1(1)_{\text{VC}}
		& \simeq 1 + \haS [\Cv1 - \Ca1] + \ec - \eb(2\hat\eta(1) -1)  \nn\\
		& \qquad +2\ec^2 \hat\varphi_0'(1) + 2\eb^2[\hat\varphi_0'(1)-\hat{e}_3(1)] -2\ec\eb [\hat{e}_3(1)+2\hat\varphi_0'(1)]\nn\\
		& \qquad + \haS\ec\big[ \Cv1 - \Ca1 - 2\hat\eta(1)\Cv3\big] \nn\\
		& \qquad + \haS\eb\big[ (2\hat\eta(1) - 1)\big( \Ca1 - \Cv1\big) - 2\hat\eta(1) \Cv2 \big]\,, \\
	R_2(1)_{\text{VC}}
		& \simeq	1 + \haS[\Ca3 + \rDs \Ca2]  \\
		& \qquad -\ec\big[(1+\rDs) \hat\eta(1) + \rDs\big] - \eb(2\hat\eta(1) -1) \nn\\
		& \qquad -\ec^2\big[2\rDs\hat\varphi_0'(1)  + (1+\rDs)\hat{e}_3(1)\big] + 2\eb^2\big[\hat\varphi_0'(1) - \hat{e}_3(1)\big] \nn\\
		& \qquad - \ec\eb\big\{(1-\rDs)[2\hat\eta(1) -1 + 8\hat\varphi_0'(1)]/2 + (3 + \rDs)\hat{e}_3(1)\big\}\nn\\
		& \qquad + \haS\ec\big\{(2\hat\eta(1) + 1)\rDs\Ca2 + [3\hat\eta(1) + 2 - \rDs(\hat\eta(1) + 1)]\Ca3  \big\} \nn\\
		& \qquad + \haS\eb\big\{ [\rDs(2-3\hat\eta(1)) +\hat\eta(1) - 1]\Ca2 -(2\hat\eta(1) -1)\Ca3  \big\}\,, \nn\\
	R_0(1)_{\text{VC}}
		& \simeq 1 + \frac{1 - \rDs}{1 + \rDs}\bigg\{ -\haS[\Ca2 + \Ca3] + \ec(2\hat\eta(1) +1) + \eb(2\hat\eta(1) -1) \\
		& \qquad + 2\ec^2[\hat{e}_3(1) + \hat\varphi_0'(1)] + 2\eb^2[\hat{e}_3(1) - \hat\varphi_0'(1)] + 4\ec\eb\hat{e}_3(1) \nn\\
		& \qquad + \haS \big(\Ca2 + \Ca3 \big)\big[\eb(2\hat\eta(1) -1) - \ec(2\hat\eta(1) +1) \big]\bigg\}\,,\nn
\end{align}
\end{subequations}
in which all $C_X$ functions are evaluated at $w=1$.
Here, one finds a downward shift in $\mathcal{F}(1)$ towards the LQCD prediction
by requiring $\hat c_0(1) < 0$.
However, unlike in the RC expansion, the same downward shift from the second-order power correction enters into both $\mathcal{F}(1)$ and $\mathcal{G}(1)$--- both 
have the same $2(\ec-\eb)^2 \hat{c}_0(1)$ term---resulting in a large downward shift in $\mathcal{G}(1)$.
It is again useful to examine the numerical forms for the truncated expressions,
\begin{subequations}
\begin{align}
	\mathcal{G}(1)_{\text{VC}}
		& \simeq 1.108 + 0.055 \hat{c}_0(1) - 0.171\hat\eta(1) + 0.047(\hat\varphi_0'(1)-\hat{e}_3(1))\,,\\
	\mathcal{F}(1)_{\text{VC}}
		& \simeq 0.966 + 0.055\hat{c}_0(1) \,,\\
	R_1(1)_{\text{VC}}
		& \simeq 1.407 - 0.130\hat\eta(1) + 0.047\hat\varphi_0'(1)-0.038\hat{e}_3(1)\,,\\
	R_2(1)_{\text{VC}}
		& \simeq 0.992 -0.465\hat\eta(1) -0.069\hat\varphi_0'(1)-0.015\hat{e}_3(1)\,,\\
	R_0(1)_{\text{VC}}	
		& \simeq 1.105  + 0.274\hat\eta(1) + 0.044\hat\varphi_0'(1) - 0.017\hat{e}_3(1) \,,
\end{align}
\end{subequations}
in which we use $\mbS = 4.71\,\GeV$, $\dmbc = 3.40\,\GeV$ and $\lam1 \to 0$.
Assuming $\hat\eta(1)$ falls in the range given in Eq.~\eqref{eqn:eta1m} when second-order corrections are included, 
the downward shift in $\mathcal{G}(1)$ must be counter-balanced by the $\hat\varphi_0'(1)-\hat{e}_3(1)$ term,
resulting in a large upward shift in $R_1(1)$, potentially such that $R_1(1) > 1.40$,
in some tension with the prior fit results in Eq.~\eqref{eqn:R1R2fits}.
However, a proper assessment of the consistency of the VC limit with data 
requires a full fit beyond zero recoil.

\section{Fits}
\label{sec:fitres}

\subsection{Experimental data}
\label{sec:explqcd}
Prior to 2018 only two measurements were available for (isospin-averaged) $\Bbar \to D \ell\nu$ and $\Bbar^0 \to D^{*+} \ell \nu$, 
Ref.~\cite{Glattauer:2015teq} (`Belle 15') and Ref.~\cite{Abdesselam:2017kjf} (`Belle 17'), respectively, 
that provided kinematic distributions for the recoil and decay angles fully corrected for detector effects. 
These analyses, by providing `unfolded' kinematic distributions, 
permit fits outside of the experimental frameworks, 
using different parametrizations of the $\Bbar \to \Dx$ form factors than what was used in each analysis.

Subsequently, the Belle Collaboration performed an untagged analysis of $\Bbar^0 \to D^{*+} \ell \nu$~\cite{Belle:2018ezy} (`Belle 19'),
providing response functions and efficiencies, 
into which alternate form factor parametrizations may be folded,
in order to generate predictions for bin yields in the kinematic distributions. 
In this work, we will consider combined fits that include the 2017 and 2019 $\Bbar \to D^* \ell \nu$ analyses either together or separately.
A summary of the experimental inputs for each fit scenario is shown in Table~\ref{tab:FitKey}.
We use masses $m_{\Bbar^0} = 5.28$\,GeV, $m_{D} = 1.87$\,GeV and $m_{D^{*+}} = 2.01$\,GeV.

In most of our fits, we only fit to the $w$ spectra of Refs.~\cite{Abdesselam:2017kjf} and \cite{Belle:2018ezy}, 
as there is little information constraining the form factors encoded in the projections of the angular distributions. 
We combine and unfold the reported results for electrons and muons of Ref.~\cite{Belle:2018ezy} using the provided migration matrices and efficiency corrections. 
Systematic uncertainties are incorporated into the unfolding procedure using nuisance parameters 
that act upon the resulting yields to avoid the d'Agostini bias (cf. Eq.~(3) in Ref.~\cite{DAgostini:1993arp}). 

\subsection{Lattice QCD inputs}

For $\Bbar \to D$ decay, LQCD predictions for the SM form factors $f_+$ and $f_0$ have 
long been available at and beyond zero recoil~\cite{MILC:2015uhg,Na:2015kha}.
Their relationship to $h_+$ and $h_-$ is
\begin{equation}
	f_+ = \frac{h_+(1+\rD) - h_-(1-\rD)}{2\sqrt{\rD}}\,,\quad f_0 = \sqrt{\rD}\bigg[h_+\frac{w+1}{1+ \rD} - h_-\frac{w-1}{1-\rD}\bigg]\,.
\end{equation}
Ref.~\cite{MILC:2015uhg} is currently the most precise, 
and conveniently provides a synthetic dataset at three values of $w =1.0$, $1.08$ and $1.16$,
including statistical and systematic correlations.
These may be incorporated directly into the combined fits;
the values are shown in Table~\ref{tab:FNAL} 
(the corresponding $\mathcal{G}(1) = 2\sqrt{\rD}\,f_+(1)/(1 +\rD)$ is shown in Eq.~\eqref{eqn:F1G1}).
Without combination with any experimental data, 
the predicted LFUV ratios from Ref.~\cite{MILC:2015uhg} and \cite{Na:2015kha}, respectively, are
$R(D)_{\text{LQCD}} = 0.285(15)$ and $R(D)_{\text{LQCD}} = 0.300(8)$, leading to the FLAG average~\cite{Aoki:2021kgd}
\begin{equation}
	R(D)_{\text{LQCD}} =  0.2934(53)\,.
\end{equation}

\begin{table}[t]
\newcolumntype{C}{ >{\centering\arraybackslash $} m{2cm} <{$}}
\begin{tabular}{C|CCC}
\hline\hline
  \text{Form factor} & w = 1.0 & w = 1.08 & w = 1.16 \\ \hline
  f_{+} & 1.1994(95)  &  1.0941(104)  &  1.0047(123) \\
  f_{0} & 0.9026(72)  &  0.8609(77)  &  0.8254(94) \\
  \hline
  \text{Form factor} & w = 1.03 & w = 1.10 & w = 1.17 \\ \hline
  h_{A_1}	 & 0.877(16)	 & 0.807(15)	 & 0.745(22)\\ 
  h_{A_2}	 & -0.586(82)	 & -0.492(82)	 & -0.391(95)\\ 
  h_{A_3}	 &1.213(75)	 & 1.103(75)	 & 0.989(86)\\ 
  h_{V}	 & 1.212(44)	 & 1.079(44)	 & 0.948(54)\\ 
\hline\hline
\end{tabular}
\caption{Top: Synthetic data for the $\Bbar \to D$ form factors at $w =1.0,\,1.08,\,1.16$~\cite{MILC:2015uhg},
and for the $\Bbar \to D^*$ form factors at $w =1.03,\,1.10,\,1.17$~\cite{Bazavov:2021bax}.
The correlations can be found in Table VII of Ref.~\cite{MILC:2015uhg},
and in the ancillary files of Ref.~\cite{Bazavov:2021bax}, respectively.}
\label{tab:FNAL}
\end{table}

For $\Bbar \to D^*$, LQCD results are available at zero recoil for $h_{A_1}(1) = \mathcal{F}(1)$
with the most precise result as quoted in Eq.~\eqref{eqn:F1G1}.
For all SM $\Bbar \to D^*$ form factors at and beyond zero recoil, 
recently Fermilab/MILC~\cite{Bazavov:2021bax} has provided the first predictions (not yet published),
including synthetic data at $w=1.03$, $w=1.10$ and $w=1.17$;
the values are shown in Table~\ref{tab:FNAL}.
Results for the $\Bbar \to D^*$ form factors beyond zero recoil are also expected soon from HPQCD.

\subsection{Fitting setup and scenarios}
\label{sec:fitscens}
To determine the leading and subleading Isgur-Wise functions and $|V_{cb}|$, 
we carry out a simultaneous $\chi^2$ fit of the experimental and lattice data
(and in some scenarios include constraints from QCDSR).
To take into account the uncertainties in $m_b^{1S}$ and $\delta m_{bc}$, we introduce both as nuisance parameters into the fit, 
assuming Gaussian constraints (see Eq.~\eqref{eqn:1Sinputs} for their value and uncertainties). 
The constraints from LQCD are incorporated into the fit assuming multivariate Gaussian errors. 
The $\chi^2$ function is numerically minimized and uncertainties are evaluated using the asymptotic approximation 
by scanning the $\chi^2$ contour to find the $\Delta \chi^2 = 1$ crossing point, providing the $68\%$ confidence level. 

As mentioned in Sec.~\ref{sec:zrp}, 
in the $\Bbar \to \Dx$ transitions
many first-order corrections vanish at zero recoil,
such that the HQ expansion is more constrained at $w=1$ than beyond zero recoil,
prospectively leading to higher sensitivity to second-order contributions.
For this reason, when working at $\mathcal{O}(\aS, 1/m_{c,b})$
as in Ref.~\cite{Bernlochner:2017jka},
it is a well-motivated approach to consider information from `shape-only' fits.
In these fits,
information concerning the overall normalization of the $\Bbar \to \Dx$ rates---in effect,
LQCD predictions for $\mathcal{G}(1)$ and $\mathcal{F}(1)$---are not imposed,
and only the shapes of the $\Bbar \to \Dx$ spectra are used to constrain the subleading Isgur-Wise functions.
Two different variations of this approach were considered: 
one in which no lattice information was used
(denoted `NoL' in Ref.~\cite{Bernlochner:2017jka});
and one in which beyond zero recoil LQCD predictions for $\Bbar \to D$ were included 
(denoted `$L_{w\ge 1}$' in Ref.~\cite{Bernlochner:2017jka}).

At zero recoil, the $\mathcal{O}(1/m_{c,b}^2)$ corrections are also more constrained.
As can be seen in Eqs.~\eqref{eqn:BDFFs}, \eqref{eqn:BDsFFs} and~\eqref{eqn:Lhat12def}, 
at $\mathcal{O}(1/m_{c,b}^2, \theta^2)$ all corrections to the matrix elements at zero recoil are determined just by $\lam{1,2}$,
while beyond zero recoil effects from $\hvphq$ can become important.
Along similar lines, 
in Fig.~\ref{fig:F1G1} we see that the $\mathcal{O}(\aS^2)$ corrections 
are relevant at zero recoil.
It therefore remains interesting to consider similar `shape-only' fits, 
that probe the structure of second-order power corrections at $\mathcal{O}(\theta^2)$ beyond zero recoil,
and we therefore also include the correction in Eq.~\eqref{eqn:hA1as2shift} in all our fits.

In Ref.~\cite{Bernlochner:2017jka} additional constraints were considered for the subleading Isgur-Wise functions,
$\hat\eta(1)$, $\hat\chi_2(1)$ and $\hat\chi_{2,3}'(1)$, 
arising from QCDSR calculations~\cite{Ligeti:1993hw, Neubert:1992wq,Neubert:1992pn},
including renormalization improvement factors~\cite{Neubert:1993mb} 
(which, for improvement at renormalization scale $\mu$, 
mandates the inclusion of a compensating $-C^Q_g(\mu)$ factor 
in the $a_Q$ coefficient~\eqref{eqn:aQ}, 
that enters at $\mathcal{O}(\aS/m_{c,b})$.)
Because these calculations are model dependent,
Ref.~\cite{Bernlochner:2017jka} assigned inflated uncertainties in the fit inputs, 
typically much larger than the fit uncertainties
(and much larger than the size of $\mathcal{O}(\aS/m_{c,b})$ corrections).\footnote{%
See footnote~\ref{foot:uncert}.\label{foot:uncert2}}
The QCDSR inputs were taken to be~\cite{Bernlochner:2017jka}
\begin{gather}
	\label{eqn:inputs}
	\hat{\chi}_2(1) = -0.06 \pm 0.02\,, \qquad
 	 \hat{\chi}_2'(1) = 0 \pm 0.02\,, \qquad
	\hat{\chi}_3'(1) = 0.04 \pm 0.02\,, \nn\\*
	\hat\eta(1)  = 0.62 \pm 0.2\,,  \qquad  \hat\eta'(1) = 0 \pm 0.2\,. 
\end{gather}
To compare with prior fits using QCDSR (denoted with a `$+\text{SR}$' suffix in Ref.~\cite{Bernlochner:2017jka}), 
we perform some fits also with these QCDSR constraints.
However, the future arc of precision data-driven fits bends away from the ongoing inclusion of inputs
with poorly quantifiable theory uncertainties,
and therefore we focus on fits without QCDSR constraints.

The various fit scenarios and their inputs considered in this work are summarized in Table~\ref{tab:FitKey}. 
They comprise the following:
\begin{enumerate}[wide, labelwidth=0pt, labelindent=0pt, label = \textbf{(\roman*)}, noitemsep, topsep =0pt, itemjoin =\quad, series = fits]
\item Our baseline fit scenario uses all published LQCD data---i.e., 
except for the not-yet-published $\Bbar \to D^*$ form factors beyond zero recoil~\cite{Bazavov:2021bax}---plus
all available experimental data from Belle. This fit is denoted `\BaseFit', adapting from the notation in Ref.~\cite{Bernlochner:2017jka}.
\item We also perform fit (i), with the relative normalization between the $D$ and $D^*$---in 
effect the relation between $\mathcal{G}(1)$ and $\mathcal{F}(1)$---allowed to float, 
so that only shape information is used. This fit is denoted as `\ShapeFit'. 
\item We further consider a fit using all available LQCD data, denoted by `\AllLQCDFit'.
\item A fit that includes only experimental data, but no LQCD inputs, labelled `\NoLQCDFit'. 
\end{enumerate}
In addition, we consider the same \BaseFit fit, with the following variations:
\begin{enumerate}[wide, labelwidth=0pt, labelindent=0pt, label = \textbf{(\roman*)}, noitemsep, topsep =0pt, itemjoin =\quad, resume = fits]
\item Using only either 2017 or 2019 $\Bbar \to D^*$ data from Belle, denoted with a `17' or `19' suffix, respectively
\item Including QCDSR as discussed above, denoted with a `+SR'
\item Including LQCD data beyond zero recoil for $h_{A_1}$ alone, denoted `\hAOneFit'.
This fit provides an interesting contrast to the \AllLQCDFit fit.
\end{enumerate}
To further characterize the role of the second order power corrections, finally we consider:
\begin{enumerate}[wide, labelwidth=0pt, labelindent=0pt, label = \textbf{(\roman*)}, noitemsep, topsep =0pt, itemjoin =\quad, resume = fits]
\item A fit at first-order in the HQ expansion, 
similar to the abovementioned `$L_{w\ge1}$+SR' fit of Ref.~\cite{Bernlochner:2017jka}, 
which we denote here with a `NLO' suffix. 
\end{enumerate}

\begin{table}[t]
\newcolumntype{C}{ >{\centering\arraybackslash } m{2cm} <{}}
\newcolumntype{D}{ >{\centering\arraybackslash } m{1.25cm} <{}}
\resizebox{\textwidth}{!}{
\begin{tabular}{clc|c|cccc|c|DDD}
	\hline\hline
	&\multirow{2}{*}{Fit}  &  \multirow{2}{*}{Order} & \multirow{2}{*}{ \makecell{Floating\\ norm.}} & \multicolumn{4}{c|}{Lattice QCD}   &  \multirow{2}{*}{QCDSR} &  \multicolumn{3}{c}{Belle Data}\\
	& &  &  &  $f_{+,0}(w \ge 1)$ &  $\mathcal{F}(1)$  & $h_{A_1}(w >1)$ & $h_{A_{2,3}, V}(w>1)$ & & '15 & '17 &  '19  \\ \hline
	\rowcolor[gray]{.9}[6pt][6pt] 
	& \BaseFit				& $\aS/m_{Q}, 1/m_{Q}^2$ & --- &$\checkmark$ & $\checkmark$ & --- & --- & --- & $\checkmark$ & $\checkmark$ & $\checkmark$ \\
	& \ShapeFit			& $\aS/m_{Q}, 1/m_{Q}^2$ & $\checkmark$ &$\checkmark$ & $\checkmark$ & --- & --- & --- & $\checkmark$ & $\checkmark$ & $\checkmark$ \\
	& \NoLQCDFit & $\aS/m_{Q}, 1/m_{Q}^2$ & --- & --- & --- & --- & --- & --- & $\checkmark$ & $\checkmark$ & $\checkmark$ \\
	& \AllLQCDFit 			& $\aS/m_{Q}, 1/m_{Q}^2$ & --- &$\checkmark$ & $\checkmark$ & $\checkmark$ & $\checkmark$ & --- & $\checkmark$ & $\checkmark$ & $\checkmark$ \\
	&  \BaseFit[NLO] 			& $\aS, 1/m_{Q}$ & $\checkmark$ & $\checkmark$ & $\checkmark$ & --- & --- & $\checkmark$ & $\checkmark$ & $\checkmark$ & $\checkmark$ \\
	&  \BaseFit[+SR] 			        & $\aS/m_{Q}, 1/m_{Q}^2$ & --- &$\checkmark$ & $\checkmark$ & --- & --- & $\checkmark$ & $\checkmark$ & $\checkmark$ & $\checkmark$ \\
	&  \BaseFit[17] 				& $\aS/m_{Q}, 1/m_{Q}^2$ & --- &$\checkmark$ & $\checkmark$ & --- & --- & --- & $\checkmark$ & $\checkmark$ & --- \\
	&  \BaseFit[19] 				& $\aS/m_{Q}, 1/m_{Q}^2$ & --- &$\checkmark$ & $\checkmark$ & --- & --- & --- & $\checkmark$ & ---  & $\checkmark$ \\
	&  \hAOneFit 		& $\aS/m_{Q}, 1/m_{Q}^2$ & --- &$\checkmark$ & $\checkmark$ & $\checkmark$ & --- & --- & $\checkmark$ & $\checkmark$ & $\checkmark$ \\
	\hline\hline
\end{tabular}}
\caption{Summary of theory and data inputs for each fit scenario. Our baseline fit scenario is highlighted in gray. 
Note the $\mathcal{O}(\aS^2)$ shift in Eq.~\eqref{eqn:hA1as2shift} is also imposed in all second-order fits.}
\label{tab:FitKey}
\end{table}

\subsection{Nested hypothesis tests}

Before proceeding to obtain results for our various fit scenarios,
we employ a nested hypothesis test (NHT)-based prescription to determine 
the optimal number of parameters for the \BaseFit fit scenario.
Such a prescription not only allows systematic determination of those parameters to which the current data has sensitivity,
but also prevents overfitting.
The optimal parameter set obtained through this prescription depends on the precision of the available experimental data,
such that the prescription permits systematic improvements as future data becomes available.

We use here a variation of the prescription developed in Ref.~\cite{Bernlochner:2019ldg}.
The core idea of an NHT is to test a $N$-parameter fit hypothesis 
versus alternative fit hypotheses that use one additional parameter.
The difference in $\chi^2$,
\begin{equation}
	\Delta \chi^2 = \chi^2_{N} - \chi^2_{N+1} \,.
\end{equation} 
provides a convenient test statistic, 
because it is distributed as a $\chi^2$ in the large $N$ limit~\cite{wilks1938} with a single degree of freedom.
We choose $\Delta \chi^2 = 1$ as the decision boundary: the $(N+1)$-parameter hypothesis is then rejected in favor 
of the $N$-parameter fit at $68\%$ CL.

As in Ref.~\cite{Bernlochner:2019ldg}, we apply the NHT starting from a suitably small initial number of parameters.
In this case, based on the parameters entering at zero-recoil (see Sec.~\ref{sec:zrrce}) 
we pick all the HQ mass parameters, the leading Isgur-Wise parameters,
and $\hat\eta(1)$. Thus, the initial parameters are
\begin{equation}
	|V_{cb}|\,;\quad \mbS\,, \quad \dmbc\,, \quad\rho_1\,, \quad \lam2\,; \quad  \rhossq\,, \quad \cs\,;\quad  \hat\eta(1)\,.
\end{equation}
We then incrementally add all combinations of the remaining seven candidate parameters
\begin{equation}
	\hat\eta'(1)\,, \quad \hat{\chi}_2(1)\,, \quad \hat{\chi}'_{2}(1)\,, \quad \hat{\chi}'_{3}(1)\,, \quad \hvphp(1)\,, \quad \hat\beta_{2}(1)\,, \quad \hat\beta'_{3}(1)\,,
\end{equation}
one by one.
This generates a `graph' of fit hypotheses, 
with each node of the graph representing a possible set of fit parameters, 
and each edge denoting the addition of one of the candidate parameters.
Over the graph, we identify a `terminating node'---a
parameter set---as a fit hypothesis that is preferred over all hypotheses that nest it.
In order to avoid runaways in fit parameters, 
we constrain $\hvphp(1)$, $\hat\beta_{2}(1)$, and $\hat\beta'_{3}(1)$ to be at most $\mathcal{O}(1)$ 
(in practice less than approximately $9.$)
in a terminating node.
We further require that no two parameters are more than approximately $95\%$ correlated,
in order to avoid flat directions and consequent overfitting and/or non-Gaussian uncertainties.
The terminating node with the fewest parameters (and hence the largest number of degrees of freedom) 
and lowest $\chi^2$ is then selected as the optimal fit. 

Under this prescription, we find sixteen terminating nodes.
Of these, we observe eight nodes involve either $\hat\beta_{2}(1)$ or  $\hat\beta'_{3}(1)$, 
and are the same as the remaining eight under the approximate replacement 
$\sim \ec\beta_2(1) \to \chi_2(1)$ or $\sim \ec\beta_3'(1) \to \chi_3'(1)$.
As discussed in Sec.~\ref{sec:zerorec}, because we do not expect sensitivity to 
$\mathcal{O}(1/m_cm_b, 1/m_b^2)$ or $\mathcal{O}(\aS/m_{c,b})$ corrections in the currently-available data,
we expect that $\chi_{2}(1)$ and $\chi_{3}'(1)$ can reabsorb $\ec\beta_{2}(1)$ and $\ec\beta_{3}'(1)$, respectively, 
as in Eq.~\eqref{eqn:b2b3abs}.
In effect, $\chi_{2}(1) + \ec \beta_{2}(1)$ and $\chi_{3}'(1) + \ec \beta_{3}'(1)$ should be approximately flat directions in the fit,
which comports with the behavior seen in the terminating nodes.
The fit parameters and corresponding fit results for the remaining eight terminating nodes, 
labelled `$S1$' through `$S8$', are shown in Table~\ref{tab:nhtfits}.
These fits are excellent, with $\chi^2/\text{ndf} \simeq 1$ for all terminating nodes.

\begin{table}[t]
\newcolumntype{C}{ >{\raggedleft\arraybackslash $} m{2.5cm} <{$}}
\newcolumntype{D}{ >{\centering\arraybackslash $} m{2.5cm} <{$}}
\renewcommand{\arraystretch}{0.9}
\resizebox{\textwidth}{!}{
\begin{tabular}{D|>{\columncolor[gray]{.9}[6pt][6pt]}CC>{\columncolor[gray]{.95}[6pt][6pt]}CCCCCC}
\hline\hline
\text{Params}	 & S1	 & S2	 & S3	 & S4	 & S5	 & S6	 & S7	 & S8  \\ 
 \hline
|V_{cb}|\times 10^3	 & 38.70(62)	 & 38.90(64)	 & 38.70(68)	 & 38.70(68)	 & 38.70(69)	 & 38.70(67)	 & 38.80(68)	 & 38.70(69)  \\ 
\rhossq	 & 1.10(4)	 & 1.15(4)	 & 1.19(5)	 & 1.15(5)	 & 1.15(4)	 & 1.10(7)	 & 1.12(8)	 & 1.10(4)  \\ 
\cs	 & 2.39(18)	 & 2.44(19)	 & 2.16(24)	 & 2.25(23)	 & 2.29(29)	 & 2.38(19)	 & 2.41(20)	 & 2.40(29)  \\ 
\hat\chi_2(1)	 & -0.12(2)	 & -0.14(3)	 & \nax	 & \nax	 & -0.12(5)	 & \nax	 & -0.13(4)	 & -0.12(5)  \\ 
\hat\chi_2'(1)	 & \nax	 & \nax	 & -0.15(8)	 & -0.08(7)	 & -0.07(11)	 & \nax	 & \nax	 & 0.00(10)  \\ 
\hat\chi_3'(1)	 & \nax	 & \nax	 & 0.04(1)	 & 0.04(1)	 & \nax	 & 0.04(1)	 & \nax	 & \nax  \\ 
\hat\eta(1)	 & 0.34(4)	 & 0.33(4)	 & 0.34(4)	 & 0.34(4)	 & 0.34(4)	 & 0.34(4)	 & 0.34(4)	 & 0.34(4)  \\ 
\hat\eta'(1)	 & \nax	 & 0.12(10)	 & 0.14(11)	 & \nax	 & 0.15(11)	 & -0.15(14)	 & 0.05(19)	 & \nax  \\ 
\mbS~[\GeV]	 & 4.71(5)	 & 4.71(5)	 & 4.70(5)	 & 4.70(5)	 & 4.71(5)	 & 4.71(5)	 & 4.71(5)	 & 4.71(5)  \\ 
\dmbc~[\GeV]	 & 3.41(2)	 & 3.41(2)	 & 3.41(2)	 & 3.41(2)	 & 3.41(2)	 & 3.41(2)	 & 3.41(2)	 & 3.41(2)  \\ 
\hat\beta_2(1)	 & \nax	 & \nax	 & \nax	 & \nax	 & \nax	 & \nax	 & \nax	 & \nax  \\ 
\hat\beta_3'(1)	 & \nax	 & \nax	 & \nax	 & \nax	 & \nax	 & \nax	 & \nax	 & \nax  \\ 
\hvphp(1)	 & 0.25(21)	 & \nax	 & \nax	 & 0.24(21)	 & \nax	 & 0.53(31)	 & 0.17(40)	 & 0.25(21)  \\ 
\lam2~[\GeV^2]	 & 0.12(2)	 & 0.12(2)	 & 0.12(2)	 & 0.12(2)	 & 0.12(2)	 & 0.12(2)	 & 0.12(2)	 & 0.12(2)  \\ 
\rho_1~[\GeV^3]	 & -0.36(24)	 & -0.35(24)	 & -0.37(24)	 & -0.36(24)	 & -0.37(24)	 & -0.36(24)	 & -0.36(24)	 & -0.36(24)  \\ 
\hline
\chi^2 	 &29.8	 & 30.0	 & 28.9	 & 29.3	 & 29.5	 & 29.6	 & 29.8	 & 29.8  \\ 
\text{ndf} 	 &31	 & 31	 & 30	 & 30	 & 30	 & 30	 & 30	 & 30  \\ 
\rho^2 	 &1.35(5)	 & 1.37(5)	 & 1.34(6)	 & 1.34(6)	 & 1.34(6)	 & 1.34(6)	 & 1.36(6)	 & 1.35(6)  \\ 
c 	 &2.41(17)	 & 2.43(17)	 & 2.14(22)	 & 2.26(21)	 & 2.29(28)	 & 2.40(17)	 & 2.42(17)	 & 2.42(27)  \\ 
\hline\hline
\end{tabular}}
\caption{Fit values and parameters for each terminating node of the nested hypothesis test graph.
The node $S1$ (dark gray) is chosen as the optimal fit hypothesis.
To characterize possible model dependence in the parameter truncation, we also consider $S3$ (light gray).
The last four rows show the corresponding values for the fit $\chi^2$, number of degrees of freedom, 
and the slope and the curvature of $\xi(w)$ at zero recoil.}
\label{tab:nhtfits}
\end{table}

Per our prescription, the terminating node with the fewest parameters and lowest $\chi^2$ is $S1$,
highlighted in gray in Table~\ref{tab:nhtfits}.
In this fit, $\hvphp(1)$ is distinguished from zero at the $1\sigma$ level, 
while all chromomagnetic terms are compatible with zero, except $\hat\chi_2(1)$, 
which is somewhat smaller than $\hat\eta(1)$.
This is in line with expectations from the operator product conjecture (see Sec.~\ref{sec:OPconj}) that leads to the RC expansion.
It is notable that $S2$ exchanges $\hvphp(1)$ with $\hat\eta'(1)$, producing a comparably good fit:
while the zero-recoil second-order corrections from $\lam{1,2}$ are important, 
the fit appears not to distinguish contributions from first- versus second-order Isgur-Wise functions beyond zero recoil.
This matches the expectations of Ref.~\cite{Bernlochner:2017jka}.

\begin{figure}[t]
	\subfigbottomskip=-10pt
	\subfigcapskip=-10pt
	\subfigure[][\label{fig:FFR1}]{
        		\centering
     	  	\includegraphics[width=0.48\textwidth]{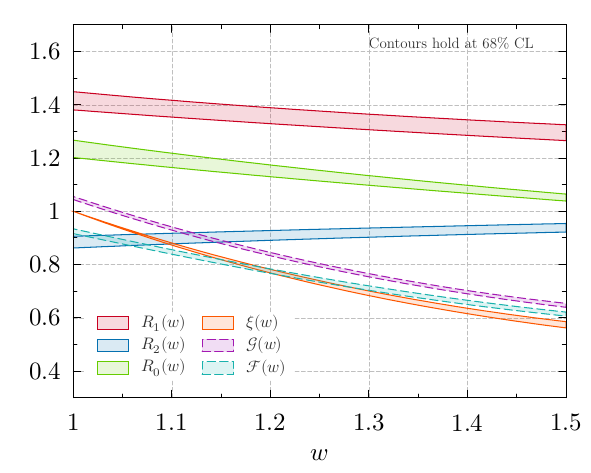}
       }
	\hfill
	\subfigure[][\label{fig:FFR2}]{
        		\centering
		\includegraphics[width=0.48\textwidth]{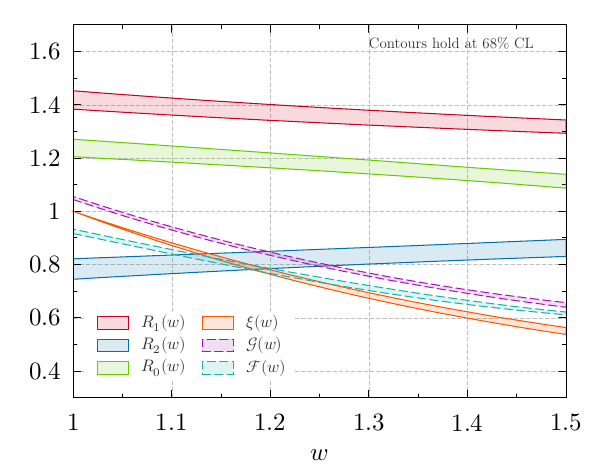}
      }
	\caption{Form factors and form-factor ratios as functions of the recoil parameter for \subref{fig:FFR1} the $S1$ and \subref{fig:FFR2} the $S3$ fits.}
	\label{fig:FFR}
\end{figure}

To characterize the behavior of the selected fit hypothesis $S1$, 
in Fig.~\ref{fig:FFR1} we show the various form factor ratios $R_{1,2,0}(w)$,
along with the leading-order Isgur-Wise function $\xi(w)$, 
and the form factors $\mathcal{F}(w)$ and $\mathcal{G}(w)$.
The uncertainties in all the form factor ratios are well controlled.
The small uncertainties in $\mathcal{F}(w)$ and $\mathcal{G}(w)$
are directly determined by the precision of the LQCD and experimental data.
For comparison, we also show in Fig.~\ref{fig:FFR2} the same ratios and form factors for the $S3$ hypothesis,
which has the lowest $\chi^2$ of those fits with $30$ degrees of freedom.
The $S3$ fit results exhibit slightly larger uncertainties,
while $\mathcal{F}(w)$ and $\mathcal{G}(w)$ remain almost entirely unchanged,
as expected, and $\xi(w)$ deviates from $S1$ only very slightly at high recoil.
Most notable is an overall downward shift in $R_2$, and a small disagreement in $R_0$ at high recoil,
both at the $1\sigma$ level or so (depending on correlations).
This is perhaps not a surprise, because $h_{A_2}$ and $h_{A_3}$ are sensitive to $\hat\eta(1)$ at $\mathcal{O}(\ec)$ while $h_{A_1}$ is not, 
and therefore they can be more sensitive to variations in the fits.
Nonetheless, it is reasonable to expect that such moderate shifts in $R_2$ may be absorbed for example by 
the corresponding small variations in $\xi(w)$ and $R_0(w)$,
such that other physical observables hardly change.

\begin{figure}[t]
	\subfigbottomskip=-10pt
	\subfigcapskip=-10pt
	\subfigure[][\label{fig:nhts1}]{
        		\centering
     	  	\includegraphics[width=0.45\textwidth]{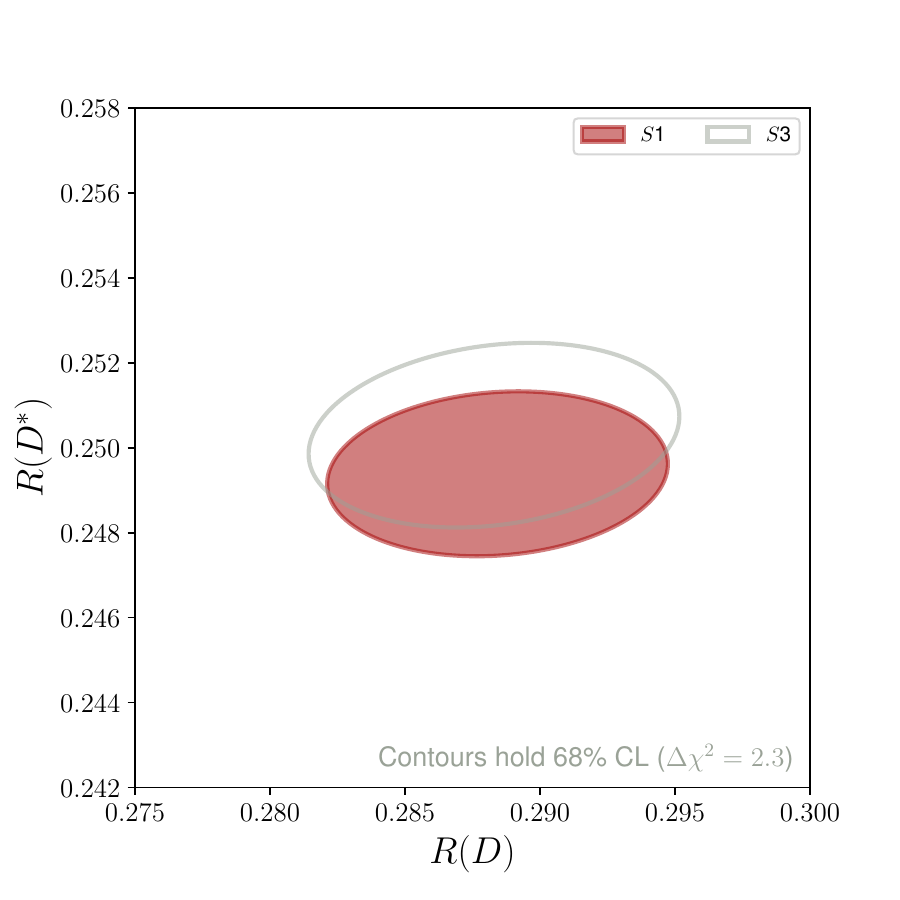}
       }
	\hfill
	\subfigure[][\label{fig:nhts2}]{
        		\centering
		\includegraphics[width=0.45\textwidth]{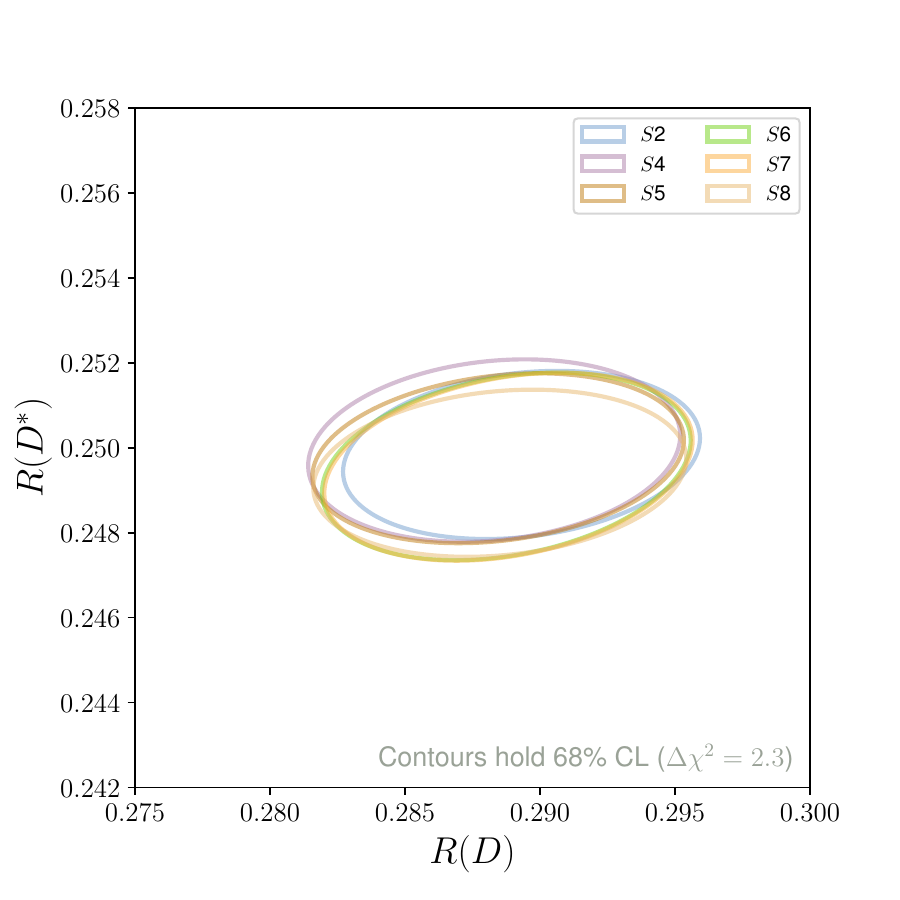}
      }
	\caption{The predicted values for $R(\Dx)$ for \subref{fig:nhts1} the selected hypotheses $S1$ and $S3$ and \subref{fig:nhts2} all other hypotheses.}
	\label{fig:nhts}
\end{figure}

In Fig.~\ref{fig:nhts1} we show the $R(\Dx)$ predictions for the $S1$ and $S3$ scenarios,
finding good agreement. 
In Fig.~\ref{fig:nhts2} we show the $R(\Dx)$ predictions for the other six terminating nodes:
The $R(\Dx)$ predictions are very similar between all hypotheses,
providing good evidence that the truncation of the fit parameters determined by our NHT prescription
has not introduced a model dependence associated with the choice of parameters into the fit.

\subsection{Fit results}
\label{sec:fitresults}

Using the selected $S1$ fit hypothesis for the \BaseFit scenario as our baseline,
in Table~\ref{tab:fitresults} we present the fit results for the various scenarios discussed
in Sec.~\ref{sec:fitscens} and summarized in Table~\ref{tab:FitKey}.
In Table~\ref{tab:recpar} we show for each scenario the corresponding recovered parameters:
 $R(D)$, $R(D^*)$, and their correlation; 
 $\mathcal{F}(1)$ and $\mathcal{G}(1)$ and their correlation;
 and the zero-recoil values for the form-factor ratios $R_1(1)$, $R_2(1)$, and $R_0(1)$. 
 Figure~\ref{fig:fit_details} shows the fitted experimental and LQCD data, 
 and the predicted differential spectra for 
 $\Bbar \to \Dx \tau \nu$ as a function of $w$ for the baseline fit.
We note the fitted value for $\hat\eta(1)$ in the \BaseFit scenario is in excellent agreement with Eq.~\eqref{eqn:eta1m},
as recovered from $\mathcal{O}(\aS,1/m_{c,b})$ fits.
The $\mathcal{O}(\aS^2)$ correction in Eq.~\eqref{eqn:hA1as2shift} is included in all second-order fits
(and must be included when using these fit results).
In all these fit scenarios, the fitted values for $|V_{cb}|$ are in good agreement with the \BaseFit result,
which follows from the compatibility of the $\mathcal{F}(1)$ and $\mathcal{G}(1)$ results across all fit scenarios.

\begin{table}[t]
\newcolumntype{C}{ >{\raggedleft\arraybackslash $} m{2.1cm} <{$}}
\newcolumntype{D}{ >{\centering\arraybackslash $} m{2.5cm} <{$}}
\renewcommand{\arraystretch}{0.9}
\resizebox{\textwidth}{!}{
\begin{tabular}{D|CCCCCCCCC}
\hline\hline
\text{Parameters} 	 &\makecell[c]{\text{\BaseFit}}	 & \makecell[c]{\text{\ShapeFit}}	 & \makecell[c]{\text{\NoLQCDFit}}	 & \makecell[c]{\text{\AllLQCDFit}}	 & \makecell[c]{\text{\BaseFit[NLO]}}	 & \makecell[c]{\text{\BaseFit[+SR]}}	 & \makecell[c]{\text{\BaseFit[17]}}	 & \makecell[c]{\text{\BaseFit[19]}}	 & \makecell[c]{\text{\hAOneFit}}  \\ 
 \hline
|V_{cb}|\times 10^3	 & 38.70(62)	 & 39.10(66)	 & 37.70(110)	 & 38.40(60)	 & 39.40(68)	 & 38.80(66)	 & 38.60(103)	 & 38.90(66)	 & 38.80(64)  \\ 
\mathcal{G}(1)	 & \nax	 & 1.06(1)	 & \nax	 & \nax	 & 1.06(1)	 & \nax	 & \nax	 & \nax	 & \nax  \\ 
\mathcal{F}(1)	 & \nax	 & 0.90(1)	 & \nax	 & \nax	 & 0.90(1)	 & \nax	 & \nax	 & \nax	 & \nax  \\ 
\rhossq	 & 1.10(4)	 & 1.08(5)	 & 1.31(17)	 & 1.05(4)	 & 1.12(4)^\dagger	 & 1.13(6)	 & 1.23(6)	 & 1.10(4)	 & 1.11(4)  \\ 
\cs	 & 2.39(18)	 & 2.24(19)	 & 1.95(33)	 & 2.38(15)	 & \nax	 & 2.44(19)	 & 2.81(21)	 & 2.36(18)	 & 2.45(18)  \\ 
\hat\chi_2(1)	 & -0.12(2)	 & -0.09(3)	 & -0.19(15)	 & -0.11(2)	 & -0.05(2)	 & -0.06(2)	 & -0.21(3)	 & -0.11(2)	 & -0.12(2)  \\ 
\hat\chi_2'(1)	 & \nax	 & \nax	 & \nax	 & \nax	 & 0.00(2)	 & -0.00(2)	 & \nax	 & \nax	 & \nax  \\ 
\hat\chi_3'(1)	 & \nax	 & \nax	 & \nax	 & \nax	 & 0.02(1)	 & 0.04(2)	 & \nax	 & \nax	 & \nax  \\ 
\hat\eta(1)	 & 0.34(4)	 & 0.29(7)	 & 0.10(39)	 & 0.27(4)	 & 0.31(3)	 & 0.35(4)	 & 0.34(4)	 & 0.34(4)	 & 0.34(4)  \\ 
\hat\eta'(1)	 & \nax	 & \nax	 & \nax	 & \nax	 & 0.02(8)	 & -0.01(20)	 & \nax	 & \nax	 & \nax  \\ 
\mbS~[\GeV]	 & 4.71(5)	 & 4.72(5)	 & 4.71(5)	 & 4.76(3)	 & 4.70(5)	 & 4.70(5)	 & 4.70(5)	 & 4.70(5)	 & 4.70(5)  \\ 
\dmbc~[\GeV]	 & 3.41(2)	 & 3.40(2)	 & 3.40(2)	 & 3.40(2)	 & 3.40(2)	 & 3.41(2)	 & 3.41(2)	 & 3.41(2)	 & 3.41(2)  \\ 
\hat\beta_2(1)	 & \nax	 & \nax	 & \nax	 & \nax	 & \nax	 & -0.14(89)	 & \nax	 & \nax	 & \nax  \\ 
\hat\beta_3'(1)	 & \nax	 & \nax	 & \nax	 & \nax	 & \nax	 & -0.15(33)	 & \nax	 & \nax	 & \nax  \\ 
\hvphp(1)	 & 0.25(21)	 & 0.15(21)	 & -2.29(122)	 & 0.31(27)	 & \nax	 & 0.29(42)	 & 0.19(20)	 & 0.25(21)	 & 0.23(21)  \\ 
\lam2~[\GeV^2]	 & 0.12(2)	 & 0.11(2)	 & 0.11(2)	 & 0.11(2)	 & \nax	 & 0.12(2)	 & 0.12(2)	 & 0.12(2)	 & 0.12(2)  \\ 
\rho_1~[\GeV^3]	 & -0.36(24)	 & 0.04(50)	 & -0.02(50)	 & -0.16(27)	 & \nax	 & -0.39(24)	 & -0.38(24)	 & -0.36(24)	 & -0.39(26)  \\ 
\hline
\chi^2 	 &29.8	 & 26.5	 & 20.3	 & 49.4	 & 33.1	 & 31.7	 & 49.1	 & 18.3	 & 31.2  \\ 
\text{ndf} 	 &31	 & 29	 & 31	 & 42	 & 33	 & 31	 & 51	 & 21	 & 33  \\ 
\rho^2 	 &1.35(5)	 & 1.31(5)	 & 1.35(10)	 & 1.32(4)	 & 1.25(3) & 1.37(5)	 & 1.48(6)	 & 1.34(5)	 & 1.36(4)  \\ 
c 	 &2.41(17)	 & 2.28(18)	 & 1.89(31)	 & 2.43(15)	 & 1.90(7)	 & 2.45(17)	 & 2.74(17)	 & 2.38(17)	 & 2.46(16)  \\ 
\lam1~[\GeV^2] 	 &-0.42(10)	 & -0.20(25)	 & -0.23(24)	 & -0.30(11)	 &  \nax	 & -0.43(9)	 & -0.43(9)	 & -0.42(10)	 & -0.43(11)  \\ 
\hline\hline
\end{tabular}}
\caption{Fit results. Below the line we show the corresponding values for the fit $\chi^2$, number of degrees of freedom, 
the (negative) slope and curvature of $\xi(w)$ at zero recoil, and the value of $\lam1$ via Eq.~\eqref{eqn:1Slams}.
The slope for \BaseFit[NLO], marked by a `$\dagger$', corresponds to the slope $\bar{\rho}^2_*$ defined as in Ref.~\cite{Bernlochner:2017jka}.
Note the $\mathcal{O}(\aS^2)$ shift in Eq.~\eqref{eqn:hA1as2shift} is imposed in all second-order fits, 
and must be included when using these fit results.}
\label{tab:fitresults}
\end{table}

\begin{table}[t]
\newcolumntype{C}{ >{\raggedleft\arraybackslash $} m{1.5cm} <{$}}
\newcolumntype{D}{ >{\centering\arraybackslash $} m{2.5cm} <{$}}
\renewcommand{\arraystretch}{1.1}
\resizebox{0.95\textwidth}{!}{
\begin{tabular}{D|CCCCCCCCC}
\hline\hline
\text{Scenario}	 & \makecell[c]{\text{$R(D)$}}	 & \makecell[c]{\text{$R(D^*)$}}	 & \makecell[c]{\text{$\rho_{R(D), R(D^*)}$}}	 & \makecell[c]{\text{$\mathcal{F}(1)$}}	 & \makecell[c]{\text{$\mathcal{G}(1)$}}	 & \makecell[c]{\text{$\rho_{\mathcal{F}(1), \mathcal{G}(1)}$}}	 & \makecell[c]{\text{$R_1(1)$}}	 & \makecell[c]{\text{$R_2(1)$}}	 & \makecell[c]{\text{$R_0(1)$}} \\ 
\hline
\text{\BaseFit}	 & 0.288(4)	 & 0.249(1)	 & 0.121~~	 & 0.917(10)	 & 1.050(6)	 & 0.507~~	 & 1.43(4)	 & 0.89(3)	 & 1.23(3) \\ 
\text{\ShapeFit}	 & 0.290(4)	 & 0.249(1)	 & 0.069~~	 & 0.938(23)	 & 1.055(11)	 & 0.854~~	 & 1.40(4)	 & 0.90(3)	 & 1.20(4) \\ 
\text{\NoLQCDFit}	 & 0.278(7)	 & 0.248(3)	 & 0.662~~	 & 0.935(23)	 & 1.088(64)	 & 0.385~~	 & 1.43(6)	 & 0.92(10)	 & 1.16(11) \\ 
\text{\AllLQCDFit}	 & 0.285(4)	 & 0.249(1)	 & 0.146~~	 & 0.929(10)	 & 1.054(6)	 & 0.480~~	 & 1.38(2)	 & 0.94(2)	 & 1.18(2) \\ 
\text{\BaseFit[NLO]}	 & 0.296(3)	 & 0.249(1)	 & 0.347~~	 & 0.966(1)	 & 1.053(5)	 & 0.418~~	 & 1.34(3)	 & 0.87(3)	 & 1.18(2) \\ 
\text{\BaseFit[+SR]}	 & 0.289(4)	 & 0.250(1)	 & 0.265~~	 & 0.916(10)	 & 1.048(6)	 & 0.500~~	 & 1.43(4)	 & 0.86(10)	 & 1.24(3) \\ 
\text{\BaseFit[17]}	 & 0.287(4)	 & 0.258(3)	 & 0.572~~	 & 0.916(10)	 & 1.050(6)	 & 0.524~~	 & 1.43(3)	 & 0.96(3)	 & 1.23(3) \\ 
\text{\BaseFit[19]}	 & 0.289(4)	 & 0.249(1)	 & 0.121~~	 & 0.917(10)	 & 1.050(6)	 & 0.505~~	 & 1.43(4)	 & 0.89(3)	 & 1.24(3) \\ 
\text{\hAOneFit}	 & 0.288(4)	 & 0.250(1)	 & 0.148~~	 & 0.916(11)	 & 1.049(6)	 & 0.556~~	 & 1.43(4)	 & 0.89(3)	 & 1.24(3) \\ 
\hline\hline
\end{tabular}}
\caption{Recovered parameters for each fit scenario of Table~\ref{tab:FitKey}.}
\label{tab:recpar}
\end{table}

\begin{figure}[h!]
	\subfigbottomskip=-10pt
	\subfigcapskip=-10pt
	\subfigure[][\label{fig:RateD}]{
        		\centering
     	  	\includegraphics[width=0.42\textwidth]{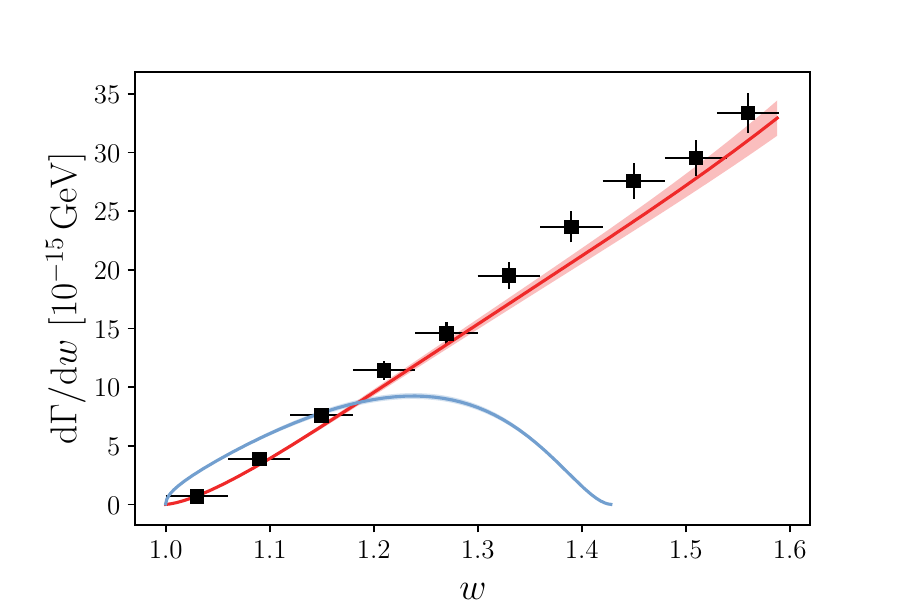}
       }
	\hfill
	\subfigure[][\label{fig:RateDsTagged}]{
        		\centering
		\includegraphics[width=0.45\textwidth]{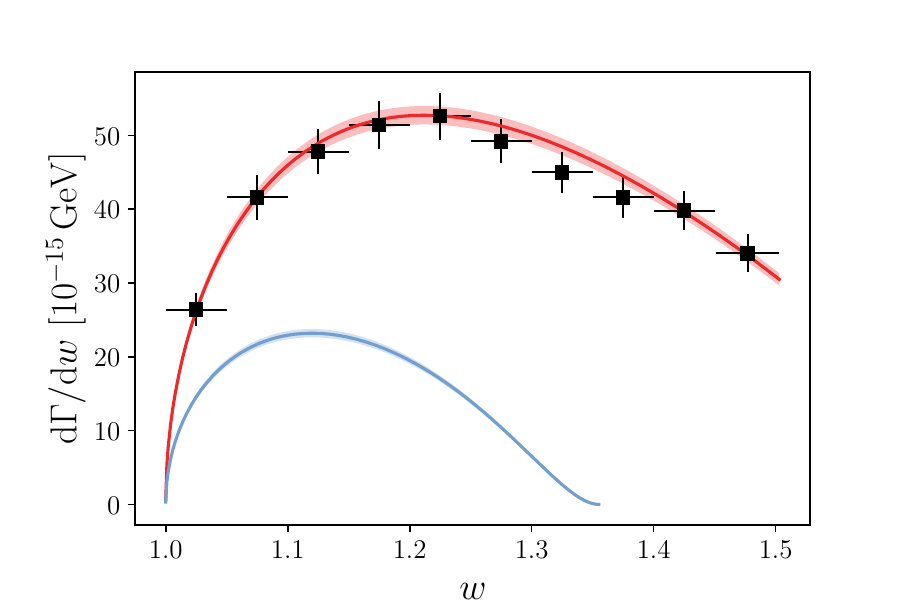}
      }
	\subfigure[][\label{fig:RateDsUntagged}]{
        		\centering
		\includegraphics[width=0.45\textwidth]{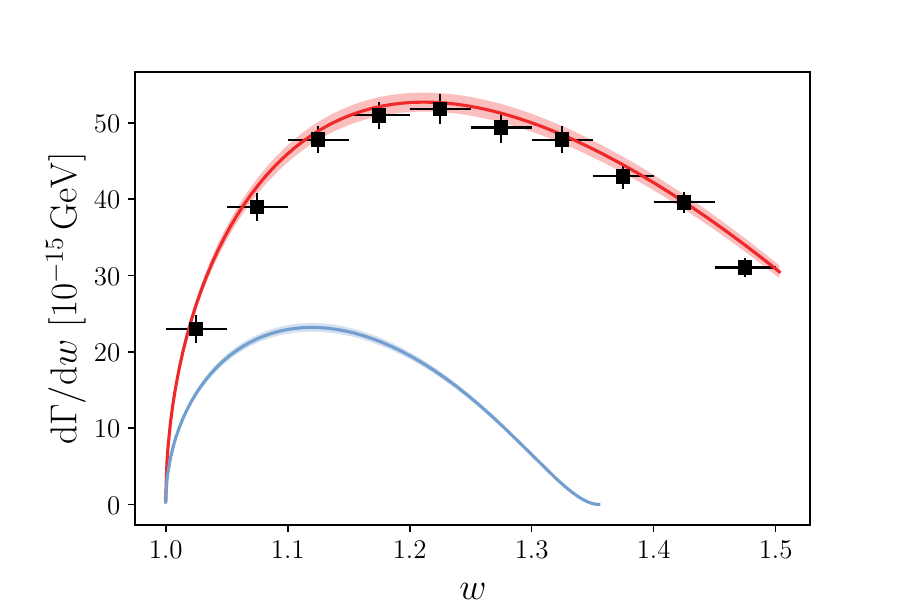}
      }\hfill
	\subfigure[][\label{fig:fp}]{
        		\centering
		\includegraphics[width=0.45\textwidth]{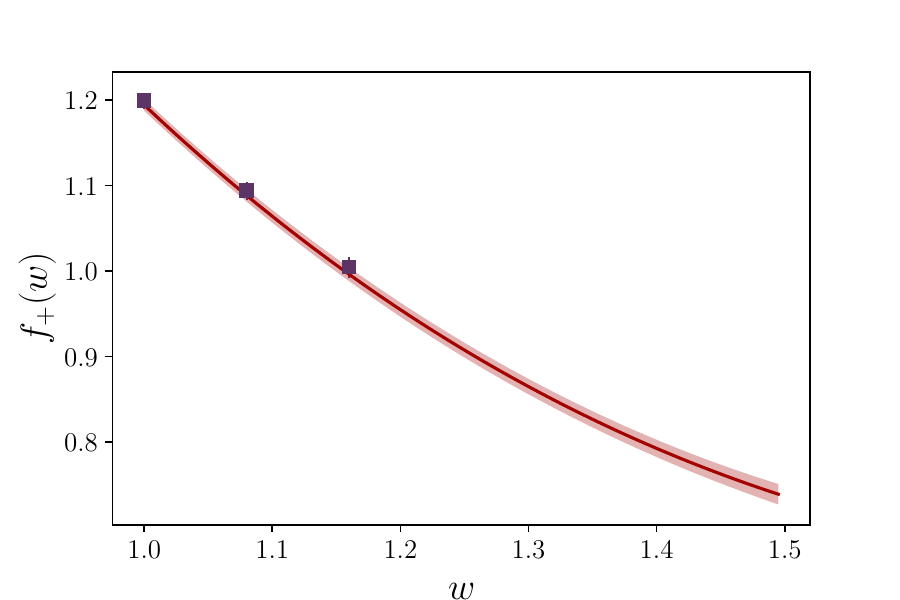}
      }   
	\subfigure[][\label{fig:f0}]{
        		\centering
		\includegraphics[width=0.45\textwidth]{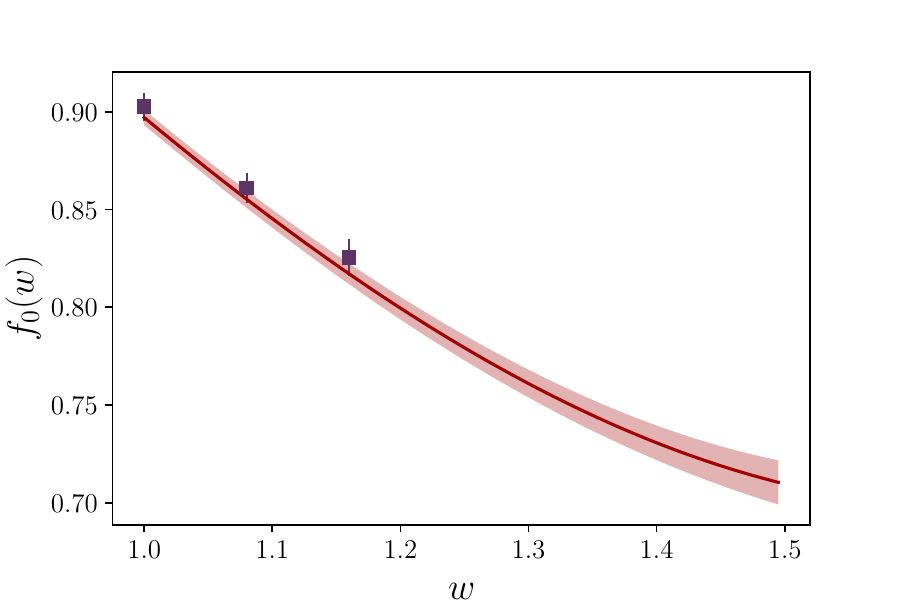}
      } \hfill    
	\subfigure[][\label{fig:hA1}]{
        		\centering
		\includegraphics[width=0.45\textwidth]{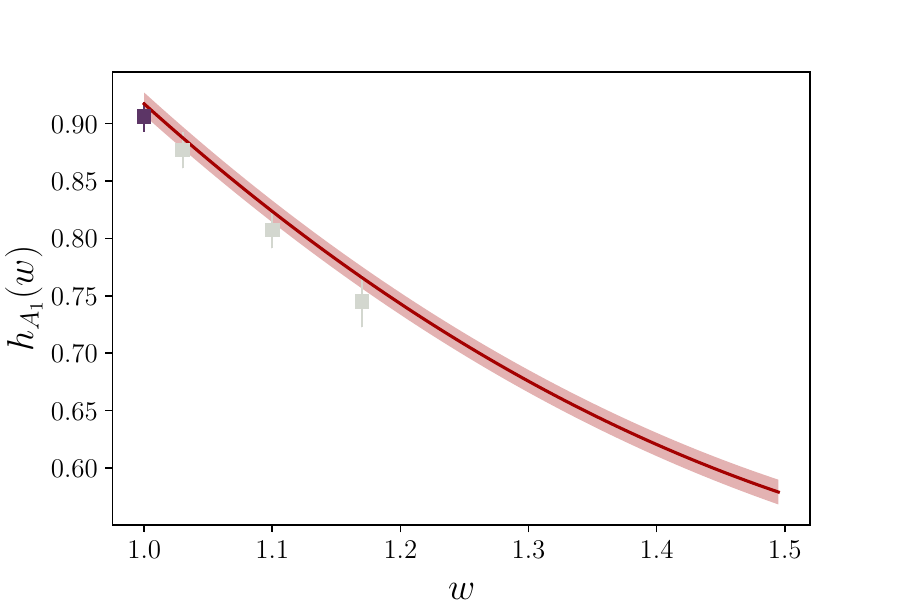}
      }                  
	\caption{The spectra and form factors (red bands) recovered from the \BaseFit fit scenario in the RC expansion, 
	compared to the fitted experimental data (black markers) and LQCD data (plum markers): 
	\subref{fig:RateD} $d\Gamma[\Bbar \to D \ell \nu]/dw$ (Belle 2015); 
	\subref{fig:RateDsTagged} $d\Gamma[\Bbar \to D^* \ell \nu]/dw$ (Belle 2017);
	\subref{fig:RateDsUntagged} $d\Gamma[\Bbar \to D^* \ell \nu]/dw$ (Belle 2019);
	\subref{fig:fp} $f_+(w)$;
	\subref{fig:f0} $f_0(w)$;
	and \subref{fig:hA1} $h_{A_1}(w)$.
	Also shown are the corresponding $\Bbar \to \Dx \tau \nu$ spectra (blue bands).
	For $h_{A_1}(1) = \mathcal{F}(1)$ the zero recoil prediction of Ref.~\cite{Aoki:2021kgd} is used. 
	The beyond zero recoil lattice points for $h_{A_1}$ from Ref.~\cite{Bazavov:2021bax}, 
	which are not included in this fit, are shown as gray markers.}
	\label{fig:fit_details}
\end{figure}

The \ShapeFit scenario results are similar to those of the \BaseFit scenario, 
with the exception of a larger uncertainty in $\hat\eta(1)$ and $\rho_1$,
but both are compatible with the \BaseFit fit scenario results.
The resulting $\mathcal{F}(1)$ and $\mathcal{G}(1)$ recovered from the \ShapeFit fit 
(see Table~\ref{tab:recpar}) are $\mathcal{F}(1) = 0.938(22)$ and $\mathcal{G}(1) = 1.055(11)$,
whose uncertainties are similarly larger but remain compatible with the LQCD data.
We see from this that the differential shape information has mild constraining power on parameters 
entering the zero recoil predictions,
such that the tension of the first-order prediction $\mathcal{F}(1)^{\text{NLO}}_{\text{HQET}} \simeq 0.966(2)$ 
(see Eq.~\eqref{eqn:F1NLO}) with the LQCD prediction is relaxed.

The \NoLQCDFit scenario fit also results in larger uncertainties, as expected,
but is compatible with the baseline fit.
Put in other words, the included LQCD data is in agreement with the experimental data,
in the context of an RC expansion-based parametrization.
By contrast, the \AllLQCDFit scenario, which uses LQCD predictions for $\Bbar \to D^*$ beyond zero recoil, 
produces a fit of notably poorer quality,
due to known tensions between the LQCD beyond zero-recoil $\Bbar \to D^*l\nu$ predictions~\cite{Bazavov:2021bax}
and experimental measurements.
However, including just the beyond zero-recoil LQCD data for $h_{A_1}$, 
as done in the \hAOneFit scenario,
produces fits in good agreement with the \BaseFit results.
We characterize the behavior of the \AllLQCDFit  versus \hAOneFit scenarios further in Appendix~\ref{app:FNAL_Comparison}.
Finally, the \BaseFit[+SR] scenario fit results
are compatible with those of \BaseFit,
suggesting that QCDSR inputs are not incompatible with current data.
The QCDSR inputs appear to simply allow additional sub(sub)leading Isgur-Wise function parameters to be constrained,
beyond those already selected by the NHT prescription.

\begin{figure}[t]
	\subfigbottomskip=-10pt
	\subfigcapskip=-10pt
	\subfigure[][\label{fig:RDRDstaster1}]{
        		\centering
     	  	\includegraphics[width=0.45\textwidth]{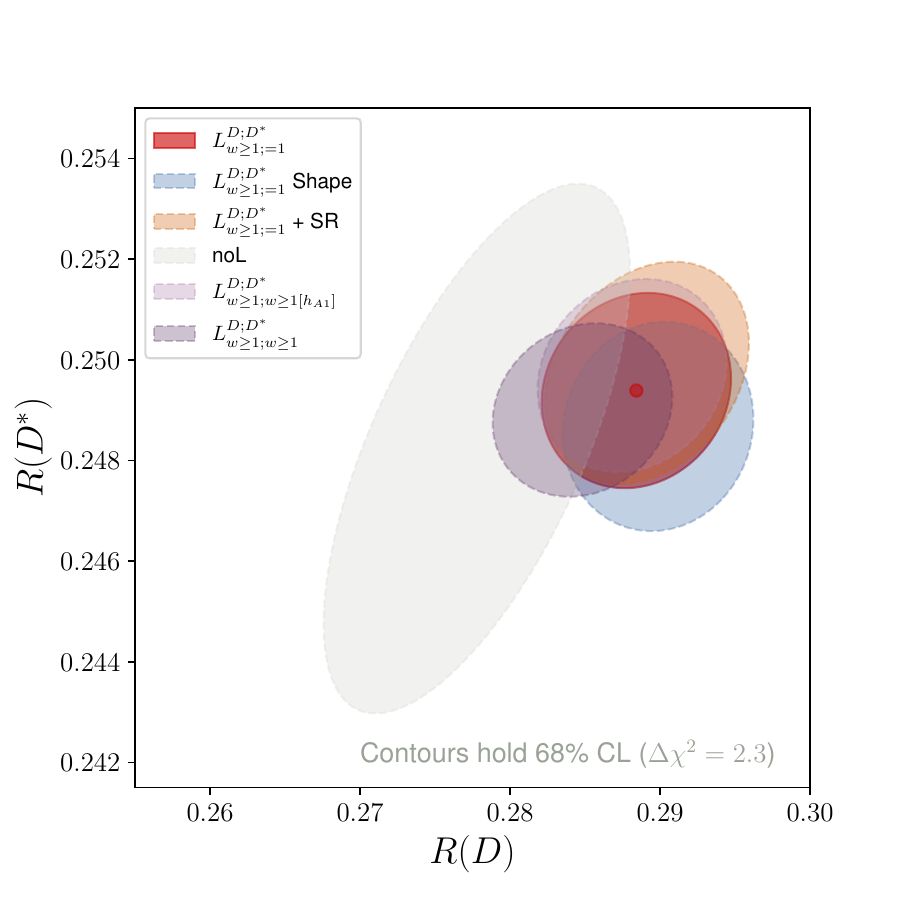}
       }
	\hfill
	\subfigure[][\label{fig:RDRDstaster2}]{
        		\centering
		\includegraphics[width=0.45\textwidth]{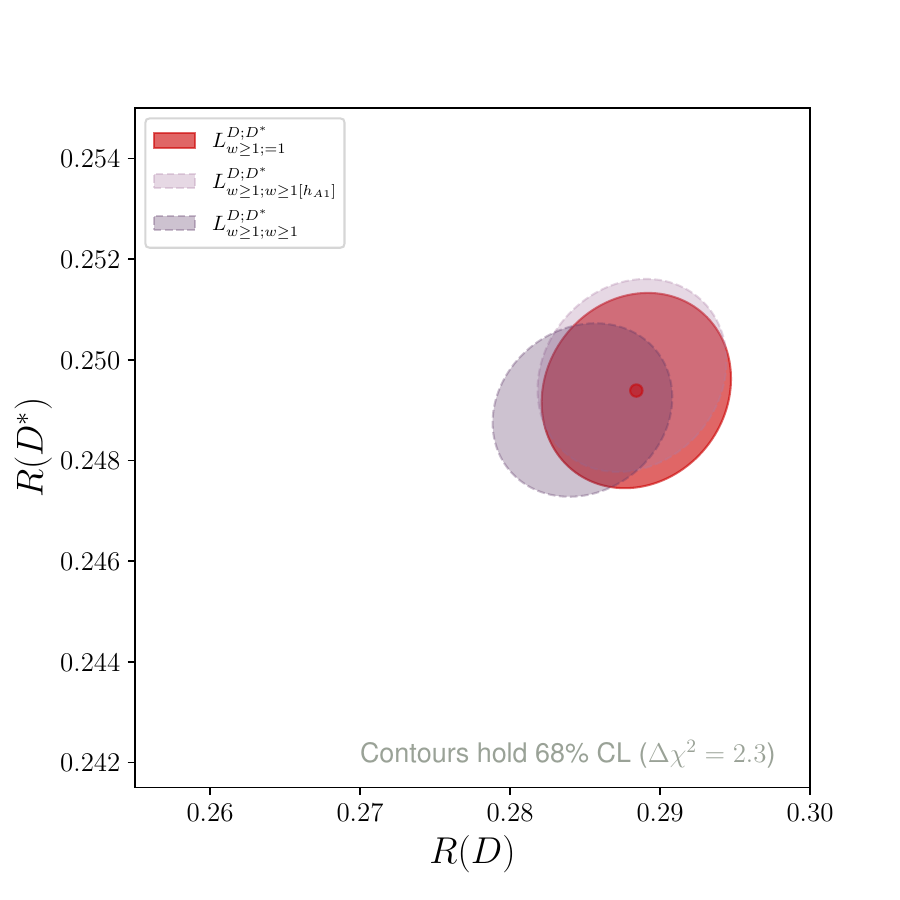}
      }
	\caption{Fit results for $R(\Dx)$ for the 9 scenarios of Table~\ref{tab:FitKey}.}
	\label{fig:RDRDsScen}
\end{figure}

In Fig.~\ref{fig:RDRDsScen} we show a comparison of the $R(\Dx)$ predictions for the
\BaseFit, \ShapeFit, \NoLQCDFit, \BaseFit[+SR],  \AllLQCDFit, and \hAOneFit scenarios.
We see that these predictions are all in agreement, with the exception of the \NoLQCDFit results
that mildly shift further down in $R(D)$ by about $1\sigma$ but with notably larger uncertainties.
We note in Fig.~\ref{fig:RDRDstaster2} that the \hAOneFit scenario is in excellent agreement with the \BaseFit.

While the parameters of the \BaseFit[NLO] fit scenario appear naively compatible with the baseline results,
the zero-recoil slope and curvature are significantly different.
We see in Table~\ref{tab:recpar} that the main effect is a large shift in $R(D)$ versus the baseline fit.
As we will discuss in Sec.~\ref{sec:biases} below,
the source of this shift can be attributed to a slope-curvature constraint imposed on $\mathcal{G}(w)$.
Apart from this effect, 
the similar size of the uncertainties in the \BaseFit[NLO] fit results (as also seen in $R(\Dx)$) versus the baseline
suggests that while the $\mathcal{O}(1/m_c^2)$ RC corrections 
ameliorate the tension in the first-order prediction $\mathcal{F}(1)^{\text{NLO}}_{\text{HQET}}$ in Eq.~\eqref{eqn:F1NLO}
compared to LQCD data (see also Fig.~\ref{fig:F1G1}),
they do not otherwise introduce large uncertainties into physical observables.
This is precisely the expected behavior that motivated the $\mathcal{O}(\aS,1/m_{c,b})$ `shape only' fits of Ref.~\cite{Bernlochner:2017jka}.

Between \BaseFit[17] and \BaseFit[19] we further note significant differences in the slope and curvature,
suggesting a mild tension between these two datasets. 
We explore the implications of this further in Sec.~\ref{sec:biases} below.

\subsection{Biases and the major axis of doom}
\label{sec:biases}

The astute reader will have noted that our correlated $R(\Dx)$ predictions from the \BaseFit scenario,
\begin{equation}
	\label{eqn:baseRDDs}
	R(D)  = 0.288(4)\,, \quad R(D^*) = 0.249(1)\,,  \quad \rho = 0.12,
\end{equation}
differ by approximately $2.7\sigma$ from our prior predictions~\cite{Bernlochner:2017jka}, $R(D) = 0.298(3)$,  $R(D^*) = 0.261(4)$ with correlation $0.19$.
However, the origin of this difference is not the inclusion of second-order power corrections;
in particular it is not due to any hint of unexpectedly large $\mathcal{O}(1/m_c^2)$ corrections (which was hypothesized in prior literature~\cite{Bigi:2017jbd}). 
Rather, we identify two sources of external biases that are mainly responsible for this shift.

\begin{figure}[t]
	\includegraphics[width = .6\textwidth]{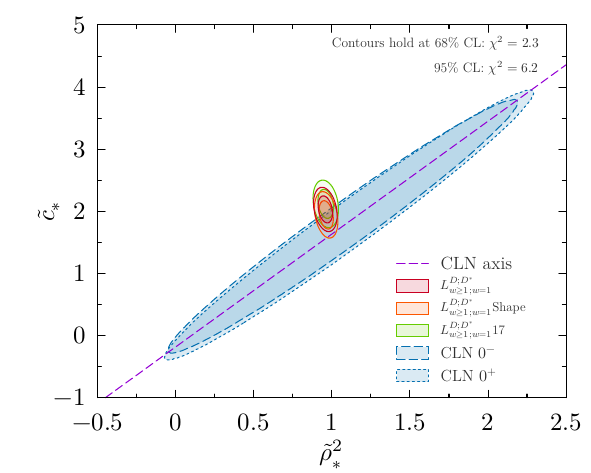}
	\caption{The allowed-region ellipses (blue) arising from dispersive bounds plus unitarity constraints applied to the $\Bbar \to D$ form factor $\mathcal{G}(w)$~\cite{Caprini:1997mu}.
	The major axis of the tighter ellipse, corresponding to the $J^P= 0^-$ current, is shown by the dashed purple line.
	Also shown are the recovered CLs for the \BaseFit, \ShapeFit, and \BaseFit[17] fit scenarios (red, orange, and green ellipses, respectively).}
	\label{fig:rhosqc}
\end{figure}

The first of these is a so-called major-axis approximation introduced in Ref.~\cite{Caprini:1997mu},
which is a core feature of the CLN parametrization. 
In Ref.~\cite{Caprini:1997mu}, the application of dispersive bounds from unitarity constraints
to the $\Bbar \to D$ form factor $\mathcal{G}(w)$ was shown to constrain the allowed region 
in the $\trhossq - \tcs$ plane (slope and curvature, defined in Eq.~\eqref{eqn:Gexp}) in the form of two 
elongated overlapping ellipses for the $J^P = 0^-$ and $0^+$ currents, respectively.
QCDSR results were applied to the first-order HQET corrections, in order to relate bounds on the $J^P = 0^-$ current to $\mathcal{G}(w)$.
These ellipses, which incorporated also estimates of theoretical uncertainties in the first-order corrections, 
are reproduced in Fig.~\ref{fig:rhosqc} in blue.\footnote{%
Our $\tcs = 2c_1$ in Ref.~\cite{Caprini:1997mu}.}
Ref.~\cite{Caprini:1997mu} approximated these allowed regions simply by the major axis of the most constraining ellipse
(perhaps because the size of the minor axes of these ellipses were far smaller than the experimental uncertainties in $\tcs$ at the time),
shown by the purple dashed line in Fig.~\ref{fig:rhosqc}.
This imposes the relationship between $\tcs$ and $\trhossq$ in Eq.~\eqref{eqn:tcscons}, 
leading to a polynomial form $\mathcal{G}(w)/\mathcal{G}(w_0) = 1 - 8 a^2 \trhossq z_* + (57.\trhossq - 7.5) z_*^2  + \ldots$, 
and, after application of HQET relations at $\mathcal{O}(\aS, 1/m_{c,b})$,
to similar polynomial forms in $z_*$ for $h_{A_1}(w)/h_{A_1}(w_0)$.
The CLN parametrization, and all parametrizations derived from it, implicitly apply this constraint on the $\trhossq-\tcs$ plane.

The experimental data and LQCD predictions have reached a level of precision, however,
such that the size of the $\trhossq - \tcs$ allowed region recovered from fits is now comparable to the minor axes. 
To see this, we show in Fig.~\ref{fig:rhosqc} the recovered $68\%$ and $95\%$ CLs
for the \BaseFit, \ShapeFit, and \BaseFit[17] scenarios, by red, orange and green ellipses, respectively.
Constraining $\trhossq$ and $\tcs$ to the major axis is barely compatible with these fits at 95\% CL.  
Thus, imposing the CLN constraint in Eq.~\eqref{eqn:tcscons} introduces fit biases into the analysis of current data.

\begin{figure}[t]
	\subfigbottomskip=-10pt
	\subfigcapskip=-10pt
	\subfigure[][\label{fig:revolutionBLPRXP}]{
		\includegraphics[width = .48\textwidth]{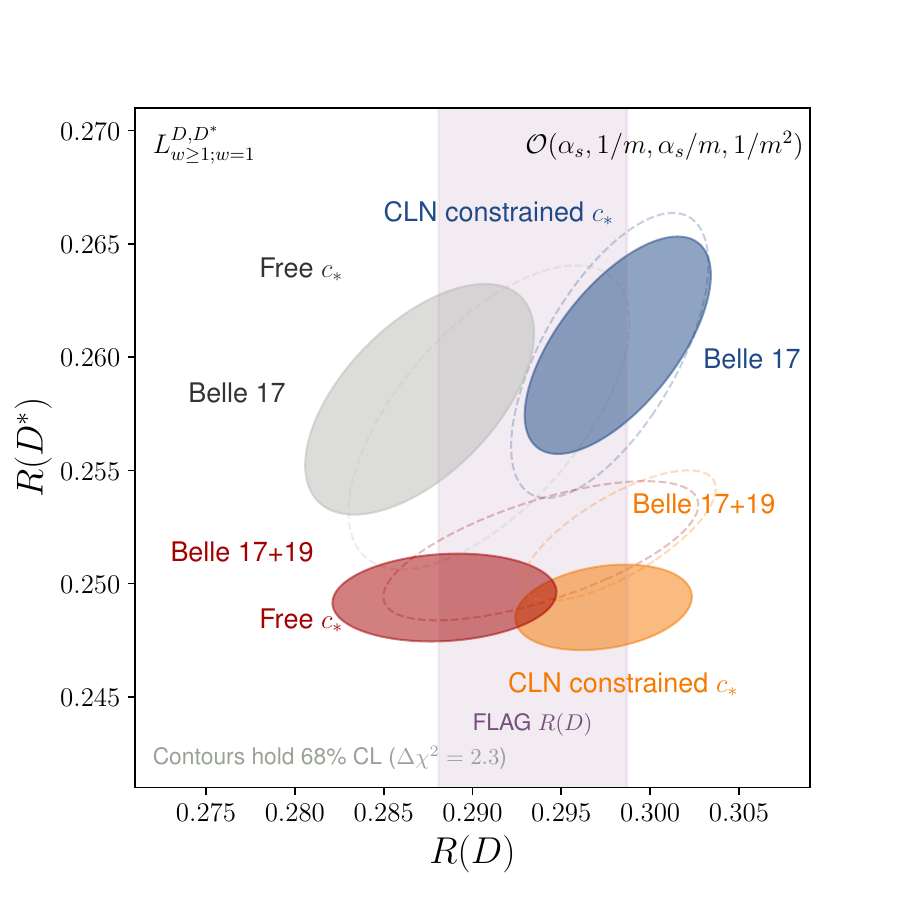}
	}\hfil
	\subfigure[][\label{fig:revolutionBLPR}]{
		\includegraphics[width = .48\textwidth]{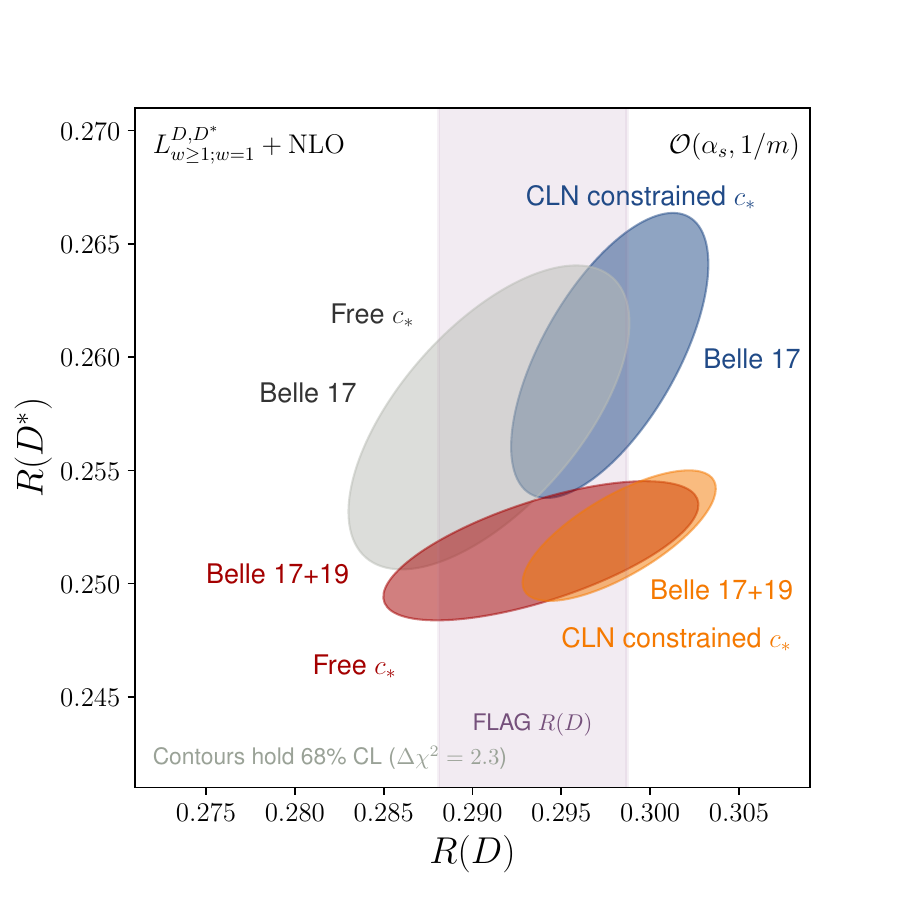}
	}\hfil
	\caption{
	Left~\subref{fig:revolutionBLPRXP}:
	The \BaseFit $R(\Dx)$ predictions using different sets of inputs and assumptions. 
	The blue versus gray (orange versus red) ellipses demonstrate the shift in the predicted $R(D)$ when the CLN constraint on $c_*$ is applied versus lifted, respectively,
	to the \BaseFit[17] (\BaseFit) scenario.
	The inclusion of the Belle 2019 data results in a reduction of the central value and uncertainty of $R(D^*)$
	(gray versus red or blue versus orange ellipses).  
	The light-purple band shows the $R(D)$ LQCD prediction~\cite{Aoki:2021kgd}.
	Right~\subref{fig:revolutionBLPR}: The same pattern for the \BaseFit[NLO] scenario fits,
	that incorporate only $\mathcal{O}(\aS,1/m_{c,b})$ corrections and (in this case) do not incorporate QCDSR inputs.
	For comparison, in the left figure we show the \BaseFit[NLO] CLs by dashed ellipses.}
	\label{fig:revolution}.
\end{figure}

To demonstrate this explicitly, 
we show in Fig.~\ref{fig:revolutionBLPRXP} the recovered $R(D)$\,--\,$R(D^*)$ CLs 
arising from applying the CLN constraint~\eqref{eqn:tcscons} 
to the \BaseFit[17] scenario (blue ellipse) versus the \BaseFit[17] scenario without such a constraint (gray ellipse).
We do the same for the \BaseFit scenario (red ellipse), i.e., using all Belle data, versus the \BaseFit scenario without such a constraint (orange ellipse).
One observes a significant shift in $R(D)$, commensurate with a bias introduced into $\mathcal{G}(w)$:
$R(D^*)$ remains unaffected because the parameters entering the first-order power corrections may 
compensate for the bias when translated to $h_{A_i}$ and $h_V$.
In Fig.~\ref{fig:revolutionBLPR} we show the same comparison, 
but for a \BaseFit[NLO]-type scenario,
that incorporates only $\mathcal{O}(\aS,1/m_{c,b})$ corrections and (in this case) no QCDSR inputs.
One sees that a similar downward shift in $R(D)$ occurs
independently of whether second-order power corrections were included.
A similar result occurs for \BaseFit[NLO] with QCDSR inputs.
To guide the eye, in Fig.~\ref{fig:revolutionBLPRXP} we show the \BaseFit[NLO]-type fits by dashed ellipses.
The fits using first-order versus second-order HQET corrections are respectively compatible
for both the CLN constrained and unconstrained scenarios,
substantiating that the second-order power corrections do not appear to play a major role in these shifts.

The second source of external bias is, as already noted in Sec.~\ref{sec:fitresults},
a tension in the fits to the Belle 2017 versus the Belle 2017 + 2019 $\Bbar \to D^* \ell \nu$ datasets.
The inclusion of the more precise latter dataset results in an additional reduction of both the central value and the uncertainty of the predicted $R(D^*)$.
In Fig.~\ref{fig:revolution}, this corresponds to the shift from the gray to red or blue to orange ellipses.
Once again, comparing Figs~\ref{fig:revolutionBLPRXP} and~\ref{fig:revolutionBLPR} 
we see that the shift is relatively insensitive to the inclusion of the second-order power corrections.
Overall, the fits using first-order versus second-order HQET corrections are respectively compatible
for each set of inputs or assumptions.
Thus the significant shifts in the $R(\Dx)$ predictions can be attributed mainly to two sources of bias: 
the CLN constraint enforcing a linear relationship between $\trhossq$ and $\tcs$, 
and the tension between the Belle datasets.

Because of the tension in the $R(D^*)$ predictions from fits using either the Belle 2017 or 2019 dataset, 
we adopt a scale factor for its uncertainty, $\sqrt{\chi^2}$ for 2 experiments~\cite{Zyla:2020zbs},
in order to account for the differences between the two datasets.
From the results in Table~\ref{tab:recpar}, this leads to the $R(D^*)$ prediction 
\begin{equation}
	R(D^*) = 0.249(3)\,,
\end{equation}
in which the scale factor is $2.6$. 

\subsection{Vanishing chromomagnetic limit fits}

\begin{figure}[t]
	\subfigbottomskip=-10pt
	\subfigcapskip=-10pt
	\subfigure[][\label{fig:F1G1NoChromo}]{
		\includegraphics[width = .48\textwidth]{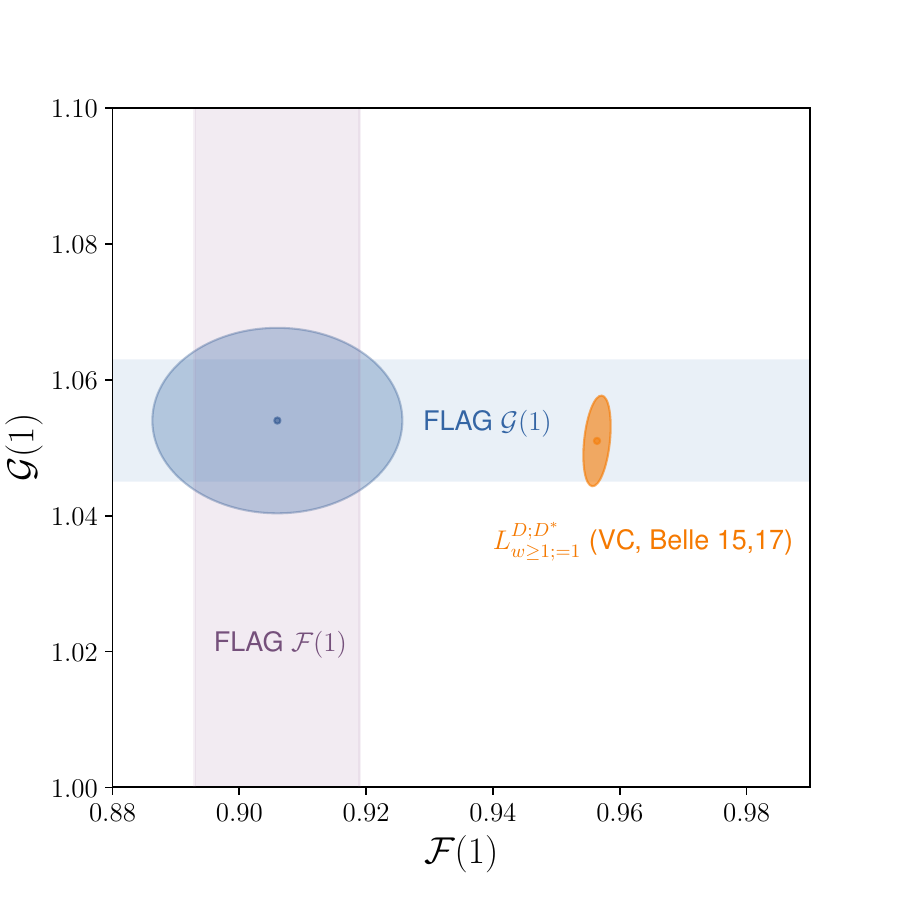}
	}\hfil
	\subfigure[][\label{fig:RDxNoChromo}]{
		\includegraphics[width = .48\textwidth]{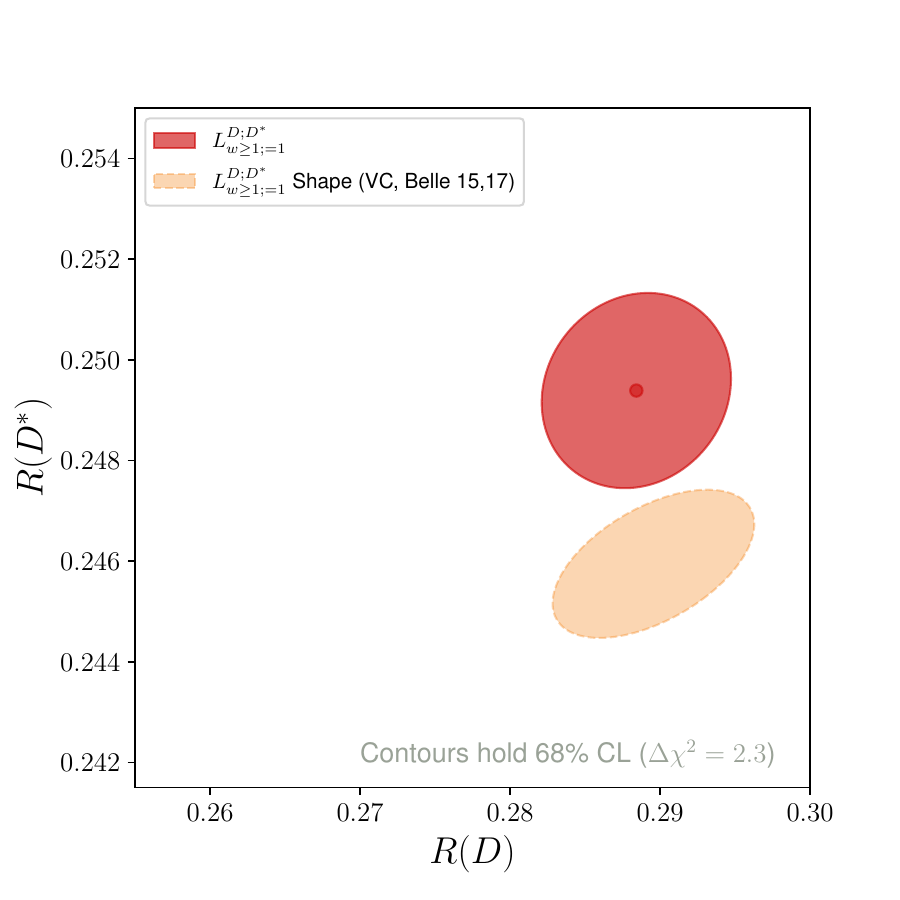}
	}\hfil
	\caption{Left \subref{fig:F1G1NoChromo}: the recovered values of $\mathcal{G}(1)$ and $\mathcal{F}(1)$ from the \BaseFit fit scenario 
	in the VC limit are compared to the LQCD predictions of Ref.~\cite{Aoki:2021kgd}. 
	Right \subref{fig:RDxNoChromo}: The $R(\Dx)$ predictions from the \ShapeFit fit scenario in the VC limit (orange ellipse) 
	are compared to the \BaseFit fit in the RC expansion (red ellipse).}
	\label{fig:NoChromo}
\end{figure}

Applying the VC limit to the fit scenarios in Table~\ref{tab:FitKey} instead of the RC expansion, 
we find poor fits for the \BaseFit scenario and for all its variations that include zero-recoil normalization constraints.
Typically, the $\chi^2/\text{ndf}$ corresponds to a $p$ value of less than one percent. 
This is caused by the tensions in the predicted value of $\mathcal{F}(1)$ from the VC expansion versus LQCD data: 
to describe the experimental spectra at nonzero recoil, the fit parameter $\hat{c}_0(1)$ is pushed to small values, which result in $\mathcal{F}(1) \simeq 0.96$. 
This is far from the LQCD constraint $\mathcal{F}(1)_{\text{LQCD}} = 0.906(13)$~\cite{Aoki:2021kgd}, yielding a large contribution to the fit $\chi^2$. 
This behavior also matches the approximate expectations discussed in Sec.~\ref{sec:zrVClimit}:
The zero-recoil structure of the first- and second-order power corrections in the VC limit appears inconsistent with the LQCD data~\eqref{eqn:F1G1}
and the recovered ratios $R_{1,2}(1)$ from first-order fits. 
The recovered values for $\mathcal{F}(1)$ and $\mathcal{G}(1)$ for the \BaseFit scenario are shown in Fig.~\ref{fig:F1G1NoChromo}.

We next consider the \ShapeFit scenario, that relaxes the zero-recoil normalization constraints. 
In this scenario, the VC limit parametrization achieves excellent fits. 
To avoid overfitting, we again apply our NHT prescription, 
considering all combinations of the candidate parameter set 
$\hat\eta(1)$, $\hat\eta'(1)$, $\hat\phi_0'(1)$, $\hat{e}_3(1)$, $\hat{e}_3'(1)$ and $\hat{c}_0(1)$.
The prescription identifies three terminating nodes.
Of these, two are the same as the third under the approximate replacements 
$\sim \ec \hat{e}_3(1) \to \hat\eta(1)$ or $\sim \ec \hat{e}_3'(1) \to \hat\eta'(1)$,
matching our expectation in Sec.~\ref{sec:FFsVClimit} 
that $\hat\eta$ can reabsorb $\ec \hat{e}_3$ as in Eq.~\eqref{eqn:e3etaabs},
because there is no sensitivity to $\mathcal{O}(1/m_cm_b, 1/m_b^2)$ 
or $\mathcal{O}(\aS/m_{c,b})$ corrections in the available data.
We show the resulting fit parameters for this node in Table~\ref{tab:VCfit},
which has $\chi^2 = 27.1$ for $29$ degrees of freedom.

\begin{figure}[t]
	\includegraphics[width = .48\textwidth]{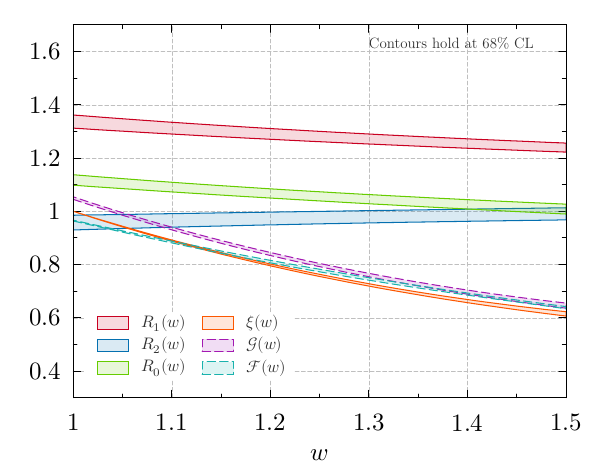}
	\caption{Form factors and form-factor ratios as functions of $w$ for the VC limit \ShapeFit fit.}
	\label{fig:FFRNoChr}
\end{figure}

The fitted value for $|V_{cb}|$ is in good agreement with the \BaseFit result for the RC expansion,
which must be the case as $\mathcal{F}(1)$ and $\mathcal{G}(1)$ are constrained to the LQCD data.
The corresponding zero recoil slope and curvature parameters are $\rho^2 = 1.20(3)$ and $c = 2.10(15)$,
which are in moderate tension with those for the \ShapeFit in the RC expansion.
A similar difference arises in $\hat\eta(1)$, 
which leads to a larger $R_2(1)$ and a smaller $R_0(1)$ as in Eqs.~\eqref{eqn:FGhqetnc}.
One sees respective up and down shifts in these form-factor ratios over the entire $w$ range, as shown in Fig.~\ref{fig:FFRNoChr}.
While $R(D)$ is mainly determined by lattice data and is unchanged versus the RC expansion fits, 
these shifts in $R_{2,0}$ result in a significantly smaller $R(D^*)$:
One finds $R(D) = 0.290(4)$ and $R(D^*) = 0.246(1)$ with correlation $0.54$.
The corresponding CL is shown in Fig.~\ref{fig:RDxNoChromo} and compared to the RC baseline fit.

\begin{table}[t]
\newcolumntype{C}{ >{\centering\arraybackslash $} m{1.3cm} <{$}}
\newcolumntype{D}{ >{\centering\arraybackslash $} m{1.8cm} <{$}}
\newcolumntype{E}{ >{\centering\arraybackslash $} m{0.65cm} <{$}}
\renewcommand{\arraystretch}{1.3}
\resizebox{\textwidth}{!}{
\begin{tabular}{DCCCCECCDDCEEE}
\hline\hline
|V_{cb}|\times 10^3	 & \mathcal{G}(1)	 & \mathcal{F}(1)	 & \rhossq	 & \cs	 & \hat\chi_1'(1)	 & \hat\eta(1)	 & \hat\eta'(1)	 & \mbS~[\GeV]	 & \dmbc~[\GeV]	 & \hat\phi_0'(1)	 & \hat{e}_3(1)	 & \hat{e}_3'(1)	 & \hat{c}_0(1)  \\ 
\hline
38.98(68)	 & 1.055(7)	 & 0.903(12)	 & 0.96(3)	 & 2.00(14)	 & \nax	 & 0.17(4)	 & -0.09(8)	 & 4.73(5)	 & 3.40(5)	 & -0.6(2)	 & \nax	 & \nax	 & \nax  \\ 
\hline\hline
\end{tabular}}
\caption{Fit results for the single terminating node of the nested hypothesis test graph for the \ShapeFit fit scenario in the VC limit.
The $\mathcal{O}(\aS^2)$ correction in Eq.~\eqref{eqn:hA1as2shift} is imposed,
and must be included when using these fit results.}
\label{tab:VCfit}
\end{table}

More concerning than the shift in $R(D^*)$, however,
is that the VC limit \ShapeFit fits have no sensitivity to $\hat{c}_0(1)$,
which solely determines the second-order power correction to $\mathcal{F}(1)$ in the VC limit 
(see Eq.~\eqref{eqn:FGhqetncF1}).
This insensitivity arises by construction: because the same $2(\ec-\eb)^2\hat{c}_0(1)$ term enters both $\mathcal{F}(1)$ and $\mathcal{G}(1)$,
and $\hat{c}_0(1)$ does not enter $R_1(1)$ or $R_2(1)$ (see Eq.~\eqref{eqn:FGhqetnc}),
relaxing the normalization constraints by definition eliminates any constraining power on $\hat{c}_0(1)$.
As a result the recovered $\mathcal{F}(1)$ is unchanged from $\mathcal{F}(1)^{\text{NLO}}_{\text{HQET}}$,
up to the small $\mathcal{O}(\aS^2)$ correction in Eq.~\eqref{eqn:F1as2shift}:
one finds the recovered $\mathcal{F}(1) = 0.957(2)$ and $\mathcal{G}(1) =1.050(5)$.
Therefore the optimal VC \ShapeFit fit does not address the tension with LQCD predictions for $\mathcal{F}(1)$.
While $\mathcal{O}(1/m_c^3)$ corrections may be of percent size,
it seems unlikely that the third-order VC limit corrections could resolve the remaining tension in $\mathcal{F}(1)$ at the $5\%$ level. 
Therefore, while the VC limit parametrization can describe the shape of the $\Bbar \to \Dx \ell \nu$ spectra, 
it is unlikely to be able to provide a full description of the data.

\subsection{Branching ratios, forward-backward asymmetries, and polarizations}

\begin{figure}[t]
	\subfigbottomskip=-10pt
	\subfigcapskip=-30pt
	\subfigure[][\label{fig:RDDs_ell}]{
		\includegraphics[width = .48\textwidth]{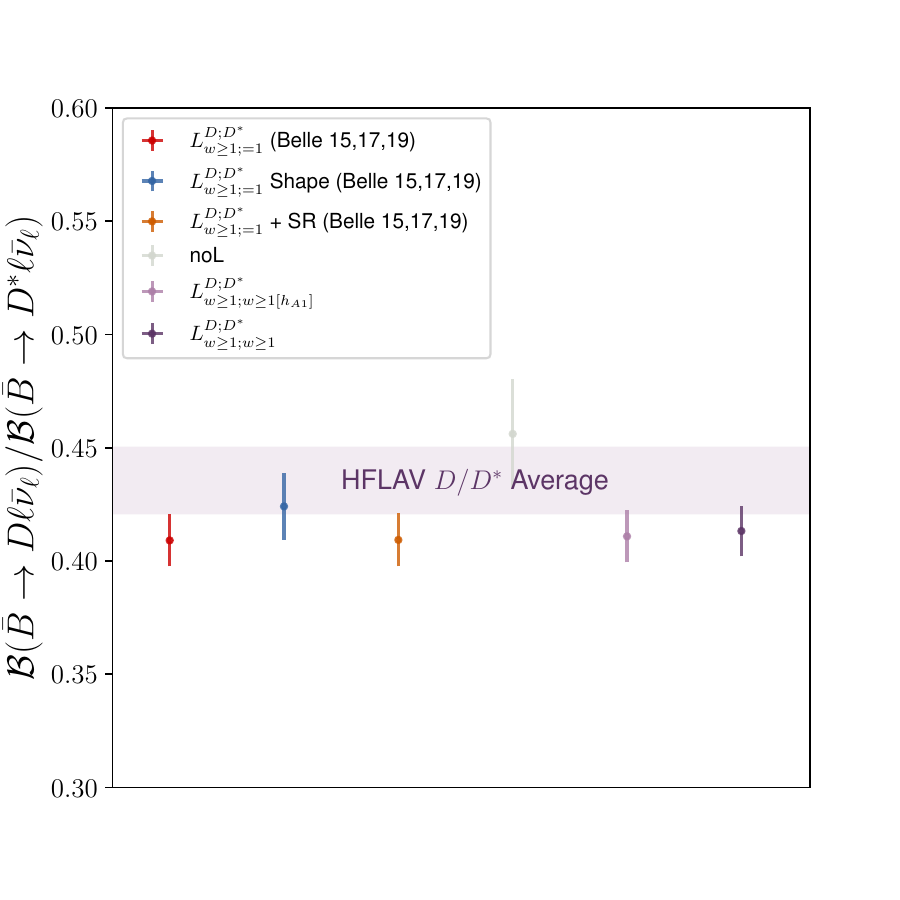}
	}\hfil
	\subfigure[][\label{fig:RDDs_tau}]{
		\includegraphics[width = .48\textwidth]{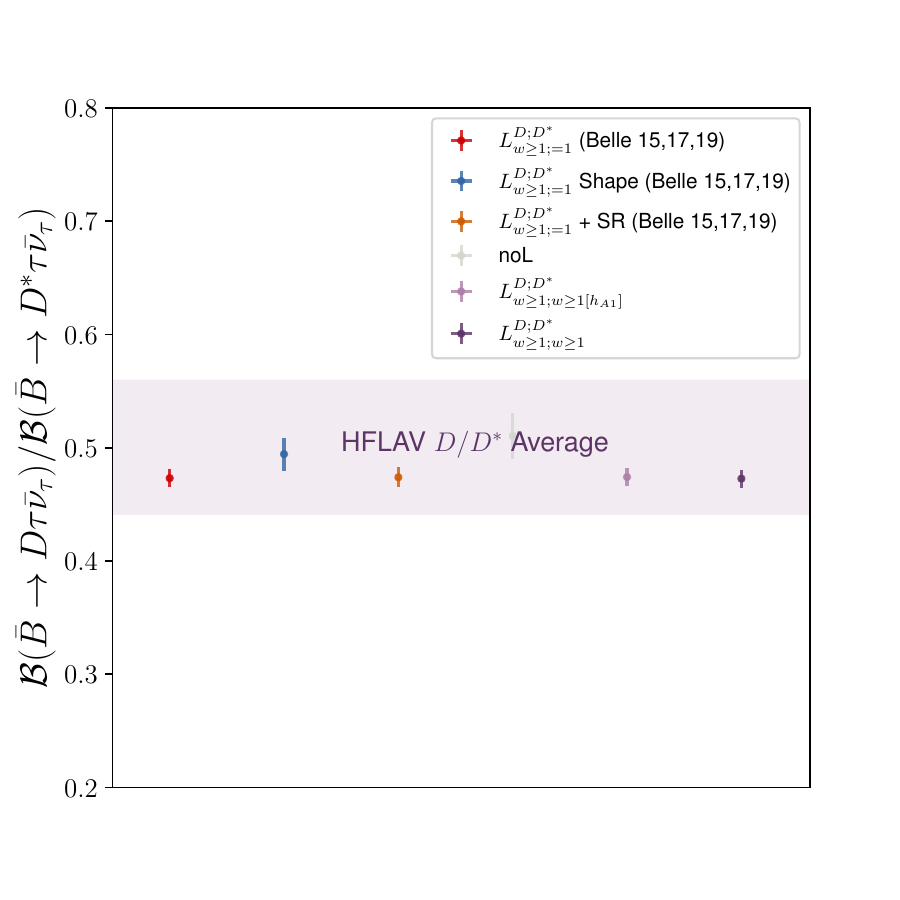}
	}
	\caption{Prediction for the ratio $R_{D/D^*}^l$ for the scenarios of Table~\ref{tab:FitKey} in the RC expansion, 
	for \subref{fig:RDDs_ell} $\ell=e$,$\mu$ and \subref{fig:RDDs_tau} $l = \tau$.}
	\label{fig:RDDs}
\end{figure}

With our fits we can produce precise predictions for several additional observables.
We quote predictions here based on our \BaseFit baseline scenario in the RC expansion, unless stated otherwise. 
First, for the $D$\,--\,$D^*$ ratio
\begin{equation}
	R_{D/D^*}^l \equiv \frac{\mathcal{B}[\Bbar \to D l \nu]}{\mathcal{B}[\Bbar \to D^* l \nu]} \,,
\end{equation}
we find 
\begin{equation}
	\label{eq:S1_RDDs}
 	 R_{D/D^*}^{e,\mu} = 0.417(12) \quad \text{and} \quad R_{D/D^*}^{\tau} = 0.479(8) \, .
\end{equation}
The light lepton value $R_{D/D^*}^{e,\mu}$ can be compared to the world averages of Ref.~\cite{Amhis:2019ckw}.
Averaging the branching fractions from both $B^0$ and $B^+$ decays assuming isospin (including a correction for their relative lifetimes), 
one obtains 
\begin{equation}
	\label{eq:HFLAV_RDDs}
	R_{D/D^*}^{e,\mu \, \text{HFLAV}} = 0.436(15) \,,
\end{equation}
in agreement with the baseline result~\eqref{eq:S1_RDDs} at the $1\sigma$ level.
The predictions for the other scenarios listed in Table~\ref{tab:FitKey} are compared with Eq.~\eqref{eq:HFLAV_RDDs} in Fig.~\ref{fig:RDDs_ell}.
They all show good agreement with the world average.

Similarly, one can convert the measured world averages of $R(D)$ and $R(D^*)$~\cite{Amhis:2019ckw} 
into a ratio of branching fractions, $R_{D/D^*}^{\tau}$. 
Using the average in Eq.~\eqref{eq:HFLAV_RDDs}, we find $R_{D/D^*}^{\tau \, \text{HFLAV}} = 0.50(6)$.
The $R_{D/D^*}^{\tau}$ predictions for the various fit scenarios in the RC expansion are compared to this value in Fig.~\ref{fig:RDDs_tau},
all showing good agreement with the world average.

The forward-backward asymmetry is defined by 
\begin{equation}
	A_{\text{FB}}[m_l] = \frac{1}{\Gamma[\Bbar \to D^ *l \nu] }\bigg[\int_{0}^1 -  \int_{-1}^0 \, \bigg] d \cos \theta_l \frac{d \Gamma[\Bbar \to D^* l \nu]}{d \cos \theta_l}\,,
\end{equation}
with $\theta_l$ denoting the polar helicity angle between the lepton momentum $\vec{p}_l$ and $-\vec{p}$ in the $l\nu$ rest frame.
For decays involving light leptons, this asymmetry has attracted recent attention~\cite{Belle:2018ezy,Bobeth:2021lya,Bhattacharya:2022cna}. 
We find
\begin{align}
  A_{\text{FB}}(m_l = 0) & =  0.244(4)   \, , &
  A_{\text{FB}}(m_e) & =  0.244(4) \, , \nn \\*
  A_{\text{FB}}(m_\mu) & = 0.239(4)  \, , &
  A_{\text{FB}}(m_\tau) & = 0.065(2)  \, .
\end{align}
In addition, we can also predict the difference $ \Delta A_{\text{FB}} $ and sum  $ \overline A_{\text{FB}}$, 
\begin{subequations}
\begin{align}
   \Delta A_{\text{FB}} \equiv  A_{\text{FB}}(m_\mu) -   A_{\text{FB}}(m_e)  & =  -0.0057(1)   \, , \\
    \overline A_{\text{FB}} \equiv \frac{1}{2} \big[A_{\text{FB}}(m_\mu) + A_{\text{FB}}(m_e)\big]   & =  0.242(4)   \, .
\end{align}
\end{subequations}
To compare to the experimental value of $ \Delta A_{\mathrm{FB}}$ from Ref.~\cite{Belle:2018ezy}, 
one needs to also include a small phase-space cut $q^2 > 0.08 \, \GeV^2$, 
for which the quoted measurement is not corrected. This results in 
\begin{align}
  A_{\text{FB}}( m_e, q^2 > 0.08 \, \GeV^2) & =  0.246(4) \, , 
  & A_{\text{FB}}(m_\mu, q^2 > 0.08 \, \GeV^2) & = 0.241(4)  \, , \nn\\
   \Delta A_{\text{FB}} (q^2 > 0.08 \, \GeV^2)& =  -0.0050(1)   \, , 
 &   \overline A_{\text{FB}}(q^2 > 0.08 \, \GeV^2)& =  0.243(4)  \, .
\end{align}
This exhibits a tension of  $\sim 4\sigma$ compared to $\Delta A_{\text{FB}}^{\text{Belle}}(q^2 > 0.08 \, \GeV^2) = 0.035(9)$~\cite{Bobeth:2021lya}. 

We also derive predictions for the LFUV ratios constrained to the $\tau$ phase space
\begin{align}
	\widetilde R(\Dx) = \frac{ \int_{1}^{w_\tau}  dw \big[d \Gamma[\Bbar \to \Dx \tau \nu] / d w\big]  }{
	 \int_{1}^{w_\tau}  dw \big[d \Gamma[\Bbar \to \Dx \ell \nu] / d w\big] }\,,
\end{align}
with $w_\tau = (m_B^2 - m_{D^{(*)}}^2 - m_\tau^2 ) / ( 2 m_B m_{D^{(*)}})$. 
These ratios feature a larger cancellation of experimental and theory uncertainties~\cite{Freytsis:2015qca,Bernlochner:2016bci,Isidori:2020eyd}. We find
\begin{align}
\widetilde R(D) =  0.571(4) \,, \qquad  \widetilde R(D^*) = 0.339(1) \,,
\end{align}
with a correlation of $\rho = 0.24$. 

For the longitudinal $D^*$ polarization fraction, 
\begin{align}
 F_{L,l} (D^*) \equiv \frac{  \Gamma_{\lambda = 0} [\Bbar \to D^* l \nu] }{ \Gamma [\Bbar \to D^* l \nu]  } \, ,
\end{align}
in which $\lambda = \pm,0$ labels the $D^*$ spin in the helicity basis,
we find for the light leptons $\ell = e,\mu$ and for the $\tau$
\begin{align}
 	F_{L,\ell} (D^*)  = 0.516(3) \quad \text{and} \quad  F_{L,\tau} (D^*)  = 0.454(3) \, ,
\end{align}
respectively.
The difference for the light leptons, $\Delta F_{L} (D^*) \equiv   F_{L,\mu} (D^*)  -  F_{L,e} (D^*) = 0.00012(1)$.
We find for the $\tau$ polarization, 
\begin{align}
 P_{\tau} (\Dx) = \frac{  \big[ \Gamma_{\lambda_\tau = +}  - \Gamma_{\lambda_\tau = -} \big] [\Bbar \to \Dx \tau \nu] }{ \Gamma [\Bbar  \to \Dx \tau \nu]  } \, ,
\end{align}
in which $\lambda_\tau = \pm$ labels the $\tau$ spin in the helicity basis,
\begin{align}
 P_{\tau} (D)   = 0.323(3) \quad \text{and} \quad   P_{\tau} (D^*)   = -0.494(5) \,.
\end{align}
Finally, for the electron-muon universality ratios
\begin{equation}
	R_{e/\mu}(\Dx) \equiv \frac{ \Gamma[\Bbar \to \Dx e \nu] }{ \Gamma [\Bbar \to \Dx \mu \nu] }\,,
\end{equation}
we find
\begin{equation}
	R_{e/\mu}(D) = 1.0028(1)\,, \quad \text{and} \quad R_{e/\mu}(D^*) = 1.0041(1)\,,
\end{equation}
where the latter is notably different from the prediction in Ref.~\cite{Bobeth:2021lya}, $R_{e/\mu}(D^*) = 1.0026(1)$.
For the \BaseFit[NLO] fit scenario 
we find similarly $R_{e/\mu}(D) = 1.0026(1)$ and $R_{e/\mu}(D^*) = 1.0041(1)$.

\section{Summary}

We developed a supplemental power counting for HQET,
based on counting insertions of the transverse residual momentum, $\slashed{D}_\perp$, within HQET correlators:
the residual chiral expansion.
We conjectured that higher-order terms within this power counting may be suppressed,
and showed how this leads to a dramatic simplification of the second-order power corrections in HQET, 
when truncating at $\mathcal{O}(\theta^2)$ in the RC expansion.
In doing so, we presented a review of the formal elements of the general HQ expansion,
that are required when working at second order and beyond.
Though these formal developments are not new per se, 
we are unaware of a self-contained and self-consistent presentation of these elements in the literature.

We proceeded to derive the $\mathcal{O}(\theta^2)$ second-order power corrections 
to the $\Bbar \to \Dx$ form factors both within and beyond the Standard Model,
including also the $\mathcal{O}(\aS/m_{c,b})$ corrections. 
At second order only three Isgur-Wise functions are required to describe the $\mathcal{O}(\theta^2)$ second-order power corrections, 
and only one when considering only $\mathcal{O}(1/m_c^2)$ terms.
Moreover, at zero recoil, almost all second-order corrections are fully determined  by the HQ mass parameters $\lam{1,2}$.
Similarly, we derived all second-order power corrections to the $\Bbar \to \Dx$ form factors in the vanishing chromomagnetic limit.
Our results provide the first check of Ref.~\cite{Falk:1992wt}---which 
used somewhat different conventions---for 
those terms entering at second-order in the RC expansion or the in VC limit.
We find agreement except for the sign of the $\mathcal{O}(1/m_cm_b)$ wavefunction $\hM{10}$ and an exchange of $\hM{17}$ with $\hM{19}$ and $\hM{22}$ with $\hM{23}$.
Whether these errors also affect the additional terms that arise when including all second-order corrections remains to be checked.

Based on the RC expansion results, we developed a form factor parametrization, 
applying the $1S$ short distance mass scheme that is self-consistent at second order in the HQ expansion.
These results are encoded in the \texttt{Hammer} library~\cite{bernlochner_florian_urs_2022_5828435, Bernlochner:2020tfi}.
We showed that the resulting zero-recoil predictions for $\Bbar \to \Dx$ form factors, $\mathcal{G}(1)$ and $\mathcal{F}(1)$, 
are in good agreement with zero recoil LQCD data, 
in particular resolving the prior tension of the LQCD data with the first-order prediction for $\mathcal{F}(1)$.

Confronting our parametrization of the form factors with experimental and LQCD data, 
we identified optimal parameter sets for the RC expansion under a nested hypothesis test prescription.
We found that the RC expansion can achieve excellent agreement with the data, with relatively few parameters, 
and without using any QCDSR or LCSR model-dependent inputs.
The VC limit parametrization produces poor fits due to its restricted structure at zero recoil,
but using only shape information yields good fits.

We recovered for our best fit
\begin{equation}
		R(D) = 0.288(4)	\,, \quad R(D^*) = 0.249(3)\,, \quad \text{and} \quad |V_{cb}| = 38.7(6)\times 10^{-3}\,,
\end{equation}
in which the $R(D^*)$ prediction contains a scale factor of $2.6$ 
to account for tensions in the predictions from the Belle 2017 versus 2019 $\Bbar \to D^* l \nu$ datasets.
The inclusion of zero-recoil second-order power corrections in the RC expansion was crucial to good fits, 
but the inclusion of second-order power corrections beyond zero recoil was not.
This supports the approach used in Ref.~\cite{Bernlochner:2017jka},
which used only the shape of the differential distributions to constrain the subleading Isgur-Wise functions, 
under the premise that second-order corrections are important only at zero recoil.
We found that the simplified linear CLN slope-curvature relation advocated in Ref.~\cite{Caprini:1997mu} is in tension with the data,
and leads to large upward biases in $R(D)$ predictions obtained in previous analyses.
Our fitting prescription is systematically improvable with more precise future data, 
that will simultaneously allow further tests of the RC expansion, as well as improved determination of $|V_{cb}|$,
the LFUV ratios $R(\Dx)$ and other observables.

\acknowledgments
We thank Nathaniel Craig, Marat Freytsis, Mike Luke, and Mark Wise for discussions and comments on the manuscript.
ZL and MP thank the Aspen Center for Physics (supported by the NSF Grant PHY-1607611) for its hospitality, where part of this work was performed.
FB is supported by DFG Emmy-Noether Grant No. BE 6075/1-1 and BMBF Grant No. 05H21PDKBA. FB thanks LBNL for its hospitality.
MP is supported by the U.S. Department of Energy, Office of Science, Office of High Energy Physics, under Award No. DE-SC0011632 and by the Walter Burke Institute for Theoretical Physics.
MTP is supported by the Argelander Starter-Kit Grant of the University of Bonn and BMBF Grant No. 05H21PDKBA.
ZL and DJR are supported by the Office of High Energy Physics of the U.S. Department of Energy under Contract No. DE-AC02-05CH11231.

\clearpage
\appendix
\makeatletter
\let\save@section\section
\renewcommand{\section}[1]{
	\save@section{#1}
	\addtocontents{toc}{\protect\vspace*{-10pt}}
}
\makeatother

\section{Hadron mass expansion}
\label{app:hadmass}

The spectroscopy of heavy hadrons in HQET is typically understood (see e.g. Ref.~\cite{Manohar:2000dt})
via the claim that the leading-order contribution to the hadron mass in the effective theory should be $m_H - m_Q$,
because of the mass subtraction in the field redefinition~\eqref{eqn:masssub} acting similarly on the Hamiltonian:
$\mathcal{H} \to \mathcal{H} - m_Q$.
Higher-order corrections can then be obtained from the expansion of the HQET Hamiltonian 
$\mathcal{H} - m_Q = \mathcal{H}_{\text{light}} + \delta\mathcal{H}$.
Here $\mathcal{H}_{\text{light}}$ encodes the dynamics of the light degrees of freedom---the brown muck---in the hadron.
For a hadron whose brown muck is in a state of definite spin-parity, $s^{\pi_\ell}$,
this generates a contribution $\LamB(s^{\pi_\ell})$ to the HQET hadron mass
(i.e., a different parameter for each spin-parity state of the brown muck).
The remaining term encodes the power corrections to the HQET Hamiltonian $\delta \mathcal{H} = -\sum_{n=1}\mathcal{L}_i/(2m_Q)^n$.

It is instructive to understand the HQ expansion of the hadron masses
via the matrix element matching~\eqref{eqn:QCDmatch} between QCD and HQET.
To do this, it is crucial to note~\cite{Luke:1990eg} that one is ultimately interested 
in operators of the form $\Qbar'(x) \Gamma Q(x) \, L(x)$, 
in which $L(x)$ is a current coupling to the heavy quark current
(such as a lepton bilinear in the case of semileptonic decays). 
Momentum conservation implies that $iD_\mu L = -(p-p')_\mu L$,
in which $p$ ($p'$) is the momentum of the hadron containing $Q$ ($Q'$).
In this work, we always choose the heavy quark velocity to be that of the hadron containing it, i.e., 
\begin{equation}
	\label{eqn:hadv}
	p = m_{H} v\,.
\end{equation} 
Thus, because the total derivative of the interaction operator should vanish, and because of the field redefinition~\eqref{eqn:masssub}, 
it follows that 
\begin{multline}
	\label{eqn:diffeq}
	i \partial_\mu \big\langle H'(p') \big| \Qbar^{v'}_+(x) \Gamma Q^v_+(x)  \big| H(p) \big\rangle \ceq \big[(m_H - m_Q) v_\mu - (m_{H'} - m_{Q'}) v'_\mu\big] 
			\\ \times \big\langle H'(p') \big| \Qbar^{v'}_+(x) \Gamma Q^v_+(x) \big| H(p) \big\rangle\,,
\end{multline}
in which the `$\ceq$' denotes equality under composition with $L(x)$.
Defining the operator $\genbar{X}_{H}$ as the annihilator of the light degrees of freedom in the hadron, 
it similarly follows that 
\begin{equation}
	\label{eqn:decconH}
	i \partial_\mu \big\langle 0 \big| \genbar{X}_{H}(x) Q^v_+(x)  \big| H(p) \big\rangle \ceq (m_H - m_Q) v_\mu \big\langle 0 \big| \genbar{X}_{H}(x) Q^v_+(x)  \big| H(p) \big\rangle\,.
\end{equation}

Let us define the power expansion of the mass splitting $m_H - m_Q = \sum_{n=0} \Delta m^{H}_{n+1}/(2m_Q)^n$.
The equation of motion for $Q^v_+$ applied to Eq.~\eqref{eqn:decconH} implies that at leading order
\begin{equation}
	\label{eqn:LamBdef}
	\Delta m^{H}_1 = \frac{\langle 0 | \genbar{X}_{H}(x) i v \ccdot \overleftarrow{D} Q^v_+(x)  | H \rangle}{\langle 0 | \genbar{X}_{H}(x) Q^v_+(x)  | H \rangle} \equiv \LamB\,,
\end{equation}
which can be thought of as the energy of the brown muck in the heavy quark limit.
One may further deduce from Eq.~\eqref{eqn:diffeq} that the individual matrix elements must take the form, 
under contraction with $v$ or $v'$,\footnote{%
Hereafter we assume that $\Hb$ and $\Hc$ (or $H$ and $H'$) belong to the same HQET multiplet, 
so that $\LamB$ and the other mass parameters $\lambda_{1,2}$, defined below, are the same for both hadrons.
For decays to excited states, for which this is not the case, two sets of parameters are required: 
The leading terms of Eqs.~\eqref{eqn:diffeqLR} generalize to
$(v_\mu w - v'_\mu)(\LamB w- \LamB')/(w^2-1)$ and $(v_\mu  - v'_\mu w)(\LamB' w- \LamB)/(w^2-1)$, respectively. }
\begin{subequations}
\label{eqn:diffeqLR}
\begin{align}
	v^{(\prime) \mu}\big\langle H'(p') \big| \Qbar^{v'}_+ \Gamma i D_\mu Q^v_+ \big| H(p) \big\rangle 
	& \ceq v^{(\prime) \mu} \bigg[\LamB \frac{v_\mu w \!-\! v'_\mu}{w+1} + v_\mu\! \sum_{n=1} \frac{\Delta m^{H}_{n+1}}{(2m_Q)^n}\bigg] \big\langle H'(p') \big| \Qbar^{v'}_+ \Gamma Q^v_+ \big| H(p) \rangle\,, \\
	v^{(\prime) \mu}\big\langle H'(p') \big| \Qbar^{v'}_+  i \overleftarrow{D}_\mu \Gamma Q^v_+ \big| H(p) \big\rangle 
	& \ceq  v^{(\prime) \mu} \bigg[\LamB \frac{v_\mu \!-\! v'_\mu w }{w+1} - v'_\mu\! \sum_{n=1} \frac{\Delta m^{H'}_{n+1}}{(2m_{Q'})^n}\bigg] \big\langle H'(p') \big| \Qbar^{v'}_+ \Gamma Q^v_+ \big| H(p) \rangle\,.
\end{align}
\end{subequations}
These ensure that the equations of motion for $Q^v_+$ and $\Qbar^{v'}_+$ are satisfied at leading order,
and the forward-scattering matrix element of the HQET operator is subleading, i.e.,
\begin{equation}
	\big\langle H \big| \Qbar^v_+ i\vcD Q^v_+ \big| H \big\rangle  \ceq \sum_{n=1} \frac{\Delta m^{H}_{n+1}}{(2m_Q)^n} \big\langle H \big| \Qbar^v_+ Q^v_+ \big| H \rangle\,. \label{eqn:Hmass}
\end{equation}
With reference to Eq.~\eqref{eqn:QCDmatch}, at first order the left side of Eq.~\eqref{eqn:Hmass} matches onto HQET as
\begin{equation}
	\frac{\big\langle H \big| \Qbar^v_+ i\vcD Q^v_+ \big| H \big\rangle}{m_H}  \ceq  \big\langle H^v \big| - \mathcal{L}_1/(2m_Q) \big| H^v \big\rangle + \ldots\,,
\end{equation}
where we made use of the contact term~\eqref{eqn:cont} plus the $Q^v_+$ equation of motion.
The hadronic mass can then be expanded to second order as\footnote{%
In the notation of Ref.~\cite{Falk:1992wt}, the higher-order power corrections to the mass are denoted as $\Delta m_H^2$. 
However, because this closely resembles the notation for the mass splitting within a HQET doublet, we use a different notation.}
\begin{equation}
	\label{eqn:HmassHQEapp}
	m_{H} = m_Q + \LamB + \frac{\Delta m^{H}_2}{2m_Q} + \ldots\,, \qquad \Delta m^{H}_2 \simeq \frac{\langle H^v | (-\mathcal{L}_1) | H^v \rangle}{\langle H^v | \Qbar^v_+ Q^v_+ | H^v \rangle}\,.
\end{equation}
In our normalization~\eqref{eqn:HQETnorm}, $\langle H^v | \Qbar^v_+ Q^v_+ | H^v \rangle = 2$.
It is conventional to define the parameters
\begin{subequations}
\label{eqn:deflam12}
\begin{align}
	 \lam1 & = -\frac{\langle H^v | \Qbar_+^v D^2 Q^v_+ | H^v \rangle}{\langle H^v | \Qbar^v_+ Q^v_+ | H^v \rangle} \,,\\
	 d_{H} \lam2 & = -\frac{\langle H^v |  \Qbar_+^v (g \sigma_{\ab} G^{\ab}/2) Q^v_+ | H^v \rangle}{\langle H^v | \Qbar^v_+ Q^v_+ | H^v \rangle} \,,
\end{align}
\end{subequations}
where $d_H$ is a spin combinatoric factor, specific to a given hadronic state.
The mass correction then becomes 
\begin{equation}
	\label{eqn:masslam12}
	\Delta m^{H}_2 = - \lam1 - d_{H}a_Q(\mu)\lam2(\mu)\,,
\end{equation}
in which we have explicitly restored the scale-dependent renormalization of the chromomagnetic operator.
For a pseudoscalar ($P$) and vector ($V$) meson,
which fill a HQ spin symmetry doublet with brown muck spin-parity $s^{\pi_\ell} = \frac{1}{2}^-$, 
the factor $d_P = 3$ and $d_V = -1$, respectively.
(For the ground-state baryon, the brown muck is in a $s^{\pi_\ell} = 0^+$ state, and the chromomagnetic parameter $\lam2$ vanishes.)

\section{Schwinger-Dyson relations} 
\label{app:SDrel}

The mass parameters play an important role in Schwinger-Dyson style relations (also called modified Ward identities) between matrix elements 
entering at different orders in the HQ expansion in Eq.~\eqref{eqn:QCDmatch}.
Writing the HQET current operator $J_{\Gamma+}(z) = \cbvp(z) \Gamma \bv(z)$, 
then from Eqs~\eqref{eqn:diffeq} and~\eqref{eqn:HmassHQEapp} the derivative of the QCD matrix element with respect to $z$,
\begin{align}
	i\partial^z_\mu \big\langle \Hc \big| J_{\Gamma+}(z) \big| \Hb \big\rangle 
	& \ceq \bigg\{\bigg[\LamB + \frac{\Delta m^{\Hb}_2}{2m_b}\bigg] v_\mu 
		-  \bigg[\LamB + \frac{\Delta m^{\Hc}_2}{2m_c}\bigg] v'_\mu\bigg\} 
		\big\langle \Hc \big| J_{\Gamma+}(z) \big| \Hb \big\rangle + \ldots\,,\label{eqn:diffJeq}
\end{align}
expanding the mass splitting to $\mathcal{O}(1/m_Q)$.
However, with reference to Eq.~\eqref{eqn:QCDmatch}, to first order the matrix element
\begin{align}
\frac{\big\langle \Hc \big| J_{\Gamma+}(z) \big| \Hb \big\rangle}{\sqrt{m_{\Hb} m_{\Hc}}}
	& = \big\langle \Hc^{v'} \big| \frac{1}{\mathcal{Z}}\int \mathcal{D} \cbvp \mathcal{D} c^{v'}_+ \mathcal{D} \bbar^{v}_+ \mathcal{D} \bv \nn \\
	& \qquad \qquad \times \exp\bigg\{i \!\int\! d^4x \big[\mathcal{L}'_{\text{HQET}} + \mathcal{L}_{\text{HQET}} \big](x) \bigg\} J_{\Gamma+} (z) \big| \Hb^v \big\rangle\nn\\
	& \simeq \big\langle \Hc^{v'} \big| J_{\Gamma+}(z) \big| \Hb^v \big\rangle \label{eqn:Jme} \\
	& + \frac{1}{2m_c} \big\langle \Hc^{v'} \big|  \mathcal{L}'_1 \circ \cbvp(z) \Gamma \bv(z) \big| \Hb^v \big\rangle
	 + \frac{1}{2m_b} \big\langle \Hc^{v'} \big| \cbvp(z) \Gamma \bv(z) \circ \mathcal{L}_1 \big| \Hb^v \big\rangle\,, \nn 
\end{align}
and, similarly, the derivative of the matrix element,
\begin{align}
i\partial^z_\mu \frac{\big\langle \Hc \big| J_{\Gamma+}(z) \big| \Hb \big\rangle}{\sqrt{m_{\Hb} m_{\Hc}}}
	& \simeq \big\langle \Hc^{v'} \big|  \cbvp(z) i\overleftarrow{D}^z_\mu \, \Gamma \, \bv(z)  \big| \Hb^v \big\rangle 
	+ \big\langle \Hc^{v'} \big|  \cbvp(z) \, \Gamma i\overrightarrow{D}^z_\mu  \bv(z) \big| \Hb^v \big\rangle \nn \\*
	& + \frac{1}{2 m_c} \big\langle \Hc^{v'} \big|  \mathcal{L}'_1 \circ \cbvp(z) \big[i\overleftarrow{D}^z_\mu \, \Gamma + \Gamma i\overrightarrow{D}^z_\mu\big] \bv(z) \big| \Hb^v \big\rangle \nn \\*
	& + \frac{1}{2 m_b} \big\langle \Hc^{v'} \big| \cbvp(z) \big[i\overleftarrow{D}^z_\mu \, \Gamma + \Gamma i\overrightarrow{D}^z_\mu\big]  \bv(z) \circ  \mathcal{L}_1 \big| \Hb^v \big\rangle\,. \label{eqn:Jmediff}
\end{align}
Applying Eqs~\eqref{eqn:Jme} and~\eqref{eqn:Jmediff} to the relation~\eqref{eqn:diffJeq}, 
and matching order by order in $1/m_{c,b}$, one obtains at leading order the familiar relation
\begin{equation}
	\label{eqn:SDrel}
	\big\langle \Hc^{v'} \big|  \cbvp(z) i\overleftarrow{D}^z_\mu \, \Gamma \, \bv(z)  \big| \Hb^v \big\rangle 
	+ \big\langle \Hc^{v'} \big|  \cbvp(z) \, \Gamma i\overrightarrow{D}^z_\mu  \bv(z) \big| \Hb^v \big\rangle
	\ceq \LamB(v-v')_\mu \big\langle \Hc^{v'} \big| J_{\Gamma+}(z) \big| \Hb^v \big\rangle\,.
\end{equation}
The left-hand side comprises uncontracted versions of the $\mathcal{O}(1/m_{c,b})$ current corrections generated by $\mathcal{J}_1$, 
that is, with the $i\Dslash$ replaced by $iD_\mu$.
At first order in $1/m_c$, contracting with $v'_\mu$ and using the derivative contact term~\eqref{eqn:cont}, 
one finds (cf. Appendix C of Ref.~\cite{Falk:1992wt})
\begin{subequations}
\label{eqn:subSDrelall}
\begin{multline}
	\label{eqn:subSDrel}
	\big\langle \Hc^{v'} \big|  \cbvp(z) \Overleftarrow{\mJbar}^{\prime}_2 \, \Pi'_+ \Gamma \, \bv(z)  \big| \Hb^v \big\rangle 
	+ \big\langle \Hc^{v'} \big|  \mathcal{L}'_1 \circ \cbvp(z) \, \Gamma iv' \ccdot \overrightarrow{D}^z \bv(z) \big| \Hb^v \big\rangle \\*
	\ceq \LamB(w-1) \big\langle \Hc^{v'} \big|  \mathcal{L}'_1 \circ \cbvp(z) \Gamma \bv(z) \big| \Hb^v \big\rangle - \Delta m^{\Hc}_2 \big\langle \Hc^{v'} \big| J_{\Gamma+}(z) \big| \Hb^v \big\rangle \,.
\end{multline}
The first term on the left-hand side is the usual second-order current correction, but with a positive projector $\Pi_+'$.
The second term on the left-hand side resembles the $\mathcal{O}(1/m_c m_b)$ 
mixed Lagrangian and current corrections generated by $\mathcal{L}_1$ and $\mathcal{J}_1$, 
with the $i\Dslash$ replaced by $i v' \ccdot D$.
This relation thus constrains a combination of the matrix elements arising from
the $\mathcal{O}(1/m_c^2)$ current corrections and the $\mathcal{O}(1/m_c m_b)$ mixed corrections, 
with the second-order mass splitting times the leading-order matrix element
and $\LamB (w-1)$ times the matrix element generated by the $\mathcal{O}(1/m_c)$ Lagrangian corrections on the right-hand side.
The conjugate relation to Eq.~\eqref{eqn:subSDrel}, 
from the $\mathcal{O}(1/m_b)$ terms in the relation~\eqref{eqn:diffJeq}, is
\begin{multline}
\big\langle \Hc^{v'} \big|  \cbvp(z) \, \Gamma \Pi_+\, \Overrightarrow{\mJ}_2  \bv(z)  \big| \Hb^v \big\rangle 
		- \big\langle \Hc^{v'} \big|   \cbvp(z) i v \ccdot \overleftarrow{D}^z \Gamma \, \bv(z) \circ \mathcal{L}_1 \big| \Hb^v \big\rangle \\*
	\ceq \LamB(w-1) \big\langle \Hc^{v'} \big|  \cbvp(z) \Gamma \bv(z)\circ \mathcal{L}_1 \big| \Hb^v \big\rangle - \Delta m^{\Hb}_2 \big\langle \Hc^{v'} \big| J_{\Gamma+}(z) \big| \Hb^v \big\rangle \,.
\end{multline}
\end{subequations}

\section{Radiative corrections}
\label{app:genradcorr}

At $\mathcal{O}(\aS)$, perturbative corrections arise via the matching
\begin{equation}
	\cbar\, \Gamma\, b \to \cbvp \Big[\Gamma + \haS \sum_i C_{\Gamma_i} \Gamma_i\Big] \bv
\end{equation}
where $\haS = \aS/\pi$, 
and $\Gamma_1 = \Gamma$ and $\Gamma_{i >1}$ are a basis of operators generated by all combinations of replacements $\gamma^\mu \to v^{(\prime)\mu}$.
The functions $C_{\Gamma_i} = C_{\Gamma_i}(w,z)$ depend on the recoil parameter $w = v \ccdot v'$ and mass ratio $z=m_c/m_b$.
Specifically, at $\mathcal{O}(\aS)$ the following operators are generated 
(using the notation of Ref.~\cite{Manohar:2000dt})
\begin{align}
\label{eqn:ascurrent}
	\cbar\, b				&\to  \cbvp  \big( 1+ \haS\, C_S\big) \bv\,, \nn\\
	\cbar \gamma^5 b		&\to  \cbvp \big(1+ \haS\, C_P \big) \g^5  \bv\,,\nn\\
	\cbar \g^\mu b			&\to  \cbvp 
				\big[\big(1+ \haS\, C_{V_1} \big) \g^\mu + \haS\, C_{V_2}\, v^\mu + \haS\, C_{V_3}\, v'^\mu \big] \bv \,, \nn\\
	\cbar \g^\mu\g^5 b		&\to  \cbvp 
					\big[\big(1+ \haS\, C_{A_1} \big) \g^\mu + \haS\, C_{A_2}\, v^\mu + \haS\, C_{A_3}\, v'^\mu \big] \g^5 \bv\, \nn \\
	\cbar \sigma^{\mu\nu} b	&\to \cbvp \big[\big(1+ \haS\, C_{T_1} \big) \sigma^{\mu\nu} 
		+ \haS\, C_{T_2}\, i(v^\mu\g^\nu - v^\nu\g^\mu) + \haS\, C_{T_3}\, i(v'^\mu\g^\nu - v'^\nu\g^\mu)  \nn \\
  	& \qquad\quad + C_{T_4}(v'^\mu v^\nu - v'^\nu v ^\mu)\big] \bv\,,
\end{align}
from which one may read off the $\Gamma_i$ basis for each of the currents in Eq.~\eqref{eqn:Gdef}.

The $\mathcal{O}(\aS)$ corrections for all five currents were computed in Ref.~\cite{Neubert:1992qq};
explicit expressions are given in Ref.~\cite{Bernlochner:2017jka}.
The vector and axial-vector currents in QCD are (partially) conserved and so not renormalized, 
but the corresponding HQET currents have nonzero anomalous dimensions, 
leading to $\mu$-dependence for $C_{V_1}$ and $C_{A_1}$ for $w \neq 1$. 
The scalar, pseudoscalar, and tensor currents are renormalized in QCD, 
and thus $C_S$, $C_P$, and $C_{T_1}$ are also $\mu$-dependent. 
In the $\overline{\rm MS}$ scheme, 
the remaining $C_{\Gamma_j}$ ($j\ge 2$) are scale independent.

Because we are interested in the phenomenology of second-order power corrections, 
the $\mathcal{O}(\aS/m_{c})$ corrections should also be incorporated, 
as they may be comparable to the $1/m_c m_b$ terms. 
At this order both radiative corrections to local operators and nonlocal operators, 
arising from the operator product of the currents with the insertion of the first-order Lagrangian $\mathcal{L}_1$, enter. 
The coefficients of the local operators are fully determined by reparametrization invariance, 
while those of the nonlocal ones are just the products of the Wilson coefficients of the operators entering the operator product, expanded at $\mathcal{O}(\alpha_s)$. 
With reference to Eq.~\eqref{eqn:QCDmatch}, these additional contributions read~\cite{Neubert:1993iv, Neubert:1993mb}
\begin{align}
	\delta \big[\cbar \Gamma b\big]
	& \to \cbvp \frac{\Overleftarrow{\mJbar}_1^{\prime}}{2m_c}\Big[ \haS \sum_i C_{\Gamma_i} \Gamma_i\Big] \bv \nn \\*
	& \quad + \frac{1}{m_c} \haS \sum_i \Big[C'_{\Gamma_i}\cbvp (-iv \ccdot \overleftarrow{D}) \Gamma_i\bv + C_{\Gamma_i}\cbvp \big[\partial_{v'_\mu}\Gamma_i\big](-i \overleftarrow{D}_\mu) \bv\Big] \nn \\*
	& \quad + \frac{\mathcal{L}'_1 }{2m_c}  \circ   \cbvp \Big[  \haS\sum_i  C_{\Gamma_i} \Gamma_i\Big] \bv
	+ \frac{\haS C_g^c}{2 m_c} \Big[\cbvp \frac{g}{2} \sigma_{\ab} G^{\ab} \cvp\Big] \circ  \cbvp \Gamma \bv\,.
	\label{eqn:asec}
\end{align}
with $C'_\Gamma = \partial C_\Gamma /\partial w$.
In the last line we have written the nonlocal contributions from $\mathcal{L}'_1$
alongside the explicit $\mathcal{O}(\aS)$ correction from the renormalization of the chromomagnetic operator 
that arises from the $\mathcal{O}(1/m_c)$ operator product in Eq.~\eqref{eqn:QCDmatchexp}.

The first term on the second line may be determined by applying the relations~\eqref{eqn:diffeqLR}
after inserting the current into a correlator, and expanding to first order.
Thus the second line may be rewritten as
\begin{equation}
	 \frac{1}{2m_c} \haS \sum_i \Big[2 \LamB C'_{\Gamma_i}(w-1)\cbvp \Gamma_i\bv + 2C_{\Gamma_i}\cbvp \big[\partial_{v'_\mu}\Gamma_i\big](-i \overleftarrow{D}_\mu)\bv\Big]\,.
\end{equation}
The proportionality to $\LamB$ of the first term is explicit although all the remaining terms in 
Eq.~\eqref{eqn:asec} are also proportional to $\LamB$, as seen in the explicit evaluation of the relevant matrix elements, 
in Eqs.~\eqref{eqn:chromoNLO} and~\eqref{eqn:currNLO}.

The $\mathcal{O}(\aS/m_{b})$ terms are constructed similarly, using the Hermitian conjugates of the insertions in Eq.~\eqref{eqn:asec}.

\section{$\Bbar \to \Dx$ first and second-order power corrections}
\label{app:firstsecondorder}

The $w$-dependent $\hL{1\ldots6}{1}$ functions have well-known expressions (see e.g. Ref.~\cite{Falk:1992wt})
\begin{align}
	\label{eqn:Lhat1def}
	\hL11 &= 2\hat\chi_1 - 4(w-1) \hat\chi_2 + 12 \hat\chi_3\,, \nn\\
 	\hL21 & = 2\hat\chi_1- 4 \hat\chi_3\,, \nn\\
	\hL31  & = 4 \hat\chi_2\,, \nn\\
	\hL41 &= 2 \hat\eta - 1\,, \nn\\
	\hL51 & = -1\,, \nn\\
	\hL61 &= - 2 (1 + \hat\eta)/(w+1)\,,
\end{align}
in terms of the standard definitions of subleading Isgur-Wise functions $\hat\chi_i(w)$ and $\hat\eta(w)$ (see below).
Note that we retained explicitly the term parametrizing the correction from the kinetic energy operator, $\hat\chi_1$  (cf. Ref.~\cite{Bernlochner:2017jka}).

To compute the second-order power corrections in the RC expansion,
it is useful to briefly revisit the derivation of Eqs.~\eqref{eqn:Lhat1def}.
Under the trace formalism, we write the Lagrangian corrections
\begin{subequations}
\label{eqn:chromoNLO}
\begin{align}
	\big\langle \Hc^{v'} \big| \cbvp \Gamma \bv \circ \big[\bbv D^2 \bv \big]\big| \Hb^v \big\rangle & = \LamB \Tr[\Hbar_c(v') \Gamma H_b(v) X_0(v,v')]\,,\\*
	\big\langle \Hc^{v'} \big| \cbvp \Gamma  \bv \circ \big[\bbv \frac{g}{2}\sigma_{\ab} G^{\ab} \bv \big] \big| \Hb^v \big\rangle & = \LamB \Tr[ \Hbar_c(v')  \Gamma \Pi_+ \sigma_{\ab} H_b(v) X^{\ab}(v,v')]\,,\\
\intertext{and for the conjugate matrix elements}	
	\big\langle \Hc^{v'} \big| \big[\cbvp D^2 \cvp\big] \circ \cbvp \Gamma  \bv \big| \Hb^v \big\rangle & = \LamB \Tr[\Hbar_c(v') \Gamma H_b(v) \genbar{X}_0(v',v)]\,,\\*
	\big\langle \Hc^{v'} \big| \big[\cbvp \frac{g}{2}\sigma_{\ab} G^{\ab} \cvp\big] \circ \cbvp \Gamma  \bv \big| \Hb^v \big\rangle & =  \LamB \Tr[\Hbar_c(v') \sigma_{\ab} \Pi'_+\Gamma H_b(v) \genbar{X}^{\ab}(v',v)]\,.
\end{align}
\end{subequations}
The implicit two-point function arising from the operator product yields the $\Pi^{(\prime)}_+$ factors,
and the explicit $\LamB$ prefactor accords with our normalization convention.
The $T$ invariance of the strong interactions requires the antisymmetric tensor $X_{\ab}$ to obey  
$X_{\ab}(v,v') = \genbar{X}_{\ab}(v', v)$,
and similarly, $\genbar{X}_0(v',v) = X_0(v,v')$.
Thus the latter is a real-valued function---an Isgur-Wise function---with 
the conventional definition $X_0 = 2\chi_1(w)$,\footnote{\label{ft:chi1}%
When we encounter the second-order power corrections below, 
it will become apparent that it is unfortunate that the standard notation uses $\chi_1$ rather than $\chi_0$.}
while $X_{\ab}$ must take the form
\begin{equation}
	X_{\ab}(v,v') = -i\chi_2(w) (v-v')_{[\alpha} \g_{\beta]}  + 2\chi_3(w) \sigma_{\ab}\,,
\end{equation}
using the notation $x_{[\alpha} y_{\beta]} \equiv x_{\alpha} y_{\beta} -  y_{\alpha} x_{\beta}$,
and each of the $\chi_i$ is a real Isgur-Wise function.
There is no $v_{[\alpha} v'_{\beta]}$ term because $\Pi_+ \sigma_{\alpha\beta} \Pi_+ v^\alpha = 0$.

Three relations of particular utility are	
\begin{subequations}
\label{eqn:sigmaids}
\begin{align}
	\sigma_{\ab} H(v) \sigma^{\ab} 
		& = 6  (-\g^5)P^v - 2\vslash \slashed{\epsilon}V^v \nn\\
		& = 2 \big[d_P P^v(-\g^5) + d_V \vslash  \slashed{\epsilon} V^v\big]\,, \label{eqn:sigmaids1}\\
	\sigma_{\ab} H(v) (v-v')^{[\alpha} \g^{\beta]} \Hbar(v') 
		& = -4i \Big\{\big[(w-1)\Pi_+ + (w-2)\Pi_-\big] P^v(-\g^5) \nn\\* 
		& \qquad -\big[\Pi_- \slashed{\epsilon} + \epsilon \ccdot v'\big]V^v\Big\}\Hbar(v')\,, \label{eqn:sigmaids2}\\
	\sigma_{\ab} H(v) v_{[\alpha} v'_{\beta]} \Hbar(v') 
		& = -2i \Pi_-\Big\{(w+1)P^v(-\g^5) + \big[(w+1)\slashed{\epsilon} + 2 \epsilon \ccdot v'\big]V^v\Big\}\Hbar(v') \label{eqn:sigmaids3}\,.
\end{align}
\end{subequations}
Applying the $\Pi_+$ projector from the left in Eqs.~\eqref{eqn:sigmaids1} and~\eqref{eqn:sigmaids2}, 
inserting them into Eqs.~\eqref{eqn:chromoNLO}, 
and matching onto Eqs.~\eqref{eqn:reps}, 
one can read off the results for $\hL{1,2,3}{1}$. 

The current corrections are determined in the trace formalism by writing
\begin{subequations}
\label{eqn:currNLO}
\begin{align}
	\big\langle \Hc^{v'} \big| \cbvp \Gamma i \overrightarrow{D}_\mu \bv \big| \Hb^v \big\rangle & = -\LamB \Tr[\Hbar_c(v') \Gamma H_b(v) \Xi_\mu(v,v')]\,,\\
	\big\langle \Hc^{v'} \big| \cbvp  (-i \overleftarrow{D}_\mu) \Gamma \bv \big| \Hb^v \big\rangle & = -\LamB \Tr[\Hbar_c(v') \Gamma H_b(v) \overline{\Xi}_\mu(v',v)]\,,
\end{align}
\end{subequations}
and $\Xi$ must take the general form $\Xi_\mu = \xi_+(v+v')_\mu + \xi_-(v-v')_\mu - \xi_3 \g_\mu$.
Straightforward application of Eqs.~\eqref{eqn:diffeqLR} at leading order contracted under $v$ and $v'$---or
equivalently the leading-order Schwinger-Dyson relation~\eqref{eqn:SDrel} plus the equation of motion for $Q^v_+$---allows 
one to immediately deduce $\xi_- = \xi/2$, and $\xi_+ = -\xi_3/(w+1) + \xi  (w-1)/(2(w+1))$.
Inserting this result into Eqs.~\eqref{eqn:currNLO} and matching onto Eqs.~\eqref{eqn:reps}, 
one can read off the results for $\hL{4,5,6}{1}$, with the Isgur-Wise function $\hat\eta \equiv \xi_3/\xi$.

We turn now to the second-order current corrections from $\mathcal{J}_2$.
Just as for the first-order Lagrangian corrections,
we write these contributions under the trace formalism as
\begin{subequations}
\begin{align}
	\big\langle \Hc^{v'} \big| \cbvp \Gamma D^2 \bv \big| \Hb^v \big\rangle & = \LamB^2\, \Tr[\Hbar_c(v') \Gamma H_b(v) \Phi_0(v,v')]\,,\label{eqn:currNNLOD2}\\
	\big\langle \Hc^{v'} \big| \cbvp \Gamma  \frac{g}{2}\sigma_{\ab} G^{\ab} \bv \big| \Hb^v \big\rangle & = \LamB^2\, \Tr[ \Hbar_c(v')  \Gamma \sigma_{\ab} H_b(v) \Phi^{\ab}(v,v')]\,, \label{eqn:currNNLOG}
	\end{align}
\end{subequations}
and similarly for the conjugate matrix elements.
$T$-invariance requires that $\Phi_{\ab}(v,v') = \overline{\Phi}_{\ab}(v', v)$, so that 
\begin{equation}
	\Phi_{\ab}(v,v') = i\varphi_1(w) v_{[\alpha}v'_{\beta]}  - i\varphi_2(w) (v-v')_{[\alpha} \g_{\beta]}  + 2\varphi_3(w) \sigma_{\ab}\,,
\end{equation}
and further $\Phi_0 = 2\varphi_0(w)$.\footnote{See footnote~\ref{ft:chi1}. 
With respect to the notation of Ref.~\cite{Falk:1992wt}, 
$2\LamB^2\varphi_0 = \phi_0$, $2\LamB^2 \varphi_1 = -\phi_1$, $2\LamB^2 \varphi_2 = \phi_2$, and $4\LamB^2 \varphi_3 = \phi_3$.}
Though we have defined the functions $\Phi_0$ and $\Phi_{\ab}$ with respect to an arbitrary interaction $\Gamma$,
note that the $\mathcal{J}_2$ terms in Eq.~\eqref{eqn:QCDmatchRC} feature an additional $\Pi_-$ insertion. 
Perforce, there is no $\Pi_+$ projector in Eq.~\eqref{eqn:currNNLOG} and hence the $\varphi_1$ term does not vanish inside the trace.

Application of the relations~\eqref{eqn:sigmaids} 
allows one to read off the contributions of the $\hat\varphi_i \equiv \varphi_i/\xi$ to the $\hL{i}2$ in Eqs.~\eqref{eqn:reps}.
However, just as for the first-order current corrections, 
the $\varphi_i$ are constrained by a Schwinger-Dyson relation, Eq.~\eqref{eqn:subSDrelRC} at $\mathcal{O}(\theta^2)$.
Applying the relations~\eqref{eqn:sigmaids} for the pseudoscalar and vector, this constraint reduces to
\begin{align}
	\lam1 \xi + d_P \lam2 \xi & = \LamB^2\big[2\varphi_0 - 4\varphi_2(w-1)+ 4 d_P \varphi_3\big]\,,\nn\\*
	\lam1 \xi + d_V \lam2 \xi & = \LamB^2\big[2\varphi_0 + 4\varphi_2+  4 d_V \varphi_3\big]\,.
\end{align}
It immediately follows that
\begin{equation}
	\label{eqn:phicons}
	2\varphi_0 = \lam1 \xi/\LamB^2\,,\quad 4\varphi_3 = \lam2 \xi/\LamB^2\,, \quad \text{and} \quad \varphi_2 = 0\,.
\end{equation}
If one includes higher-order terms in the RC expansion
these relations become more complicated, but Eq.~\eqref{eqn:phicons} remains valid at zero recoil.

An additional constraint on $\varphi_1(1)$ may be derived by observing 
that one could have equivalently defined $\Phi_{\alpha \beta}$ via	
\begin{equation}
	\label{eqn:currNNLODD}
	\big\langle \Hc^{v'} \big| \cbvp \Gamma  D_{\alpha} D_{\beta} \bv \big| \Hb^v \big\rangle 
		 = \LamB^2 \Tr\big\{ \Hbar_c(v')  \Gamma H_b(v) \big[\Psi_{\alpha\beta}(v,v') + i\Phi_{\ab}(v,v')\big]\big\}\,,
\end{equation}
in which $\Psi_{\alpha\beta}$ is a symmetric tensor, 
having the general form
\begin{multline}
	\Psi_{\ab} = \psi_1g_{\ab} + \psi_2(v+v')_{\alpha}(v+v')_{\beta} + \psi_3(v-v')_{\alpha}(v-v')_{\beta} + \psi_4(v+v')_{(\alpha}(v-v')_{\beta)} \\
		 + \psi_5(v+v')_{(\alpha}\g_{\beta)} + \psi_6(v-v')_{(\alpha}\g_{\beta)}\,.
\end{multline}
Imposition of the $Q^v_+$ equation of motion requires that $\Pi_-[v^\beta (\Psi_{\ab}+i\Phi_{\ab})]\Pi'_- = 0$
(the $\Pi^{(\prime)}_-$ projectors arise because $H(v)\vslash = -H(v)$),
and it must be the case that $\Phi_0 = \Pi_- g^{\ab} \Psi_{\ab} \Pi'_-$.
Further, because via integration by parts and Eq.~\eqref{eqn:diffeq} (or via Eq.~\eqref{eqn:SDrel})
\begin{multline}
	\label{eqn:ibp}
	\big\langle H'(p') \big| \Qbar^{v'}_+ (-i \overleftarrow{D}_\mu) \Gamma  i \overrightarrow{D}_\nu Q^v_+ \big| H(p) \big\rangle 
		= -\LamB (v-v')_\mu \big\langle H'(p') \big| \Qbar^{v'}_+ \Gamma  i D_\nu Q^v_+ \big| H(p) \big\rangle \\
		- \big\langle H'(p') \big| \Qbar^{v'}_+ \Gamma   D_\mu D_\nu Q^v_+ \big| H(p) \big\rangle\,,
\end{multline} 
then the $\Qbar_+^{v'}$ equation of motion requires $\Pi_-(v^{\prime\alpha}(\Psi_{\ab}+i\Phi_{\ab}))\Pi'_- = (w-1) \Xi_\beta(v,v')$.
These three conditions together have solution for $\hat\psi_i \equiv \psi_i/\xi$
\begin{align}
\hat\psi_1 & = \frac{\lam1}{2\LamB^2}-\frac{\lam2 w}{2\LamB^2 (w+1)} - w \hvph + \frac{w-1}{2 (w+1)}\,,\nn\\
\hat\psi_2 & = - \frac{\lam1}{4 \LamB^2 (w+1)}+ \frac{\lam2 (2 w+3)}{4 \LamB^2 (w+1)^2}+ \frac{(2 w-1) \hvph}{2 (w+1)} + \frac{(w-1) (-4 \hat\eta + w-2)}{4 (w+1)^2}\,, \nn\\
\hat\psi_3 & = \frac{\lam1}{4\LamB^2 (w-1)} - \frac{\lam2 (2 w+1)}{4\LamB^2(w^2-1)} - \frac{(2 w+1) \hvph}{2(w-1)} + \frac{w+2}{4(w+1)}\,,\nn\\
\hat\psi_4 & = \frac{-2 \hat\eta + w - 1}{4 (w+1)}\,, \qquad 
\hat\psi_5 = \frac{\lam2}{2\LamB^2(w+1)} -  \frac{\hat\eta(w-1)}{2 (w+1)} \,, \qquad
\hat\psi_6 = -\hat\eta/2\,. \label{eqn:psisol}
\end{align}
At zero recoil, the symmetric tensor $\Psi_{\ab}$ must have the form $g_{\ab} - v_\alpha v_\beta$.
In the limit $w \to 1$,  $(v-v')_\alpha (v-v')_\beta/(w-1)$ is finite, 
but it cannot be written as a linear combination of the metric and $v_\alpha v_\beta$.
Hence $(w-1)\psi_3(w)$ must itself vanish at zero recoil.
It follows that
\begin{equation}
	\label{eqn:vphzerorec}
	\hvph(1) = \frac{1}{2\LamB^2}\bigg[\frac{\lam1}{3} - \frac{\lam2}{2}\bigg]\,,
\end{equation}
and the analyticity of the matrix elements near zero recoil permits us to write
\begin{equation}
	\label{eqn:phpdef}
	\hvph(w) = \frac{1}{2\LamB^2}\bigg[\frac{\lam1}{3} - \frac{\lam2}{2}\bigg] + (w-1)\hvphq(w)\,.
\end{equation}
where the function $\hvphq$ is regular. 
As in Eq.~\eqref{eqn:quotdef}, 
$\hvphq$ denotes the \emph{quotient} with respect to $w=1$,
and $\hvphq(1) =\hvphp(1)$, the gradient at zero recoil.

The Lagrangian corrections from $\mathcal{L}_2$ are represented under the trace formalism as
\begin{subequations}
\label{eqn:chromoNNLO}
\begin{align}
	\big\langle \Hc^{v'} \big| \cbvp \Gamma \bv \circ \big[\bbv g v_\beta  D_\alpha G^{\ab}  \bv \big]\big| \Hb^v \big\rangle
		 & = -\LamB^2 \Tr[\Hbar_c(v') \Gamma H_b(v) B_0(v,v')]\,,\\
	\big\langle \Hc^{v'} \big| \cbvp \Gamma  \bv \circ \big[-i\bbv v_\alpha \sigma_{\beta\gamma} D^\gamma G^{\ab} \bv \big] \big| \Hb^v \big\rangle 
		& = -\LamB^2 \Tr[ \Hbar_c(v')  \Gamma \Pi_+ \sigma_{\ab} H_b(v) B^{\ab}(v,v')]\,,
\end{align}
\end{subequations}
The tensors $B_0$ and $B_{\ab}$ must have the same forms as $X_0$ and $X_{\ab}$, 
so that writing $B_0 = 2 \beta_1$ and $B_{\ab} = -i\beta_2(w) (v-v')_{[\alpha} \g_{\beta]}  + 2\beta_3(w) \sigma_{\ab}$,
the subsubleading Isgur-Wise functions $\hat\beta_i = \beta_i/\xi$ enter $\hL{i}2$ identically to $\hat\chi_i$.
Thus we find, 
\begin{align}
	\label{eqn:Lhat2def}
	\hL12 &=  2\hat\beta_1 - 4(w-1) \hat\beta_2 + 12 \hat\beta_3\,, \nn\\
 	\hL22 & = 2\hat\beta_1- 4 \hat\beta_3\,,\nn\\
	\hL32 & = 4 \hat\beta_2\,, \nn\\
	\hL42 &=  3 \lam2/\LamB^2 + 2(w+1) \hvph \,, \nn\\
		  & = \frac{\lam1(w+1)}{3\LamB^2} - \frac{\lam2(w-5)}{2\LamB^2} + 2(w^2-1)\hvphq\,,\nn\\
	\hL52 & = \lam2/\LamB^2 +  2(w+1) \hvph \,, \nn\\
		  & = \frac{\lam1(w+1)}{3\LamB^2} - \frac{\lam2(w-1)}{2\LamB^2} + 2(w^2-1)\hvphq\,,\nn\\ 
	\hL62 & = 4\hvph = \frac{2\lam1}{3\LamB^2} - \frac{\lam2}{\LamB^2} + 4(w-1)\hvphq\,.
\end{align}

It remains now to compute the $1/m_cm_b$ current corrections from the product term $\mJbar^\prime_1 \mathcal{J}_1$.
Because of zero-recoil normalization constraints (see Sec.~\ref{sec:zerorec}) these
terms play a crucial role in the structure of the $\hL{i}{2}$.
One may immediately derive these product terms by noting the integration by parts in Eq.~\eqref{eqn:ibp}, 
evaluating the right-hand side via Eqs.~\eqref{eqn:currNLO} and~\eqref{eqn:currNNLODD}, 
and applying the solutions~\eqref{eqn:psisol}.
The results are
\begin{align}
	\label{eqn:Mhatdef}
	\hM{8} & = \frac{\lam1}{\LamB^2} +\frac{6 \lam2 }{\LamB^2(w+1)} - 2(w-1)\hvph - \frac{2 (2 \hat\eta-1) (w-1)}{w+1}\,,\nn\\
	\hM{9} & = \frac{3 \lam2 }{\LamB^2(w+1)} + 2 \hvph - \frac{(2\hat\eta-1)(w-1)}{w+1}\,,\nn\\
	\hM{10} & =\frac{\lam1}{3\LamB^2} -\frac{\lam2 (w+4)}{2\LamB^2(w+1)} + 2 (w+2) \hvphq - \frac{2 \hat\eta-1}{w+1}\,\nn\\ 
	\hM{11} & = \frac{-\lam1 +2 \lam2}{\LamB^2} +2 (w+1) \hvph\,,\nn\\
	\hM{12} & = \frac{2 \lam1}{\LamB^2} - \frac{2\lam2  (2w+1)}{\LamB^2(w+1)} - 4 w \hvph  +\frac{2 (w-1)}{w+1}\,,\nn\\
	\hM{13} & = \frac{\lam1}{3\LamB^2} -\frac{\lam2 w}{2\LamB^2(w+1)} + 2 (w+2) \hvphq + \frac{2 \hat\eta+1}{w+1}\,,\nn\\*
	\hM{14} & = \frac{2 \lam1 (w-2)}{3\LamB^2 (w+1)} -  \frac{\lam2 (w^2-2 w-4)}{\LamB^2(w+1)^2}+ \frac{4 (w^2+2) \hvphq}{w+1} + \frac{4(2\hat\eta + 1) - 2w}{(w+1)^2}\,.
\end{align}
In these results, we have applied Eq.~\eqref{eqn:phpdef} in $\hM{10,13,14}$ in order remove superficial divergences at zero recoil.

\section{$\Bbar \to \Dx$ order $\aS/m_{c,b}$ corrections}
\label{app:asmcb}

Evaluating the matrix elements of Eq.~\eqref{eqn:asec} (and the corresponding ones for the $\aS/m_b$ terms) one may compute
the $\aS/m_{b,c}$ corrections to the form factors.
Defining these corrections via $\hat h_i \to \hat h_i +\haS \, \delta \hat h_i$ with respect to the form factors in Eqs.~\eqref{eqn:BDFFs} and~\eqref{eqn:BDsFFs},
for the $\Bbar \to D$ form factors they read:
\begin{align}
	\label{eqn:BDFFs_asmcb}
	\delta\hat h_S & = (\eb+\ec)\bigg[C_S \bigg(\hL11 - \frac{w-1}{w+1} \hL41\bigg)+2(w-1)C_S'\bigg] + \Big(\eb C_g^b + \ec C_g^c\Big) \hDg1\,, \nn\\
	\delta\hat h_+ & = (\eb+\ec)\bigg\{C_{V_1} \hL11 + \frac{w+1}{2}\big(C_{V_2}+C_{V_3}\big)\bigg[\hL11-\frac{w-1}{w+1}\hL41\bigg] \nn\\*
	& \quad + 2(w-1)\bigg[C_{V_1}' + \frac{w+1}{2}\big(C_{V_2}'+C_{V_3}'\big)\bigg]\bigg\} -  (w-1) \big(\eb C_{V_2} + \ec C_{V_3}\big) \hL51 \nn\\*
	& \quad +\Big(\eb C_g^b + \ec C_g^c\Big) \hDg1\,, \nn\\
	\delta\hat h_- & = (\eb+\ec)\frac{w+1}{2}\bigg\{(C_{V_2}-C_{V_3})\bigg[\hL11 - \frac{w-1}{w+1}\hL41\bigg]+2(w-1)\big[C_{V_2}'-C_{V_3}'\big]\bigg\} \nn\\*
	 & \quad + C_{V_1}(\ec-\eb) \hL41 -  (w+1) \big(\eb C_{V_2} - \ec C_{V_3}\big) \hL51\,, \nn\\
	\delta\hat h_T & = (\eb+\ec)\Big\{\big[C_{T_1}-C_{T_2}+C_{T_3}\big] \hL11 - C_{T_1} \hL41+2(w-1)\big[C_{T_1}'-C_{T_2}'+C_{T_3}'\big]\Big\} \nn\\*
	 & \quad + (\ec-\eb)\big[C_{T_2} + C_{T_3}\big] \hL41 + 2  \big(\eb C_{T_2} - \ec C_{T_3}\big) \hL51 + \Big(\eb C_g^b + \ec C_g^c\Big) \hDg1\,.
\end{align}

For the $\Bbar \to D^*$ form factors they read:
\begin{align}
	\label{eqn:BDsFFs_asmcb}
	\delta\hat h_P 	& =  C_P \Big\{\eb  \big(\hL11-\hL41\big) + \ec  \Big[\hL21+(w-1)\hL31+\hL51-(w+1)\hL61\Big]\Big\} \nn\\*
                    		& \quad + 2(\eb+\ec)(w-1)C_P' + \Big[\eb C_g^b \hDg1 +\ec C_g^c \Big(\hDg2 + (w-1)\hL31\Big)\Big]\,,\nn\\
	\delta\hat h_V 	& = C_{V_1} \Big[\eb \big(\hL11-\hL41\big) +\ec \big(\hL21 -\hL51\big) \Big]  -  \big(\eb C_{V_2} + \ec C_{V_3}\big)\Big[\hL41 - \hL51\Big] \nn\\*
				& \quad + 2(\eb + \ec)(w-1) C_{V_1}' + \Big(\eb C_g^b\hDg1 + \ec C_g^c\hDg2\Big)\,,\nn\\
	\delta\hat h_{A_1} 	& = C_{A_1}\bigg[\eb\bigg(\hL11 - \frac{w-1}{w+1}\hL41\bigg) + \ec \bigg( \hL21 - \frac{w-1}{w+1}\hL51\bigg)\bigg] + 2(w-1)(\eb+\ec) C_{A_1}'  \nn \\*
					& \quad +  \frac{w-1}{w+1}\big(\eb C_{A_2} + \ec C_{A_3}\big)\Big[\hL41 -\hL51\Big] + \Big(\eb C_g^b \hDg1 + \ec C_g^c\hDg2\Big)\,,\nn\\
	\delta\hat h_{A_2} 	& = C_{A_2} \Big\{\eb  \big(\hL11-\hL41\big) + \ec  \Big[\hL21+(w-1)\hL31+\hL51-(w+1)\hL61\Big]\Big\} \nn\\*
					& \quad -\frac{1}{w+1} \bigg[\eb C_{A_2} \Big[\hL41 +(2w -1)\hL51 \Big]+ \ec   C_{A_3} \big(\hL41 -3\hL51 \big)\bigg]\nn\\*
					&\quad +\ec C_{A_1} \big(\hL31 + \hL61\big) +2(w-1)(\eb+\ec)C_{A_2}'+\ec C_g^c \hL31\,,\nn\\
	\delta\hat h_{A_3} 	& = C_{A_1} \Big[\eb  \big(\hL11-\hL41\big) + \ec  \big(\hL21-\hL31-\hL51+\hL61\big)\Big] \nn\\*
    					& \quad +\frac{1}{w+1} \bigg\{\eb C_{A_2} \Big[w \hL41 -(w-2)\hL51 \Big] + \ec C_{A_3} \big(w\hL41 -3w\hL51 \big)\bigg\} \nn\\*
					& \quad + C_{A_3} \Big\{\eb  \big(\hL11-\hL41\big) + \ec  \Big[\hL21+(w-1)\hL31+\hL51-(w+1)\hL61\Big]\Big\} \nn\\*
					& \quad +2(w-1)(\eb+\ec)\big(C_{A_1}'+C_{A_3}'\big)+\Big[\eb C_g^b \hDg1 + \ec C_g^c \big( \hDg2 - \hL31\big)\Big]\,,\nn\\
	\delta\hat h_{T_1} 	& = C_{T_1}\big(\eb \hL11 + \ec \hL21 \big)+\frac{w-1}{2}\big(C_{T_2}-C_{T_3}\big)\Big[\eb \big(\hL11 -\hL41 \big) + \ec \big(\hL21 -\hL51 \big)\Big] \nn\\*
					& \quad + 2(w-1)(\eb+\ec)\bigg[C_{T_1}' + \frac{w-1}{2}\big(C_{T_2}' -C_{T_3}'\big)\bigg] -  (w-1) \big(\eb C_{T_2} - \ec C_{T_3}\big) \hL51 \nn \\*
					& \quad + \Big(\eb C_g^b\hDg1 +\ec C_g^c\hDg2\Big)\,,\nn\\
	\delta\hat h_{T_2} 	& = -C_{T_1}\big(\eb \hL41 - \ec \hL51 \big)+\frac{w+1}{2}\big(C_{T_2}+C_{T_3}\big)\Big[\eb \big(\hL11 -\hL41 \big) + \ec \big(\hL21 -\hL51 \big)\Big] \nn\\*
					& \quad + (\eb C_{T_2} + \ec C_{T_3})\Big(\hL41 - w \hL51\Big) +(w^2-1)(\eb+\ec)\big(C_{T_2}' + C_{T_3}'\big)\,,\nn\\
	\delta\hat h_{T_3} 	& = - \ec C_{T_1} \big( \hL31 - \hL61\big)  + C_{T_2}\Big[\eb \big(\hL11 -\hL41 \big) + \ec \big(\hL21 -\hL51 \big)\Big] \nn\\*
    					& \quad -\frac{1}{w+1} \bigg[\eb C_{T_2} \Big[\hL41 +(2w-1)\hL51\Big]+ \ec C_{T_3} \Big(\hL41 -3\hL51\Big)\bigg] \nn\\*
					& \quad +2(w-1)(\eb+\ec)C_{T_2}' - \ec C_g^c \hL31\,.
\end{align}
As above, the derivatives $C_i' = \partial C_i / \partial w $, $C_g^Q(\mu) = -(3/2)[\ln(m_Q/\mu)-13/9]$~\cite{Eichten:1990vp}
and the $w$-dependent functions
\begin{align}
	\label{eqn:deldef}
    \hDg1 = -4(w-1)\hat \chi_2 + 12 \hat\chi_3\,, \qquad
    \hDg2 = -4\hat\chi_3  \,.
\end{align}
These arise by a simple redefinition of $\hat\chi_{2,3} \to (1+\haS C_g^Q) \hat\chi_{2,3}$ in the $1/m_{b,c}$ contributions 
to the various form factors.
Applying the redefinition in Eq.~\eqref{eqn:replNNLO}, then $\hDg1 = \hL11$ and $\hDg2 = \hL21$, up to $\mathcal{O}(\aS/m_Q^2)$ corrections. 
One can also see explicitly that $\hat\chi_1$ enters in these expressions through $\hL11$ and $\hL21$ as $2 (\eb + \ec) \hat\chi_1$,
as expected from reparametrization invariance, 
so that the redefinition in Eq.~\eqref{eqn:replNNLO} introduces only $\mathcal{O}(\aS/m_Q^2)$ corrections.

\section{Vanishing chromomagnetic limit}
\label{app:vclimit}

In the $G^{\ab} \to 0$ limit, Lagrangian and current insertions reduce to $\mathcal{L}_1 = -\Qbar^v_+ D^2 Q^v_+ $, $\mathcal{L}_2 = 0$, and $\mJ_2 = -D^2$. 
Therefore $\chi_{2,3}$, $\lam2$, and $\varphi_{1,2,3}$ all vanish. 
The Isgur-Wise function $\varphi_0$ does not vanish, however, 
because the $\mJ_2$ current correction in Eq.~\eqref{eqn:currNNLOD2} is still present.
Unlike at $\mathcal{O}(\theta^2)$ in the RC expansion,
the second-order Schwinger-Dyson relations~\eqref{eqn:subSDrelall} include mixed current-Lagrangian matrix elements.
Nonetheless, at zero recoil these additional terms manifestly vanish, 
with two consequences:
Eq.~\eqref{eqn:vphzerorec} still holds, so that one deduces that $\lam1 \to 0$ in this limit. Hence $\varphi_0(1) = \lam1/\LamB^2 \to 0$.
To make this and the analyticity of the matrix elements explicit at $w=1$, 
we write
\begin{equation}
	\label{eqn:vcphi0}
  	\varphi_0 (w) = (w-1)\, \varphi^\quot_0(w)\,,
\end{equation}
where the quotient function $\varphi^\quot_0$ is regular near zero recoil 
(see Eq.~\eqref{eqn:quotdef}; Ref.~\cite{Falk:1992wt} uses the notation $\hat\phi$).
The derivation of second-order power corrections from single insertions of $\mJ_2$, $\mJbar_2'$, 
or mixed insertions of $\mJ_1$ and $\mJbar_1'$ now proceeds as in the RC expansion above,
but imposing Eq.~\eqref{eqn:vcphi0} and that all other $\varphi_i$ functions and $\lam{1,2}$ vanish.
These matrix elements match onto $\hM{8,\ldots,14}$.

To compute corrections from the mixed current-Lagrangian terms, one considers the following matrix elements
\begin{align}
\label{eqn:currlgn}
\big\langle \Hc^{v'} \big|  \cbvp(z) \, \Gamma (i \overrightarrow{D}_\alpha)   \bv(z)  \circ \mathcal{L}_1 \big| \Hb^v \big\rangle = \LamB^2\, \Tr\big[ \Hbar_c(v')  \Gamma H_b(v) E_\alpha(v,v') \big]\,, \nn \\
\big\langle \Hc^{v'} \big|  \cbvp(z) \, (-i \overleftarrow{D}_\alpha)  \Gamma \bv(z)  \circ \mathcal{L}_1 \big| \Hb^v \big\rangle = \LamB^2\, \Tr\big[ \Hbar_c(v')  \Gamma H_b(v) E'_\alpha(v,v') \big]\,. 
\end{align}
The most general decomposition for $E^{(\prime)}_\alpha$ is
\begin{align}
E^{(\prime)}_\alpha(v,v') &= e^{(\prime)}_1(w) v_\alpha + e^{(\prime)}_2(w)  v'_\alpha + e^{(\prime)}_3(w) \gamma_\alpha\,.
\end{align}
Similar expressions hold for their conjugates (corresponding to matrix elements with insertions of $\mathcal{L}_1'$ and swapped derivatives)
with tensor $\genbar{E}^{(\prime)}_\alpha(v',v)$, analogously to Eqs.~\eqref{eqn:currNLO}.
Then Eqs.~\eqref{eqn:cont} and~\eqref{eqn:currNNLOD2} and the equations of motion respectively require
\begin{equation}
	\label{eqn:vEeqns}
	\Pi_-\big(v^\alpha E_\alpha(v,v')\big)\Pi_-' = 2\varphi_0\Pi_-\Pi_-' \,, \qquad \Pi_-\big(v^{\prime\alpha} E'_\alpha(v,v') \big)\Pi_-' = 0 \,,
\end{equation}
i.e., $e_1 +we_2 - e_3 = 2 \varphi_0$ and $w e_1' + e_2' - e_3' = 0$.
Finally, Eqs.~\eqref{eqn:diffJeq}, \eqref{eqn:Jmediff}, and \eqref{eqn:chromoNLO} at $\mathcal{O}(1/m_{c,b})$ impose
\begin{align}
	\label{eqn:EEp}
  	\Pi_-\big(E_\alpha - E'_\alpha\big)\Pi_-' = -(v_\alpha - v'_\alpha) 2\chi_1 \Pi_-\Pi_-'\,.
\end{align}
Under the projectors this equation contains three relations for the $e_i^{(\prime)}$, one of which is simply $e_3' = e_3$.
Together with the two relations in Eq.~\eqref{eqn:vEeqns}, 
and following the choice of Ref.~\cite{Falk:1992wt}, 
one may express all the remaining $e_i^{(\prime)}$ in terms of just $e_3$, leading to
\begin{align}
\label{eq:ei-relations}
    (w+1) e_1 &= e_3 - 2(\varphi^\quot_0 + w\,\chi_1)\,, &
    (w+1) e_2 &= e_3 +2 (w\, \varphi^\quot_0 +\chi_1)\,, \nn \\
    (w+1) e'_1 &= e_3 - 2(\varphi^\quot_0 -  \chi_1) \,, &
    (w+1) e'_2 &= e_3 +2 w \, (\varphi^\quot_0 - \chi_1)\,.
\end{align}
These results may be applied to the mixed current-Lagrangian matrix elements~\eqref{eqn:currlgn}, 
that generate corresponding second-order power corrections in Eq.~\eqref{eqn:QCDmatchexp}.
These in turn match onto $\hL{4,5,6}{2}$ as well as the $\hM{15,\ldots,24}$ wavefunctions, defined in Appendix~\ref{app:mbmc-ffs}.
(As mentioned below Eq.~\eqref{eqn:reps}, there are additional $\hM{i}$ wavefunctions not present in the RC expansion at $\mathcal{O}(\theta^2)$.
Several of these arise in the VC limit.
We also show in Appendix~\ref{app:mbmc-ffs} the contributions from the complete set of $\hM{i}$ wavefunctions to the $B \to \Dx$ form factors.)

The remaining nonvanishing matrix elements generating second order power corrections in Eq.~\eqref{eqn:QCDmatchexp} 
are those with two insertions of $\mathcal{L}_1^{(\prime)}$, 
\begin{align}
\label{eqn:L1L1chromo}
\big\langle \Hc^{v'} \big| \frac{1}{2} \cbvp(z) \, \Gamma \bv(z)  \circ \mathcal{L}_1 \circ \mathcal{L}_1 \big| \Hb^v \big\rangle = -\LamB^2\, \Tr\big[ \Hbar_c(v')  \Gamma H_b(v) C_0(v,v') \big]\,, \nn \\
\big\langle \Hc^{v'} \big| \mathcal{L}'_1 \circ \cbvp(z) \,  \Gamma \bv(z)  \circ \mathcal{L}_1 \big| \Hb^v \big\rangle = -\LamB^2\, \Tr\big[ \Hbar_c(v')  \Gamma H_b(v) D_0(v ,v') \big]\,, 
\end{align}
with two Isgur-Wise functions $C_0(v,v') = 2 c_0(w)$ and $D_0(v,v') = 2 d_0(w)$.
These matrix elements trivially match onto $\hL{1,2}{2}$ and $\hat M_{1,\ldots, 7}$.

The expressions for the nonvanishing $\hL{i}{1,2}$ and $\hM{i}$ in the VC limit are shown in Eqs~\eqref{eqn:hLhMVClim}.

\section{$\Bbar \to \Dx$ order $1/m_cm_b$ corrections}
\label{app:mbmc-ffs}

We provide here the complete set of the mixed $\mathcal{O}(1/m_c m_b)$ traces and their contribution to the form factors, 
according to the notation used in this paper. 
See footnote~\ref{ft:mcorrs} for a summary of (apparent) typographical errors in the expressions in Appendix~A of Ref.~\cite{Falk:1992wt}.

The full expression for the heavy quark bilinear tensor is
\begin{align}
    H^{(1,1)}_{bc}(v,v') & = \Pi_+ \Big\{ P^v \Pbar^{v'}\hM{1}(-\g^5)\g^5   + P^v\Vbar^{v'}(-\g^5) \big[ \hM{2}\slashed{\epsilon}^{\prime*} + \hM{3}\epsilon^{\prime*} \ccdot v \big]  \label{eqn:fullMi} \\*
				 & \qquad \qquad + V^{v}\Pbar^{v'}\big[\hM{2} \slashed{\epsilon} + \hM{3}\epsilon \ccdot v'\big] \g^5  
				   + V^{v}\Vbar^{v'}\big[ \hM{4} \slashed{\epsilon}\slashed{\epsilon}^{\prime*}  + \hM{5} \epsilon \ccdot \epsilon^{\prime*}  \nn\\*
	 			& \qquad \qquad  \qquad + \hM{6}\slashed{\epsilon}\epsilon^{\prime*} \ccdot v + \hM{6}\slashed{\epsilon}^{\prime*}\epsilon \ccdot v' + \hM{7} \epsilon^{\prime*} \ccdot v \epsilon \ccdot v'\big] \Big\} \Pi_+' \nn \\
     & + \Pi_- \Big\{ P^v \Pbar^{v'}\hM{8}(-\g^5)\g^5   + P^v\Vbar^{v'}(-\g^5) \big[ \hM{9}\slashed{\epsilon}^{\prime*} + \hM{10}\epsilon^{\prime*} \ccdot v \big] \nn \\*
				 & \qquad \qquad + V^{v}\Pbar^{v'}\big[\hM{9} \slashed{\epsilon} + \hM{10}\epsilon \ccdot v'\big] \g^5  
				   + V^{v}\Vbar^{v'}\big[ \hM{11} \slashed{\epsilon}\slashed{\epsilon}^{\prime*}  + \hM{12} \epsilon \ccdot \epsilon^{\prime*}  \nn\\*
	 			& \qquad \qquad  \qquad + \hM{13}\slashed{\epsilon}\epsilon^{\prime*} \ccdot v + \hM{13}\slashed{\epsilon}^{\prime*}\epsilon \ccdot v' + \hM{14} \epsilon^{\prime*} \ccdot v \epsilon \ccdot v'\big] \Big\} \Pi_-' \nn \\
	 	& +  \Pi_+ \Big\{ P^v \Pbar^{v'}\hM{15}(-\g^5)\g^5   + P^v\Vbar^{v'}(-\g^5) \big[ \hM{16}\slashed{\epsilon}^{\prime*} + \hM{17}\epsilon^{\prime*} \ccdot v \big] \nn \\*
				 & \qquad \qquad + V^{v}\Pbar^{v'}\big[\hM{18} \slashed{\epsilon} + \hM{19}\epsilon \ccdot v'\big] \g^5  
				   + V^{v}\Vbar^{v'}\big[ \hM{20} \slashed{\epsilon}\slashed{\epsilon}^{\prime*}  + \hM{21} \epsilon \ccdot \epsilon^{\prime*}  \nn\\*
	 			& \qquad \qquad  \qquad + \hM{22}\slashed{\epsilon}\epsilon^{\prime*} \ccdot v + \hM{23}\slashed{\epsilon}^{\prime*}\epsilon \ccdot v' + \hM{24} \epsilon^{\prime*} \ccdot v \epsilon \ccdot v'\big]\Big\} \Pi_-' \nn \\
	 	& +  \Pi_- \Big\{ P^v \Pbar^{v'}\hM{15}(-\g^5)\g^5   + P^v\Vbar^{v'}(-\g^5) \big[ \hM{18}\slashed{\epsilon}^{\prime*} + \hM{19}\epsilon^{\prime*} \ccdot v \big] \nn \\*
				 & \qquad \qquad + V^{v}\Pbar^{v'}\big[\hM{16} \slashed{\epsilon} + \hM{17}\epsilon \ccdot v'\big] \g^5  
				   + V^{v}\Vbar^{v'}\big[ \hM{20} \slashed{\epsilon}\slashed{\epsilon}^{\prime*}  + \hM{21} \epsilon \ccdot \epsilon^{\prime*}  \nn\\*
	 			& \qquad \qquad  \qquad + \hM{23}\slashed{\epsilon}\epsilon^{\prime*} \ccdot v + \hM{22}\slashed{\epsilon}^{\prime*}\epsilon \ccdot v' + \hM{24} \epsilon^{\prime*} \ccdot v \epsilon \ccdot v'\big] \Big\} \Pi_+' \,.\nn
\end{align}
Writing the $\mathcal{O}(\ec\eb)$ terms in the $\hat h_i$ form factors in Eqs.~\eqref{eqn:BDFFs} and~\eqref{eqn:BDsFFs} 
in the form $\ec\eb\ \delta \hat h_i$, then the inclusion of all the wavefunctions $\hat M_i$ modifies these terms such that they become
\begin{align}
\label{eqn:compmcmb}
    & \delta \hat h_S  = \hM{1} + \hM{8} - 2\frac{w-1}{w+1} \hM{15} \, , \nn \\
    & \delta  \hat h_+  = \hM{1} - \hM{8} \, , \nn \\
    & \delta \hat h_-    = 0 \, , \nn \\
    & \delta \hat h_T  = \hM{1} + \hM{8} -2 \hM{15} \, , \nn \\
    & \delta \hat h_P    = \hM{2} - \hM{9} + (w-1)(\hM{3} +\hM{10}) + \hM{16} - \hM{18} - (1+w)(\hM{17} + \hM{19}) \, , \nn \\
    & \delta \hat h_V  = \hM{2} + \hM{9} - (\hM{16} + \hM{18}) \, , \nn \\
    & \delta \hat h_{A_1}   = \hM{2} + \hM{9} - \frac{w-1}{w+1} (\hM{16} + \hM{18}) \, , \nn \\
    & \delta \hat h_{A_2}   = \hM{3} - \hM{10} + (\hM{17} - \hM{19}) \, , \nn \\
    & \delta \hat h_{A_3}  = \hM{2} + \hM{9} - (\hM{3} - \hM{10}) - (\hM{16} + \hM{18}) + (\hM{17} - \hM{19}) \, , \nn \\
    & \delta \hat h_{T_1}   =\hM{2} - \hM{9}  \, , \nn \\
    & \delta \hat h_{T_2}  = \hM{16} - \hM{18} \, , \nn \\
    & \delta \hat h_{T_3}   = (\hM{3} + \hM{10}) - (\hM{17} + \hM{19}) \, , \nn \\
    & \delta \hat h_1    = (\hM{4} - \hM{11}) + (\hM{5} - \hM{12}) \, .
\end{align}

\section{Fits to FNAL/MILC $\Bbar \to D^*$ data for $w>1$}
\label{app:FNAL_Comparison}

\begin{figure}[bp]
	\subfigbottomskip=-8pt
	\subfigcapskip=-8pt
	\subfigure[][\label{fig:fpfnal}]{
        		\centering
		\includegraphics[width=0.42\textwidth]{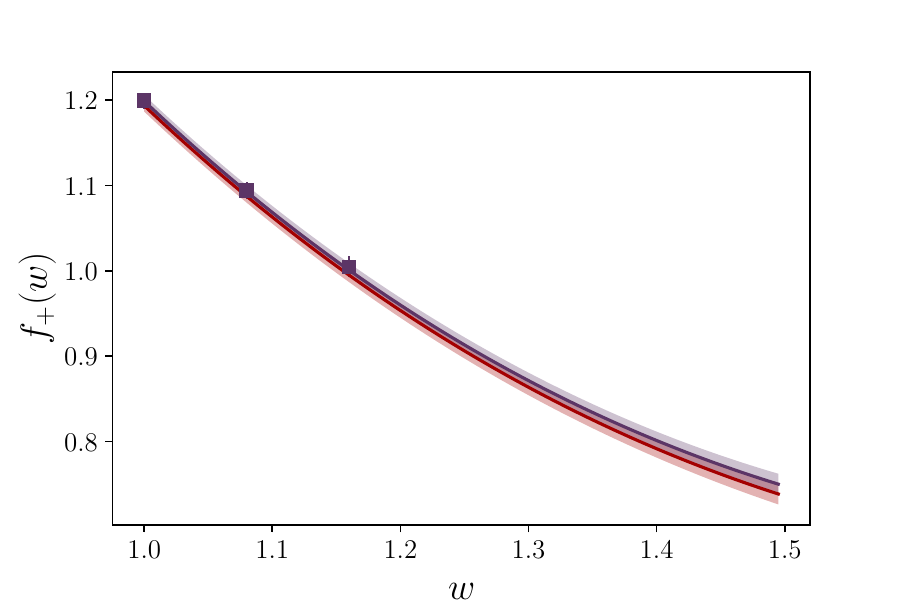}
      }
	\subfigure[][\label{fig:f0fnal}]{
        		\centering
		\includegraphics[width=0.42\textwidth]{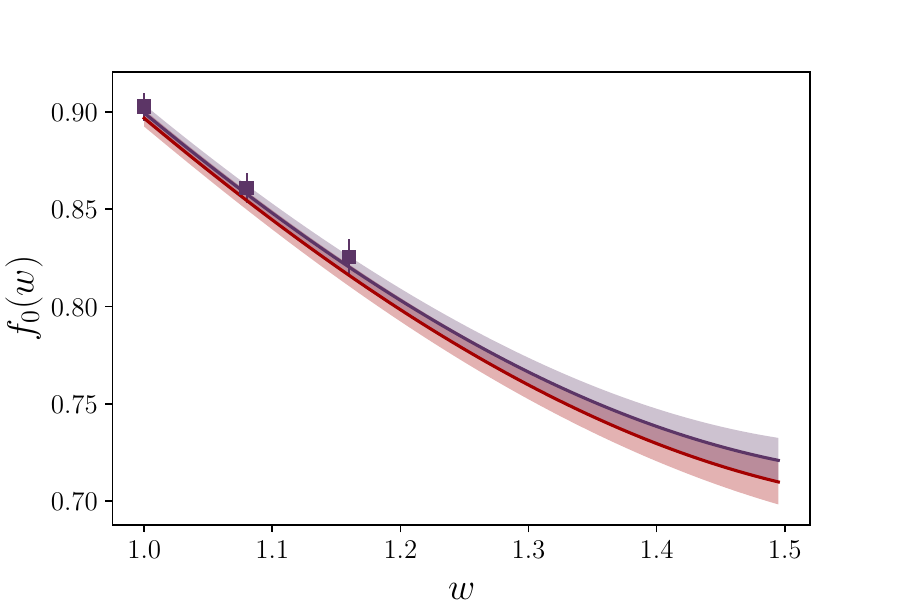}
      }
	\subfigure[][\label{fig:hA1fnal}]{
        		\centering
		\includegraphics[width=0.42\textwidth]{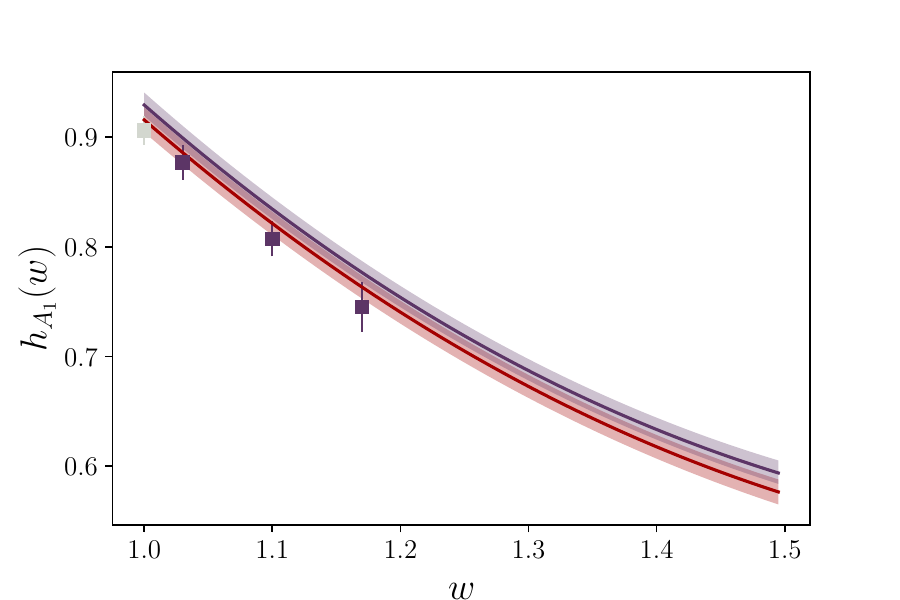}
      }
	\subfigure[][\label{fig:hA2fnal}]{
        		\centering
		\includegraphics[width=0.42\textwidth]{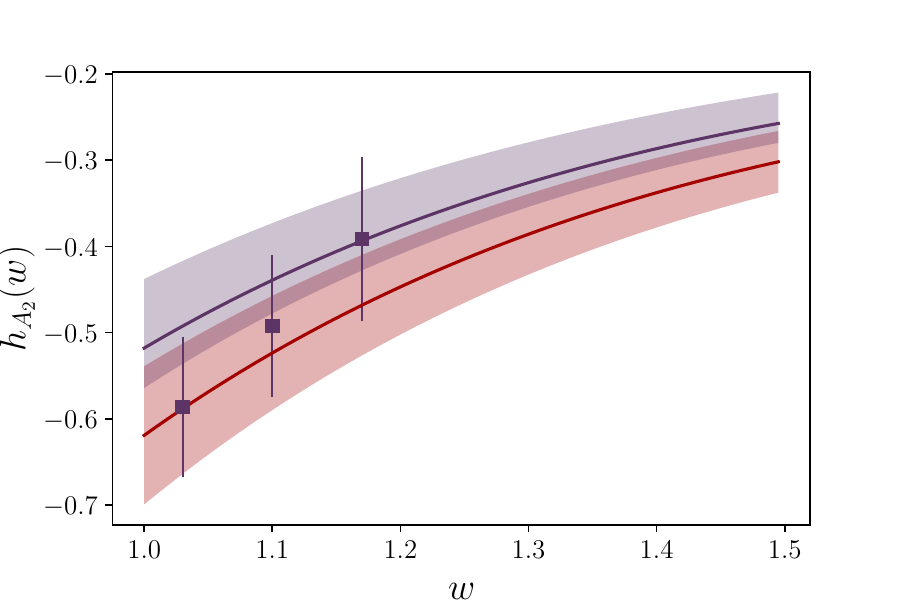}
      }
	\subfigure[][\label{fig:hA3fnal}]{
        		\centering
		\includegraphics[width=0.42\textwidth]{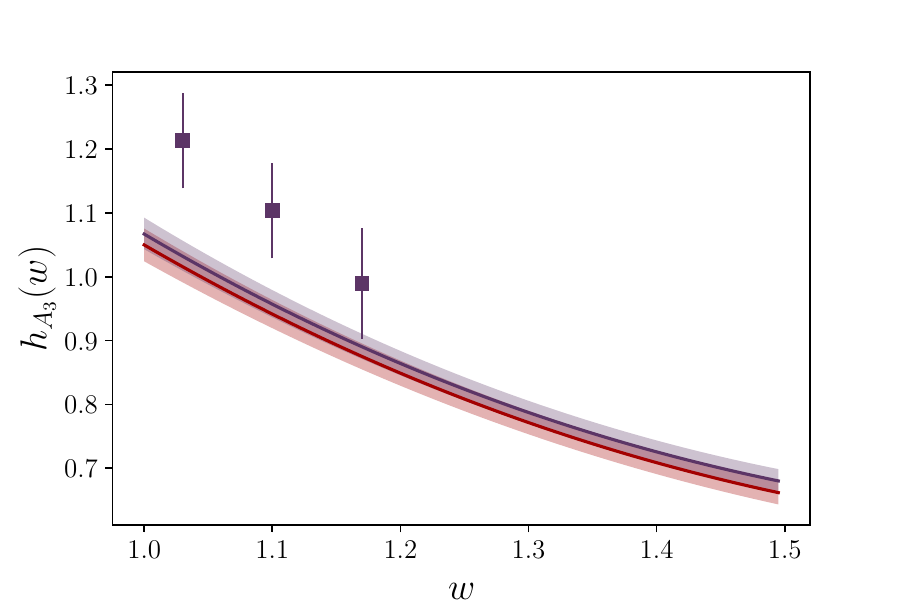}
      }
	\subfigure[][\label{fig:hVfnal}]{
        		\centering
		\includegraphics[width=0.42\textwidth]{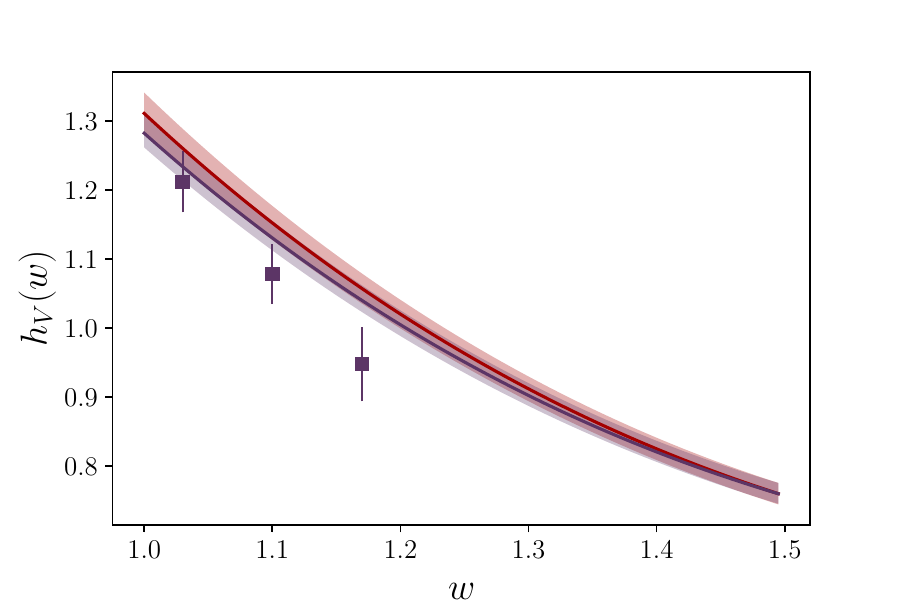}
      }
    \vspace*{6pt}
	\caption{Form factors from fits that include the $B\to D^*$ LQCD predictions~\cite{Bazavov:2021bax} for $w>1$.
	The red band shows the results for the \hAOneFit fit scenario, that uses $h_{A1}$ data only, 
	while the dark plum shows the results for the \AllLQCDFit scenario, 
	that uses the full set of synthetic LQCD points.}
	\label{fig:fit_FNAL_full}
\end{figure}

Figure~\ref{fig:fit_FNAL_full} shows details of the fits of  \hAOneFit and \AllLQCDFit.
The synthetic LQCD data points are compared to the predicted functional forms of $h_{A_{1-3},V}$ and $f_{+,0}$. 
The fits are flexible enough to describe $f_{+,0}$ and $h_{A_{1,2}}$ beyond zero recoil, 
but there is a degree of tension between $h_{A_{3},V}$ and the fit form factors. 
The $\chi^2$ of both fits are $31.2$ and $49.4$ with $33$ and $42$ degrees of freedom, respectively. 
More details can be found in Table~\ref{tab:fitresults}.

\FloatBarrier

\section{Humdrum correlations}

We report in the following tables the correlations for each fit scenario considered in Table~\ref{tab:FitKey} in the RC expansion
and the \ShapeFit fit in the VC limit.

\begin{table}[htp]
\newcolumntype{L}{ >{\arraybackslash $} l <{$}}
\newcolumntype{R}{ >{\raggedleft\arraybackslash $} m{1.5cm} <{$}}
\resizebox{0.95\textwidth}{!}{
\begin{tabular}{L|RRRRRRRRRR}
\text{\BaseFit} & |V_{cb}| & \rhossq & \cs & \mbS~[\GeV] & \dmbc~[\GeV] & \lam2~[\GeV^2] & \hat\eta(1) & \rho_1~[\GeV^3] & \hat\chi_2(1) & \hvphp(1) \\ 
\hline
|V_{cb}| & 1. & 0.256 & 0.248 & 0.030 & 0.082 & 0.033 & 0.352 & -0.463 & -0.224 & 0.148 \\ 
\rhossq & \nax & 1. & 0.357 & -0.720 & 0.107 & 0.034 & 0.421 & -0.075 & -0.473 & -0.632 \\ 
\cs & \nax & \nax & 1. & -0.460 & 0.048 & -0.056 & 0.383 & -0.076 & -0.647 & -0.112 \\ 
\mbS~[\GeV] & \nax & \nax & \nax & 1. & 0.028 & 0.008 & -0.429 & -0.007 & 0.369 & 0.362 \\ 
\dmbc~[\GeV] & \nax & \nax & \nax & \nax & 1. & 0.009 & 0.108 & 0.477 & -0.089 & 0.011 \\ 
\lam2~[\GeV^2] & \nax & \nax & \nax & \nax & \nax & 1. & -0.255 & -0.094 & -0.034 & -0.006 \\ 
\hat\eta(1) & \nax & \nax & \nax & \nax & \nax & \nax & 1. & -0.379 & -0.374 & 0.189 \\ 
\rho_1~[\GeV^3] & \nax & \nax & \nax & \nax & \nax & \nax & \nax & 1. & 0.105 & -0.279 \\ 
\hat\chi_2(1) & \nax & \nax & \nax & \nax & \nax & \nax & \nax & \nax & 1. & 0.305 \\ 
\hvphp(1) & \nax & \nax & \nax & \nax & \nax & \nax & \nax & \nax & \nax & 1. \\ 
\end{tabular}}
\caption{Parameter correlations for the \BaseFit fit scenario in the RC expansion.}
\end{table}

\begin{table}[htp]
\newcolumntype{L}{ >{\arraybackslash $} l <{$}}
\newcolumntype{R}{ >{\raggedleft\arraybackslash $} m{1.5cm} <{$}}
\resizebox{0.95\textwidth}{!}{
\begin{tabular}{L|RRRRRRRRRRRR}
\text{\ShapeFit} & |V_{cb}| & \mathcal{G}(1) & \mathcal{F}(1) & \rhossq & \cs & \mbS~[\GeV] & \dmbc~[\GeV] & \lam2~[\GeV^2] & \hat\eta(1) & \rho_1~[\GeV^3] & \hat\chi_2(1) & \hvphp(1) \\ 
\hline
|V_{cb}| & 1. & -0.215 & -0.627 & 0.078 & 0.056 & 0.059 & -0.007 & -0.012 & -0.006 & 0.021 & 0.069 & 0.049 \\ 
\mathcal{G}(1) & \nax & 1. & 0.110 & 0.128 & -0.201 & -0.007 & 0.002 & 0.001 & -0.117 & -0.006 & -0.054 & -0.380 \\ 
\mathcal{F}(1) & \nax & \nax & 1. & -0.007 & 0.131 & 0.017 & -0.004 & -0.001 & -0.020 & 0.010 & -0.150 & -0.074 \\ 
\rhossq & \nax & \nax & \nax & 1. & 0.394 & -0.564 & 0.207 & -0.033 & 0.632 & -0.484 & -0.665 & -0.406 \\ 
\cs & \nax & \nax & \nax & \nax & 1. & -0.406 & 0.138 & -0.007 & 0.427 & -0.322 & -0.645 & 0.066 \\ 
\mbS~[\GeV] & \nax & \nax & \nax & \nax & \nax & 1. & 0.009 & 0.001 & -0.225 & -0.024 & 0.267 & 0.243 \\ 
\dmbc~[\GeV] & \nax & \nax & \nax & \nax & \nax & \nax & 1. & 0.000 & 0.241 & 0.010 & -0.227 & 0.068 \\ 
\lam2~[\GeV^2] & \nax & \nax & \nax & \nax & \nax & \nax & \nax & 1. & -0.224 & -0.001 & 0.025 & 0.037 \\ 
\hat\eta(1) & \nax & \nax & \nax & \nax & \nax & \nax & \nax & \nax & 1. & -0.807 & -0.731 & 0.250 \\ 
\rho_1~[\GeV^3] & \nax & \nax & \nax & \nax & \nax & \nax & \nax & \nax & \nax & 1. & 0.656 & -0.295 \\ 
\hat\chi_2(1) & \nax & \nax & \nax & \nax & \nax & \nax & \nax & \nax & \nax & \nax & 1. & 0.061 \\ 
\hvphp(1) & \nax & \nax & \nax & \nax & \nax & \nax & \nax & \nax & \nax & \nax & \nax & 1. \\ 
\end{tabular}}
\caption{Parameter correlations for the \ShapeFit fit scenario in the RC expansion.}
\end{table}

\begin{table}[htp]
\newcolumntype{L}{ >{\arraybackslash $} l <{$}}
\newcolumntype{R}{ >{\raggedleft\arraybackslash $} m{1.5cm} <{$}}
\resizebox{0.95\textwidth}{!}{
\begin{tabular}{L|RRRRRRRRRR}
\text{\NoLQCDFit} & |V_{cb}| & \rhossq & \cs & \mbS~[\GeV] & \dmbc~[\GeV] & \lam2~[\GeV^2] & \hat\eta(1) & \rho_1~[\GeV^3] & \hat\chi_2(1) & \hvphp(1) \\ 
\hline
|V_{cb}| & 1. & 0.317 & 0.487 & 0.041 & 0.209 & 0.006 & 0.453 & -0.754 & -0.315 & -0.025 \\ 
\rhossq & \nax & 1. & 0.106 & -0.212 & 0.049 & 0.000 & 0.705 & -0.112 & -0.868 & -0.690 \\ 
\cs & \nax & \nax & 1. & -0.326 & 0.084 & 0.004 & -0.024 & -0.202 & 0.130 & 0.170 \\ 
\mbS~[\GeV] & \nax & \nax & \nax & 1. & -0.001 & 0.000 & -0.130 & -0.002 & 0.012 & -0.398 \\ 
\dmbc~[\GeV] & \nax & \nax & \nax & \nax & 1. & -0.000 & 0.122 & 0.003 & -0.059 & 0.087 \\ 
\lam2~[\GeV^2] & \nax & \nax & \nax & \nax & \nax & 1. & -0.002 & 0.001 & -0.007 & 0.024 \\ 
\hat\eta(1) & \nax & \nax & \nax & \nax & \nax & \nax & 1. & -0.354 & -0.896 & -0.206 \\ 
\rho_1~[\GeV^3] & \nax & \nax & \nax & \nax & \nax & \nax & \nax & 1. & 0.214 & -0.118 \\ 
\hat\chi_2(1) & \nax & \nax & \nax & \nax & \nax & \nax & \nax & \nax & 1. & 0.570 \\ 
\hvphp(1) & \nax & \nax & \nax & \nax & \nax & \nax & \nax & \nax & \nax & 1. \\ 
\end{tabular}}
\caption{Parameter correlations for the \NoLQCDFit fit scenario in the RC expansion.}
\end{table}

\begin{table}[htp]
\newcolumntype{L}{ >{\arraybackslash $} l <{$}}
\newcolumntype{R}{ >{\raggedleft\arraybackslash $} m{1.5cm} <{$}}
\resizebox{0.95\textwidth}{!}{
\begin{tabular}{L|RRRRRRRRRR}
\text{\AllLQCDFit} & |V_{cb}| & \rhossq & \cs & \mbS~[\GeV] & \dmbc~[\GeV] & \lam2~[\GeV^2] & \hat\eta(1) & \rho_1~[\GeV^3] & \hat\chi_2(1) & \hvphp(1) \\ 
\hline
|V_{cb}| & 1. & 0.222 & 0.160 & 0.183 & 0.085 & 0.043 & 0.294 & -0.481 & -0.148 & 0.185 \\ 
\rhossq & \nax & 1. & 0.113 & -0.523 & 0.056 & 0.032 & 0.228 & -0.036 & -0.373 & -0.681 \\ 
\cs & \nax & \nax & 1. & -0.242 & -0.004 & -0.063 & 0.215 & -0.019 & -0.541 & -0.046 \\ 
\mbS~[\GeV] & \nax & \nax & \nax & 1. & 0.142 & 0.028 & -0.200 & -0.114 & 0.170 & 0.390 \\ 
\dmbc~[\GeV] & \nax & \nax & \nax & \nax & 1. & 0.009 & 0.067 & 0.429 & -0.056 & 0.037 \\ 
\lam2~[\GeV^2] & \nax & \nax & \nax & \nax & \nax & 1. & -0.295 & -0.106 & -0.030 & 0.009 \\ 
\hat\eta(1) & \nax & \nax & \nax & \nax & \nax & \nax & 1. & -0.339 & -0.241 & 0.248 \\ 
\rho_1~[\GeV^3] & \nax & \nax & \nax & \nax & \nax & \nax & \nax & 1. & 0.065 & -0.310 \\ 
\hat\chi_2(1) & \nax & \nax & \nax & \nax & \nax & \nax & \nax & \nax & 1. & 0.308 \\ 
\hvphp(1) & \nax & \nax & \nax & \nax & \nax & \nax & \nax & \nax & \nax & 1. \\ 
\end{tabular}}
\caption{Parameter correlations for the \AllLQCDFit fit scenario in the RC expansion.}
\end{table}

\begin{table}[htp]
\newcolumntype{L}{ >{\arraybackslash $} l <{$}}
\newcolumntype{R}{ >{\raggedleft\arraybackslash $} m{1.5cm} <{$}}
\resizebox{0.95\textwidth}{!}{
\begin{tabular}{L|RRRRRRRRRRR}
\text{\BaseFit[NLO]} & |V_{cb}| & \rhossq & \mathcal{G}(1) & \mathcal{F}(1) & \hat\chi_2(1) & \hat\chi_2'(1) & \hat\chi_3'(1) & \hat\eta(1) & \hat\eta'(1) & \mbS~[\GeV] & \dmbc~[\GeV] \\ 
\hline
|V_{cb}| & 1. & 0.256 & -0.262 & -0.625 & 0.014 & 0.061 & 0.127 & 0.050 & 0.218 & 0.035 & 0.003 \\ 
\rhossq & \nax & 1. & -0.201 & -0.014 & 0.028 & -0.042 & 0.575 & 0.558 & 0.228 & -0.689 & 0.008 \\ 
\mathcal{G}(1) & \nax & \nax & 1. & 0.131 & 0.040 & 0.028 & -0.083 & -0.225 & -0.316 & -0.049 & 0.002 \\ 
\mathcal{F}(1) & \nax & \nax & \nax & 1. & -0.002 & 0.017 & 0.089 & -0.018 & -0.069 & 0.021 & 0.001 \\ 
\hat\chi_2(1) & \nax & \nax & \nax & \nax & 1. & -0.082 & 0.470 & -0.117 & -0.135 & 0.064 & -0.002 \\ 
\hat\chi_2'(1) & \nax & \nax & \nax & \nax & \nax & 1. & 0.141 & -0.091 & -0.117 & 0.056 & -0.002 \\ 
\hat\chi_3'(1) & \nax & \nax & \nax & \nax & \nax & \nax & 1. & 0.266 & 0.212 & -0.166 & 0.000 \\ 
\hat\eta(1) & \nax & \nax & \nax & \nax & \nax & \nax & \nax & 1. & -0.053 & -0.457 & -0.003 \\ 
\hat\eta'(1) & \nax & \nax & \nax & \nax & \nax & \nax & \nax & \nax & 1. & 0.174 & -0.009 \\ 
\mbS~[\GeV] & \nax & \nax & \nax & \nax & \nax & \nax & \nax & \nax & \nax & 1. & 0.000 \\ 
\dmbc~[\GeV] & \nax & \nax & \nax & \nax & \nax & \nax & \nax & \nax & \nax & \nax & 1. \\ 
\end{tabular}}
\caption{Parameter correlations for the \BaseFit[NLO] fit scenario in the RC expansion.}
\end{table}

\begin{table}[htp]
\newcolumntype{L}{ >{\arraybackslash $} l <{$}}
\newcolumntype{R}{ >{\raggedleft\arraybackslash $} m{1.5cm} <{$}}
\resizebox{0.95\textwidth}{!}{
\begin{tabular}{L|RRRRRRRRRRRRRRR}
\text{\BaseFit[+SR]} & |V_{cb}| & \rhossq & \cs & \hat\chi_2(1) & \hat\chi_2'(1) & \hat\chi_3'(1) & \hat\eta(1) & \hat\eta'(1) & \mbS~[\GeV] & \dmbc~[\GeV] & \hat\beta_2(1) & \hat\beta_3'(1) & \hvphp(1) & \lam2~[\GeV^2] & \rho_1~[\GeV^3] \\ 
\hline
|V_{cb}| & 1. & 0.392 & 0.281 & 0.003 & 0.023 & -0.010 & 0.308 & 0.078 & 0.045 & 0.068 & 0.100 & 0.184 & 0.016 & 0.040 & -0.426 \\ 
\rhossq & \nax & 1. & 0.347 & -0.053 & -0.160 & 0.155 & 0.211 & 0.305 & -0.485 & 0.055 & 0.092 & 0.192 & -0.460 & 0.035 & -0.025 \\ 
\cs & \nax & \nax & 1. & -0.050 & 0.186 & 0.147 & 0.326 & 0.143 & -0.437 & 0.037 & -0.132 & -0.051 & -0.179 & -0.049 & -0.053 \\ 
\hat\chi_2(1) & \nax & \nax & \nax & 1. & 0.001 & 0.000 & 0.001 & 0.000 & -0.003 & 0.000 & -0.130 & 0.002 & -0.005 & -0.000 & 0.001 \\ 
\hat\chi_2'(1) & \nax & \nax & \nax & \nax & 1. & -0.002 & 0.000 & -0.009 & 0.013 & -0.001 & -0.160 & -0.128 & 0.013 & -0.003 & -0.004 \\ 
\hat\chi_3'(1) & \nax & \nax & \nax & \nax & \nax & 1. & -0.004 & -0.000 & 0.009 & -0.000 & 0.003 & -0.360 & 0.014 & 0.000 & -0.002 \\ 
\hat\eta(1) & \nax & \nax & \nax & \nax & \nax & \nax & 1. & 0.012 & -0.371 & 0.103 & -0.091 & -0.035 & 0.084 & -0.254 & -0.380 \\ 
\hat\eta'(1) & \nax & \nax & \nax & \nax & \nax & \nax & \nax & 1. & 0.005 & -0.002 & -0.803 & -0.655 & -0.865 & -0.004 & 0.000 \\ 
\mbS~[\GeV] & \nax & \nax & \nax & \nax & \nax & \nax & \nax & \nax & 1. & 0.033 & 0.045 & -0.037 & 0.200 & -0.003 & -0.035 \\ 
\dmbc~[\GeV] & \nax & \nax & \nax & \nax & \nax & \nax & \nax & \nax & \nax & 1. & -0.020 & -0.008 & 0.004 & 0.009 & 0.483 \\ 
\hat\beta_2(1) & \nax & \nax & \nax & \nax & \nax & \nax & \nax & \nax & \nax & \nax & 1. & 0.896 & 0.737 & 0.005 & 0.018 \\ 
\hat\beta_3'(1) & \nax & \nax & \nax & \nax & \nax & \nax & \nax & \nax & \nax & \nax & \nax & 1. & 0.577 & 0.012 & 0.011 \\ 
\hvphp(1) & \nax & \nax & \nax & \nax & \nax & \nax & \nax & \nax & \nax & \nax & \nax & \nax & 1. & 0.001 & -0.143 \\ 
\lam2~[\GeV^2] & \nax & \nax & \nax & \nax & \nax & \nax & \nax & \nax & \nax & \nax & \nax & \nax & \nax & 1. & -0.099 \\ 
\rho_1~[\GeV^3] & \nax & \nax & \nax & \nax & \nax & \nax & \nax & \nax & \nax & \nax & \nax & \nax & \nax & \nax & 1. \\ 
\end{tabular}}
\caption{Parameter correlations for the \BaseFit[+SR] fit scenario in the RC expansion.}
\end{table}

\begin{table}[htp]
\newcolumntype{L}{ >{\arraybackslash $} l <{$}}
\newcolumntype{R}{ >{\raggedleft\arraybackslash $} m{1.5cm} <{$}}
\resizebox{0.95\textwidth}{!}{
\begin{tabular}{L|RRRRRRRRRR}
\text{\BaseFit[17]} & |V_{cb}| & \rhossq & \cs & \mbS~[\GeV] & \dmbc~[\GeV] & \lam2~[\GeV^2] & \hat\eta(1) & \rho_1~[\GeV^3] & \hat\chi_2(1) & \hvphp(1) \\ 
\hline
|V_{cb}| & 1. & 0.394 & 0.188 & 0.008 & 0.042 & -0.007 & 0.238 & -0.258 & -0.339 & 0.085 \\ 
\rhossq & \nax & 1. & 0.540 & -0.486 & 0.048 & -0.033 & 0.312 & -0.007 & -0.662 & -0.395 \\ 
\cs & \nax & \nax & 1. & -0.457 & 0.020 & -0.125 & 0.414 & -0.003 & -0.509 & -0.110 \\ 
\mbS~[\GeV] & \nax & \nax & \nax & 1. & 0.046 & 0.018 & -0.406 & -0.037 & -0.071 & 0.247 \\ 
\dmbc~[\GeV] & \nax & \nax & \nax & \nax & 1. & 0.006 & 0.101 & 0.489 & -0.015 & 0.029 \\ 
\lam2~[\GeV^2] & \nax & \nax & \nax & \nax & \nax & 1. & -0.265 & -0.086 & 0.055 & 0.013 \\ 
\hat\eta(1) & \nax & \nax & \nax & \nax & \nax & \nax & 1. & -0.373 & -0.205 & 0.269 \\ 
\rho_1~[\GeV^3] & \nax & \nax & \nax & \nax & \nax & \nax & \nax & 1. & 0.055 & -0.296 \\ 
\hat\chi_2(1) & \nax & \nax & \nax & \nax & \nax & \nax & \nax & \nax & 1. & 0.111 \\ 
\hvphp(1) & \nax & \nax & \nax & \nax & \nax & \nax & \nax & \nax & \nax & 1. \\ 
\end{tabular}}
\caption{Parameter correlations for the \BaseFit[17] fit scenario in the RC expansion.}
\end{table}

\begin{table}[htp]
\newcolumntype{L}{ >{\arraybackslash $} l <{$}}
\newcolumntype{R}{ >{\raggedleft\arraybackslash $} m{1.5cm} <{$}}
\resizebox{0.95\textwidth}{!}{
\begin{tabular}{L|RRRRRRRRRR}
\text{\BaseFit[19]} & |V_{cb}| & \rhossq & \cs & \mbS~[\GeV] & \dmbc~[\GeV] & \lam2~[\GeV^2] & \hat\eta(1) & \rho_1~[\GeV^3] & \hat\chi_2(1) & \hvphp(1) \\ 
\hline
|V_{cb}| & 1. & 0.247 & 0.217 & 0.031 & 0.073 & 0.024 & 0.335 & -0.429 & -0.184 & 0.144 \\ 
\rhossq & \nax & 1. & 0.364 & -0.725 & 0.112 & 0.032 & 0.424 & -0.070 & -0.489 & -0.633 \\ 
\cs & \nax & \nax & 1. & -0.464 & 0.053 & -0.055 & 0.385 & -0.075 & -0.655 & -0.120 \\ 
\mbS~[\GeV] & \nax & \nax & \nax & 1. & 0.019 & 0.008 & -0.432 & -0.010 & 0.392 & 0.372 \\ 
\dmbc~[\GeV] & \nax & \nax & \nax & \nax & 1. & 0.008 & 0.111 & 0.477 & -0.093 & 0.007 \\ 
\lam2~[\GeV^2] & \nax & \nax & \nax & \nax & \nax & 1. & -0.256 & -0.096 & -0.034 & -0.006 \\ 
\hat\eta(1) & \nax & \nax & \nax & \nax & \nax & \nax & 1. & -0.378 & -0.379 & 0.181 \\ 
\rho_1~[\GeV^3] & \nax & \nax & \nax & \nax & \nax & \nax & \nax & 1. & 0.100 & -0.282 \\ 
\hat\chi_2(1) & \nax & \nax & \nax & \nax & \nax & \nax & \nax & \nax & 1. & 0.313 \\ 
\hvphp(1) & \nax & \nax & \nax & \nax & \nax & \nax & \nax & \nax & \nax & 1. \\ 
\end{tabular}}
\caption{Parameter correlations for the \BaseFit[19] fit scenario in the RC expansion.}
\end{table}

\begin{table}[htp]
\newcolumntype{L}{ >{\arraybackslash $} l <{$}}
\newcolumntype{R}{ >{\raggedleft\arraybackslash $} m{1.5cm} <{$}}
\resizebox{0.95\textwidth}{!}{
\begin{tabular}{L|RRRRRRRRRR}
\text{\hAOneFit} & |V_{cb}| & \rhossq & \cs & \mbS~[\GeV] & \dmbc~[\GeV] & \lam2~[\GeV^2] & \hat\eta(1) & \rho_1~[\GeV^3] & \hat\chi_2(1) & \hvphp(1) \\ 
\hline
|V_{cb}| & 1. & 0.263 & 0.230 & 0.021 & 0.101 & 0.059 & 0.404 & -0.524 & -0.226 & 0.205 \\ 
\rhossq & \nax & 1. & 0.345 & -0.723 & 0.114 & 0.046 & 0.418 & -0.108 & -0.452 & -0.580 \\ 
\cs & \nax & \nax & 1. & -0.467 & 0.050 & -0.051 & 0.370 & -0.083 & -0.627 & -0.073 \\ 
\mbS~[\GeV] & \nax & \nax & \nax & 1. & 0.024 & 0.002 & -0.410 & 0.006 & 0.345 & 0.323 \\ 
\dmbc~[\GeV] & \nax & \nax & \nax & \nax & 1. & 0.016 & 0.126 & 0.403 & -0.093 & 0.030 \\ 
\lam2~[\GeV^2] & \nax & \nax & \nax & \nax & \nax & 1. & -0.221 & -0.122 & -0.046 & 0.007 \\ 
\hat\eta(1) & \nax & \nax & \nax & \nax & \nax & \nax & 1. & -0.449 & -0.364 & 0.259 \\ 
\rho_1~[\GeV^3] & \nax & \nax & \nax & \nax & \nax & \nax & \nax & 1. & 0.134 & -0.327 \\ 
\hat\chi_2(1) & \nax & \nax & \nax & \nax & \nax & \nax & \nax & \nax & 1. & 0.259 \\ 
\hvphp(1) & \nax & \nax & \nax & \nax & \nax & \nax & \nax & \nax & \nax & 1. \\ 
\end{tabular}}
\caption{Parameter correlations for the \hAOneFit fit scenario in the RC expansion.}
\end{table}

\begin{table}[htp]
\newcolumntype{L}{ >{\arraybackslash $} l <{$}}
\newcolumntype{R}{ >{\raggedleft\arraybackslash $} m{1.5cm} <{$}}
\resizebox{0.95\textwidth}{!}{
\begin{tabular}{L|RRRRRRRRRR}
\text{\ShapeFit} & |V_{cb}| & \mathcal{G}(1) & \mathcal{F}(1) & \rhossq & \cs & \mbS~[\GeV] & \dmbc~[\GeV] & \hat\eta(1) & \hat\eta'(1) & \hat\phi_0'(1) \\ 
\hline
|V_{cb}| & 1. & -0.256 & -0.630 & 0.259 & 0.092 & 0.126 & -0.024 & 0.004 & 0.249 & -0.027 \\ 
\mathcal{G}(1) & \nax & 1. & 0.094 & 0.066 & -0.285 & 0.021 & -0.002 & 0.122 & -0.314 & 0.317 \\ 
\mathcal{F}(1) & \nax & \nax & 1. & -0.208 & 0.107 & 0.045 & -0.007 & -0.150 & -0.105 & -0.140 \\ 
\rhossq & \nax & \nax & \nax & 1. & -0.363 & -0.026 & 0.027 & 0.704 & 0.043 & 0.735 \\ 
\cs & \nax & \nax & \nax & \nax & 1. & -0.170 & 0.051 & -0.348 & -0.045 & -0.343 \\ 
\mbS~[\GeV] & \nax & \nax & \nax & \nax & \nax & 1. & 0.025 & -0.072 & 0.264 & -0.090 \\ 
\dmbc~[\GeV] & \nax & \nax & \nax & \nax & \nax & \nax & 1. & 0.014 & -0.116 & -0.028 \\ 
\hat\eta(1) & \nax & \nax & \nax & \nax & \nax & \nax & \nax & 1. & -0.298 & 0.779 \\ 
\hat\eta'(1) & \nax & \nax & \nax & \nax & \nax & \nax & \nax & \nax & 1. & -0.277 \\ 
\hat\phi_0'(1) & \nax & \nax & \nax & \nax & \nax & \nax & \nax & \nax & \nax & 1. \\ 
\end{tabular}}
\caption{Parameter correlations for the \ShapeFit fit scenario in the VC limit.}
\end{table}

\FloatBarrier

\addtocontents{toc}{\protect\vspace*{10pt}}

\end{document}